\documentclass[pdflatex,sn-basic]{sn-jnl}

\usepackage{graphicx}
\usepackage{multirow}
\usepackage{amsmath,amssymb,amsfonts}
\usepackage{amsthm}
\usepackage{mathrsfs}
\usepackage[title]{appendix}
\usepackage{xcolor}
\usepackage{textcomp}
\usepackage{manyfoot}
\usepackage{booktabs}
\usepackage{placeins}

\begin{document}

\title[GraphCast Skill and Systematic Biases in Indian Summer Monsoon Forecasts]{GraphCast Skill and Systematic Biases in Indian Summer Monsoon Forecasts: Evaluation Against ERA5 and IMERG}

\author[1]{\fnm{Somnath} \sur{Luitel}}

\author*[2]{\fnm{Parthasarathi} \sur{Mukhopadhyay}}\email{parthasarathi64@gmail.com}

\author[1]{\fnm{Manmeet} \sur{Singh}}

\author[2]{\fnm{Sandeep} \sur{Juneja}}

\author[2]{\fnm{Saptarishi} \sur{Dhanuka}}

\affil[1]{\orgdiv{Department of Earth, Environmental, and Atmospheric Sciences}, \orgname{Western Kentucky University}, \orgaddress{\city{Bowling Green}, \state{KY}, \country{USA}}}

\affil[2]{\orgname{Ashoka University}, \orgaddress{\city{Sonipat}, \state{Haryana}, \country{India}}}

\abstract{The Indian Summer Monsoon (ISM) represents one of the most consequential and dynamically complex phenomena in the global climate system, yet its accurate prediction remains challenging for both physics-based and emerging data-driven models. This study presents a systematic climatological evaluation of Google's GraphCast, a machine-learning-based global weather prediction system, against ERA-5 reanalysis and the Integrated Multi-satellitE Retrievals for GPM (IMERG) observational precipitation dataset for the boreal summer season (June--September, JJAS) over the period 2021--2024. We analyze deterministic GraphCast forecasts initialized at 00 UTC using four 6-hourly lead times, composited to assess +24h, +48h, and +72h forecast performance across the full ISM domain. The evaluation encompasses the climatological mean state (spatial distribution of rainfall, surface temperature, and low-level winds), the distribution of rainfall by intensity class, thermodynamic drivers (tropospheric temperature gradient, DTT; apparent heat source and moisture sink, Q1 and Q2), monsoon dynamics (vertical wind shear, meridional structure), and rainfall variability across timescales (intraseasonal, synoptic, and spectral power). Results demonstrate that GraphCast reproduces the broad spatial pattern of monsoon rainfall and the timing of the annual cycle with considerable fidelity at short leads, with a near-zero domain-averaged rainfall bias over the core monsoon zone ($-$0.7 to $-$1.2~mm~day$^{-1}$) and a high area-weighted daily temporal correlation against IMERG (0.89--0.92). Rainfall temporal variance is moderately, but fairly uniformly, suppressed across timescales (regional power-spectrum variance ratio $\approx$0.63 relative to IMERG), and the rainfall intensity distribution is right-shifted relative to observations, with high-intensity events and the 95th-percentile threshold both systematically underrepresented. GraphCast reproduces the vertical structure and seasonal evolution of the apparent heat source and moisture sink (Q1, Q2) closely, while showing a degraded representation of northward-propagating intraseasonal variability. Collectively, the results point to a moderate, broadly consistent underrepresentation of rainfall intensity and variance in GraphCast's precipitation field, rather than the pronounced wet bias and variance collapse suggested by an earlier, uncorrected observational baseline, and provide benchmark diagnostics for next-generation AI weather models and their operational utility for tropical extended-range forecasting.}


\maketitle

\section*{Highlights}
\begin{itemize}
\item GraphCast 00 UTC deterministic forecasts (2021--2024) reproduce the climatological mean-state spatial pattern of Indian Summer Monsoon rainfall with a near-zero domain-averaged bias and high day-to-day temporal correlation against IMERG, alongside a moderate, fairly uniform suppression of rainfall variance across timescales (regional spectral power ratio $\approx$0.63 relative to IMERG).
\item GraphCast systematically underrepresents high-intensity rainfall events relative to IMERG across all sub-regions examined, including the 95th-percentile threshold, consistent with a deterministic-model smoothing/regression-to-the-mean signature that also explains the suppressed temporal variance.
\item The AI model closely reproduces the large-scale thermodynamic drivers of the monsoon, including the DTT annual cycle and the vertical structure and seasonal evolution of the Q1/Q2 heating and moistening profiles.
\item Intraseasonal variability and northward propagation of the BSISO are degraded at +72h lead, highlighting the current limits of purely data-driven forecasting for tropical subseasonal dynamics.
\end{itemize}

\section{Introduction}\label{sec:intro}

The Indian Summer Monsoon (ISM) is the dominant mode of precipitation variability across South Asia, delivering approximately 80\% of annual rainfall between June and September and sustaining the livelihoods of over a billion people \citep{Gadgil2003,Wang2006}. Skillful prediction of ISM onset, active-break cycles, and extreme rainfall events therefore carries substantial societal and economic relevance, bearing directly on agricultural planning, water resource management, and disaster preparedness across the region. Despite decades of advancement in numerical weather prediction (NWP), the ISM continues to pose fundamental challenges to physics-based models. These challenges arise from complex interactions among the large-scale monsoonal circulation, mesoscale orographic systems (particularly the Western Ghats and the Himalayan foothills), organized convection, and ocean--atmosphere coupling on intraseasonal timescales \citep{Sperber2013,Goswami2005}. Coupled general circulation models participating in successive phases of the Coupled Model Intercomparison Project continue to exhibit systematic dry biases over the Western Ghats, a weak-monsoon-trough signature in the Bay of Bengal, and a persistent double-ITCZ-like displacement of the rainband over the equatorial Indian Ocean \citep{Sperber2013}, biases that have proven difficult to eliminate through conventional physical parameterization improvements alone.

The recent emergence of machine-learning-based global weather prediction systems has introduced a fundamentally different modeling paradigm and has generated considerable scientific interest in whether data-driven approaches can supplant or complement conventional NWP. Pangu-Weather \citep{Bi2023} and FourCastNet \citep{Pathak2022} were among the first such systems to demonstrate that purely data-driven models, trained end-to-end on reanalysis data without any explicit representation of atmospheric physics, could match or exceed the deterministic skill of operational NWP on standard large-scale fields. Among these, Google's GraphCast \citep{Lam2023} represents a landmark development: a graph neural network trained on four decades of ERA-5 reanalysis data that produces 10-day global forecasts at 0.25\textdegree\ horizontal resolution in under one minute of wall-clock time. GraphCast has demonstrated competitive or superior performance relative to the ECMWF operational analysis on standard verification metrics such as the anomaly correlation coefficient (ACC) and root-mean-square error (RMSE) for mid-latitude geopotential height and wind fields \citep{Lam2023}. Standardized benchmarks such as WeatherBench~2 \citep{Rasp2024} have since accelerated headline-metric intercomparison of these systems, but such global-mean skill scores can mask process-level regional biases. In particular, recent work shows that current data-driven models tend to produce overly smooth forecasts with suppressed small-scale spectral energy that worsens with forecast lead time \citep{Bonavita2024}, a tendency that is of direct relevance to any regional evaluation of rainfall, a field dominated by small-scale, high-amplitude convective variability. Systematic assessments of GraphCast's fidelity in simulating the tropical monsoon climatology, including its thermodynamic structure and subseasonal variability, nonetheless remain limited, and the handful of global skill-score intercomparisons published to date say little about whether the model reproduces the specific physical mechanisms that govern monsoon rainfall.

Evaluating AI-based forecast models against observational and reanalysis benchmarks in a region-specific, process-based manner is essential for several reasons. First, neural networks trained on reanalysis data may learn statistically robust large-scale patterns while failing to capture the physical processes that generate organized convection, orographic precipitation enhancement, and moisture transport, all of which are critical to ISM dynamics. Second, AI architectures lack explicit conservation laws for mass, energy, and moisture, raising concerns about systematic biases that could accumulate over extended forecast leads \citep{Lam2023,Pathak2022}. Third, the intraseasonal variability of the monsoon, particularly the Boreal Summer Intraseasonal Oscillation \citep[BSISO;][]{Lee2013} and its northward-propagating counterpart, constitutes the primary source of predictability on 2--4 week timescales, precisely the range where AI models may offer the greatest advantage over deterministic NWP if their biases in representing this variability are well characterized.

This study addresses these gaps by conducting, to our knowledge, the first comprehensive climatological evaluation of GraphCast deterministic hindcasts initialized at 00 UTC over 2021--2024 against ERA-5 reanalysis \citep{Hersbach2020} and IMERG \citep{Huffman2020} precipitation retrievals, specifically for the Indian Summer Monsoon. Our analysis is organized around three complementary diagnostic frameworks: (i) the climatological mean state and annual cycle, (ii) the thermodynamic and dynamic drivers of monsoon variability as captured by Q1/Q2 heating profiles and the Differential Tropospheric Temperature (DTT) index, and (iii) rainfall variability across synoptic to intraseasonal timescales. By mapping GraphCast biases onto physically interpretable metrics rather than relying solely on aggregate skill scores, we aim to provide actionable benchmarks for the ongoing development of AI weather prediction systems over the South Asian monsoon domain.

The paper is organized as follows. Section~\ref{sec:data} describes the data and methods, including details of the GraphCast forecast configuration, reference datasets, and diagnostic metrics. Section~\ref{sec:meanstate} presents the climatological mean-state evaluation. Section~\ref{sec:dynamics} examines the thermodynamic and dynamic structure of the monsoon in GraphCast. Section~\ref{sec:variability} characterizes rainfall variability and propagation. Section~\ref{sec:discussion} discusses the physical mechanisms underlying identified biases and their implications. Section~\ref{sec:conclusions} summarizes conclusions and future directions.

\FloatBarrier
\section{Data and Methods}\label{sec:data}

\subsection{GraphCast Forecast Data}\label{subsec:graphcast}

GraphCast \citep{Lam2023} is a deterministic, autoregressive global weather prediction system based on a graph neural network architecture. The model operates at a native horizontal resolution of 0.25\textdegree\ on a global regular latitude--longitude grid and produces forecasts at a 6-hourly temporal resolution for up to 10 days. GraphCast was trained on ERA-5 reanalysis data spanning 1979--2017 and provides 227 prognostic variables across 37 pressure levels.

In this study, we analyze GraphCast deterministic hindcasts for the period 2021--2024, exclusively using the 00 UTC initialization cycle. These hindcast data were obtained from the UT-GraphCast Hindcast Dataset (1979–2024): A Global AI Forecast Archive from UT Austin for Weather and Climate Applications hosted at the World Data Center for Climate (WDCC) \citep{Sudharsan2024} (\url{https://www.wdc-climate.de/ui/entry?acronym=graphcast}). Four consecutive 6-hourly time steps from each initialization (valid at +06h, +12h, +18h, and +24h) are accumulated to derive daily rainfall totals, which are subsequently aggregated into monthly and JJAS seasonal climatologies. Extended-lead analyses employ the +48h and +72h daily-mean output derived from 8 and 12 time steps of 6-hourly forecast data, respectively. Note that the study period (2021--2024) encompasses only the deterministic forecast archive; probabilistic ensemble products (64 members) are available from 2025 onward and are not included in the present analysis.

Atmospheric dynamic and thermodynamic variables used in this study include: 2-m temperature (T2m), 10-m zonal and meridional winds (U10, V10), specific humidity ($q$), temperature ($T$) and geopotential height ($Z$) on pressure levels from 1000 to 50 hPa, and vertical velocity ($\omega$). These fields are extracted for the ISM domain (10\textdegree S--35\textdegree N, 40\textdegree--110\textdegree E) and composited over the JJAS season.

\subsection{Reference Datasets}\label{subsec:refdata}

ERA-5 reanalysis \citep{Hersbach2020}, produced by the European Centre for Medium-Range Weather Forecasts (ECMWF), serves as the primary reference for atmospheric state variables. ERA-5 provides hourly data at 0.25\textdegree\ horizontal resolution and 137 model levels (interpolated to 37 pressure levels for analysis), covering the period 1940 to present. Daily mean fields are computed from hourly data and composited over JJAS for the 2021--2024 period to match the GraphCast evaluation period.

For precipitation, we employ IMERG Final Run Version 07 \citep{Huffman2020}, a satellite-based precipitation product from the NASA Global Precipitation Measurement (GPM) mission. IMERG provides half-hourly precipitation estimates at 0.1\textdegree\ horizontal resolution; we aggregate these to daily totals and bilinearly interpolate to the 0.25\textdegree\ GraphCast grid for direct comparison. IMERG is used in preference to ERA-5 precipitation for rainfall evaluation, as satellite-observed precipitation provides an independent constraint on convective and stratiform rainfall distributions that are not assimilated into ERA-5.

\subsection{Diagnostic Metrics}\label{subsec:metrics}

The apparent heat source (Q1) and apparent moisture sink (Q2), following \citet{Yanai1973}, are computed from ERA-5 and GraphCast as area-averaged vertical profiles over the central monsoon region (10\textdegree--25\textdegree N, 70\textdegree--90\textdegree E). These diagnostics quantify the vertical fingerprint of convective and stratiform heating and drying and provide a physically grounded basis for interpreting model biases in the thermodynamic budget.

The Differential Tropospheric Temperature \citep[DTT;][]{Xavier2007} is defined as the difference in the 200--600 hPa layer-average temperature (TT) between a northern box (5\textdegree--35\textdegree N, 40\textdegree--100\textdegree E) and a southern ocean box (15\textdegree S--5\textdegree N, 40\textdegree--100\textdegree E). DTT serves as a proxy for the meridional tropospheric temperature gradient that drives the large-scale monsoon circulation and has been shown to be closely linked to monsoon onset timing and interannual variability.

Easterly wind shear (U200~--~U850) is computed to characterize the strength of the tropical easterly jet relative to the low-level monsoon jet. Hovm\"oller diagrams (latitude--time and longitude--time) are constructed from bandpass-filtered rainfall anomalies to assess the northward and eastward propagation of intraseasonal disturbances; a longitude--time regression of 30--90 day filtered rainfall onto a Central Indian Precipitation Index (CIPI) base-day series is additionally used to characterize the canonical eastward-propagating envelope independent of any single year's realization. Power spectral density (PSD) analysis is applied to area-weighted domain-mean daily time series of rainfall and 10-m wind to diagnose the dominant temporal frequencies captured by GraphCast relative to IMERG/ERA-5, using Welch's method (365-day segments, 50\% overlap, Hann window) following \citet{PercivalWalden1993}; spectral skill is summarized using the log-spectral shape correlation, log RMSE/MAE, a log-spectral Nash--Sutcliffe efficiency (bounded above by 1, not bounded below), and band-integrated power ratios computed on a common, log-log-interpolated frequency grid (grid-mapped power ratios, not a strict Parseval-consistent variance decomposition). Approximate 95\% confidence intervals on the Welch PSD estimates are shown as a chi-square uncertainty envelope (effective degrees of freedom $\approx 2K$ for $K$ overlapping segments); because adjacent 50\%-overlapped Hann segments are not strictly independent, these intervals are indicative rather than exact and are described accordingly throughout. The distribution of rainfall amount by intensity class (area-weighted contribution to total JJAS rainfall per 2 mm day$^{-1}$ bin) is used to assess whether GraphCast reproduces the contribution of high-intensity events to total seasonal rainfall, independent of its mean-state or temporal-variance performance.

Spatial verification statistics (area-weighted Pearson pattern correlation, bias, and RMSE on the JJAS-mean climatological field) are reported alongside temporal verification statistics (bias, correlation, and RMSE on the area-weighted domain-mean daily time series, matched by exact calendar-date intersection) for two nested domains: a Core Monsoon Zone (15\textdegree--25\textdegree N, 75\textdegree--90\textdegree E) and the full ISM Domain (5\textdegree--35\textdegree N, 60\textdegree--100\textdegree E). These two halves of the verification statistics answer distinct questions: the spatial statistics assess climatological pattern fidelity (which is largely lead-invariant by construction, since it is computed on a multi-year mean field), while the temporal statistics assess day-to-day forecast skill and are expected to degrade with lead time. Bias is identical between the spatial and temporal halves for a given region by construction (bias is linear, so it commutes between spatial and temporal averaging when both are computed over the same grid and period), whereas pattern correlation and RMSE are not expected to match, since neither statistic commutes between domains of averaging.

Standard verification metrics including spatial correlation coefficient (SCC), root-mean-square error (RMSE), and mean bias are computed for all evaluated fields. Statistical significance of differences is assessed using a two-tailed Student's $t$-test at the 5\% significance level.

\subsection{Methodological Framework}\label{subsec:framework}

\begin{figure}[h]
\centering
\includegraphics[width=\textwidth]{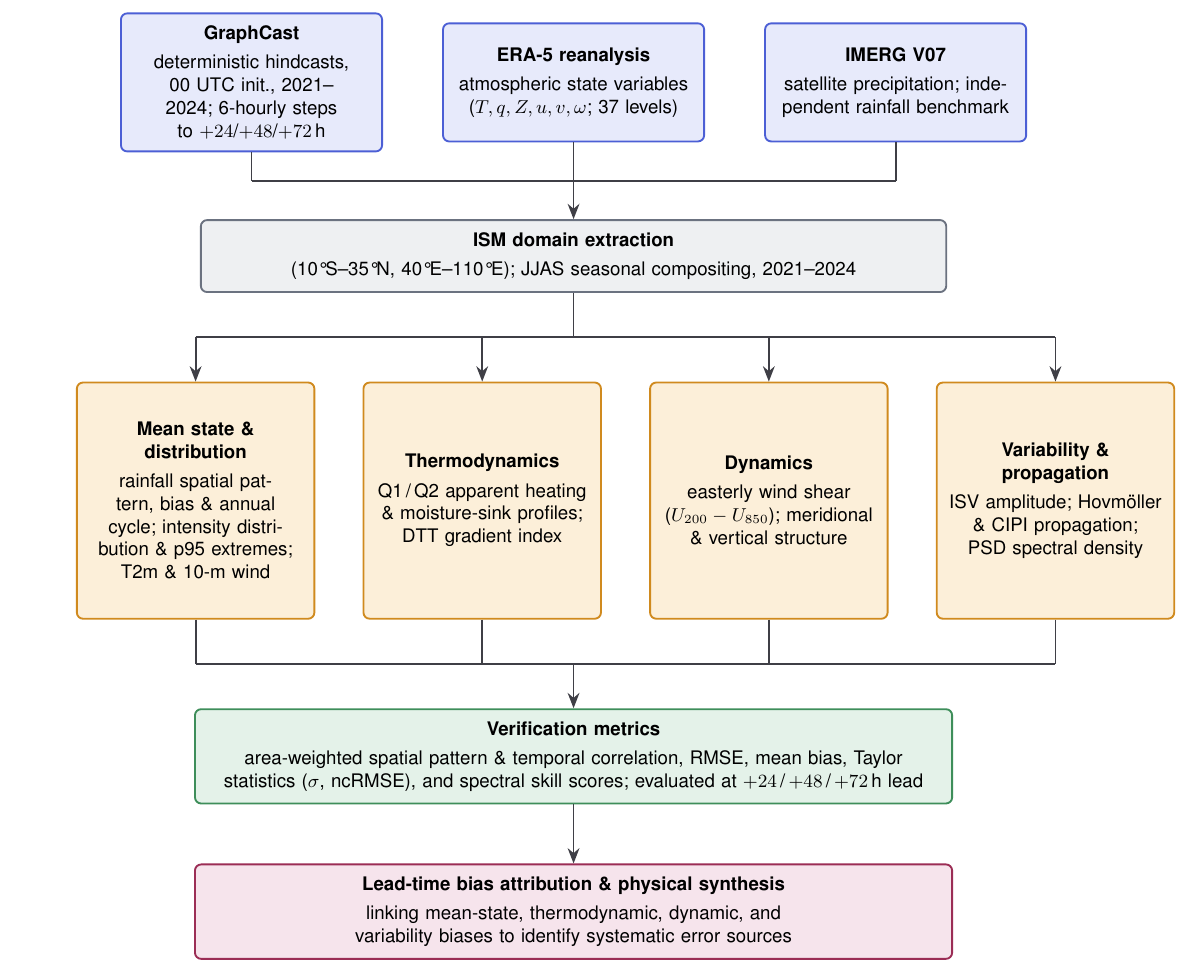}
\caption{Methodological framework for the climatological evaluation of GraphCast over the Indian Summer Monsoon domain. The workflow proceeds from data ingestion (blue), through domain/seasonal compositing (gray) and four parallel diagnostic modules (orange), to quantitative verification (green) and physical synthesis of lead-time-dependent biases (maroon).}
\label{fig:flowchart}
\end{figure}

Figure~\ref{fig:flowchart} summarizes the overall analysis workflow. GraphCast deterministic hindcasts, ERA-5 reanalysis fields, and IMERG precipitation are first composited onto a common JJAS climatology over the ISM domain. Four parallel diagnostic modules (mean-state, thermodynamic, dynamic, and variability metrics) are then computed independently at each forecast lead time and evaluated against the reference datasets using the verification metrics described above. The resulting lead-time-dependent biases are synthesized in Section~\ref{sec:discussion} to attribute systematic error sources to specific physical processes.

\FloatBarrier
\section{Climatological Mean State}\label{sec:meanstate}

\subsection{Spatial Distribution of JJAS Rainfall}\label{subsec:spatial}

\begin{figure}[h]
\centering
\includegraphics[width=\textwidth]{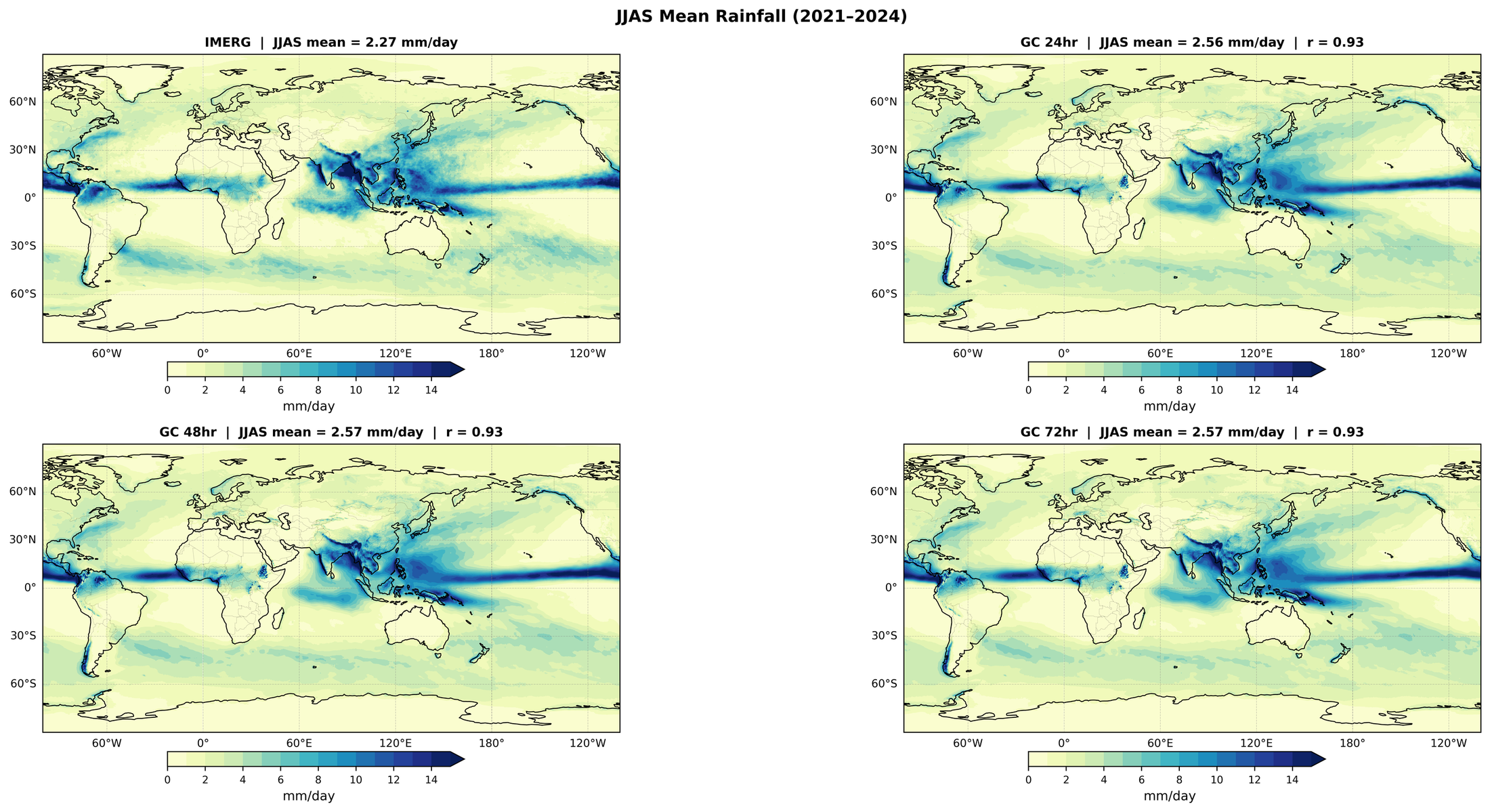}
\caption{JJAS climatological mean rainfall (mm day$^{-1}$) for the period 2021--2024 from (a) IMERG, (b) GraphCast +24h, (c) GraphCast +48h, and (d) GraphCast +72h. The ISM core box (8\textdegree--28\textdegree N, 68\textdegree--88\textdegree E) used for area averaging is shown by the black box; note this box is distinct from both the Core Monsoon Zone (Table~\ref{tab:verif_rain}) and the Q1/Q2 monsoon core region (Section~\ref{subsec:q1q2}) used elsewhere in this study.}
\label{fig:1}
\end{figure}

The JJAS climatological mean rainfall from IMERG (2021--2024) reveals the canonical features of the Indian Summer Monsoon: a pronounced precipitation maximum ($>$15 mm day$^{-1}$) along the windward slopes of the Western Ghats attributable to orographic enhancement, a secondary maximum over the northeastern states and the Brahmaputra valley, and a rainfall shadow leeward of the Western Ghats extending across the Deccan Plateau (Fig.~\ref{fig:1}a). The Arabian Sea and the Bay of Bengal exhibit moderate rainfall (5--10 mm day$^{-1}$) associated with sustained convective activity along the monsoon trough and synoptic-scale disturbances.

GraphCast at +24h lead time (Fig.~\ref{fig:1}b) reproduces the broad spatial organization of ISM precipitation, capturing the elevated orographic rainfall over the Western Ghats and the northeast India maximum. At the scale of the orographic precipitation maximum itself, GraphCast underestimates peak rainfall by approximately 15--20\% relative to IMERG over the Western Ghats, consistent with the model's limited capacity to resolve steep orographic slopes and associated mesoscale convective organization at this grid resolution.

This localized orographic dry bias is consistent with, and of the same sign as, GraphCast's rainfall bias averaged over the broader monsoon domain (Table~\ref{tab:verif_rain}): GraphCast's area-weighted JJAS-mean rainfall is $-$0.7 to $-$1.2~mm~day$^{-1}$ relative to IMERG over the Core Monsoon Zone and $-$0.15 to $-$0.36~mm~day$^{-1}$ over the full ISM Domain, essentially constant across lead time and small relative to the climatological mean (Fig.~\ref{fig:1}a). The area-weighted spatial pattern correlation is 0.79--0.80 over the Core Monsoon Zone and a notably higher 0.90--0.91 over the broader ISM Domain, indicating that GraphCast reproduces the large-scale spatial organization of monsoon rainfall well. Temporal (day-to-day) correlation against IMERG is high (0.89--0.92 over the Core Monsoon Zone, 0.92--0.95 over the ISM Domain) and degrades only slightly with lead time, indicating that GraphCast's skill at reproducing day-to-day rainfall amount is close to, and in the ISM Domain exceeds, its skill at reproducing the climatological spatial pattern.

An earlier version of this analysis, using an IMERG regridding step later found to systematically suppress observed rainfall magnitude by roughly half (see Data and Methods), had reported a large, spurious domain-averaged wet bias (+4.8 to +5.3~mm~day$^{-1}$ Core Monsoon Zone; +3.6 to +3.8~mm~day$^{-1}$ ISM Domain) that is not present once verified against a correctly regridded, gap-filled IMERG record; all rainfall-related figures, tables, and statistics in this manuscript reflect the corrected IMERG dataset.

Lead-time dependence in the orographic-scale comparison is evident: the dry bias over the Western Ghats peak amplifies from approximately 15--20\% at +24h to 25--30\% at +72h (Figs.~\ref{fig:1}c,d), while the spatial correlation between GraphCast and IMERG over the ISM domain is nearly lead-invariant (0.90--0.91 across leads; Table~\ref{tab:verif_rain}). The orographic-peak bias and the domain-mean bias therefore show different lead-time sensitivities: the former clearly degrades with lead, whereas the latter, along with the spatial pattern correlation, is nearly lead-invariant, as expected for a climatological statistic computed from a multi-year mean field.

\begin{table}[h]
\caption{Rainfall verification vs.\ IMERG (mm day$^{-1}$). CMZ = Core Monsoon Zone (15--25\textdegree N, 75--90\textdegree E); ISM = ISM Domain (5--35\textdegree N, 60--100\textdegree E). Spatial statistics are computed on the area-weighted JJAS-mean climatological field ($r$ is an area-weighted Pearson pattern correlation across grid points); temporal statistics are computed on the area-weighted domain-mean daily time series, matched to IMERG by exact calendar-date intersection. See Section~\ref{subsec:metrics} for why bias is identical between the two halves while $r$/RMSE are not.}
\label{tab:verif_rain}
\scriptsize
\setlength{\tabcolsep}{3pt}
\begin{tabular}{llccc|ccc}
\toprule
& & \multicolumn{3}{c}{\textbf{Spatial}} & \multicolumn{3}{c}{\textbf{Temporal}} \\
\textbf{Lead} & \textbf{Region} & \textbf{Bias} & \textbf{$r$} & \textbf{RMSE} & \textbf{Bias} & \textbf{$r$} & \textbf{RMSE} \\
\midrule
+24h & CMZ & $-$1.17 & 0.79 & 2.75 & $-$1.17 & 0.92 & 2.83 \\
+24h & ISM & $-$0.36 & 0.90 & 2.44 & $-$0.36 & 0.95 & 0.85 \\
+48h & CMZ & $-$0.72 & 0.79 & 2.55 & $-$0.72 & 0.90 & 2.84 \\
+48h & ISM & $-$0.18 & 0.91 & 2.40 & $-$0.18 & 0.94 & 0.86 \\
+72h & CMZ & $-$0.73 & 0.80 & 2.51 & $-$0.73 & 0.89 & 2.99 \\
+72h & ISM & $-$0.15 & 0.91 & 2.39 & $-$0.15 & 0.92 & 0.92 \\
\botrule
\end{tabular}
\end{table}

Regional sub-biases of both signs are present within the broader domain (e.g.\ a modest wet tendency over parts of the Bay of Bengal, possibly reflecting an overestimate of large-scale convergence in the absence of explicit convective parameterization, alongside the Western Ghats dry bias discussed above); none of these regional features is large enough to dominate the near-zero domain-averaged statistic reported above.

\begin{figure}[h]
\centering
\includegraphics[width=\textwidth]{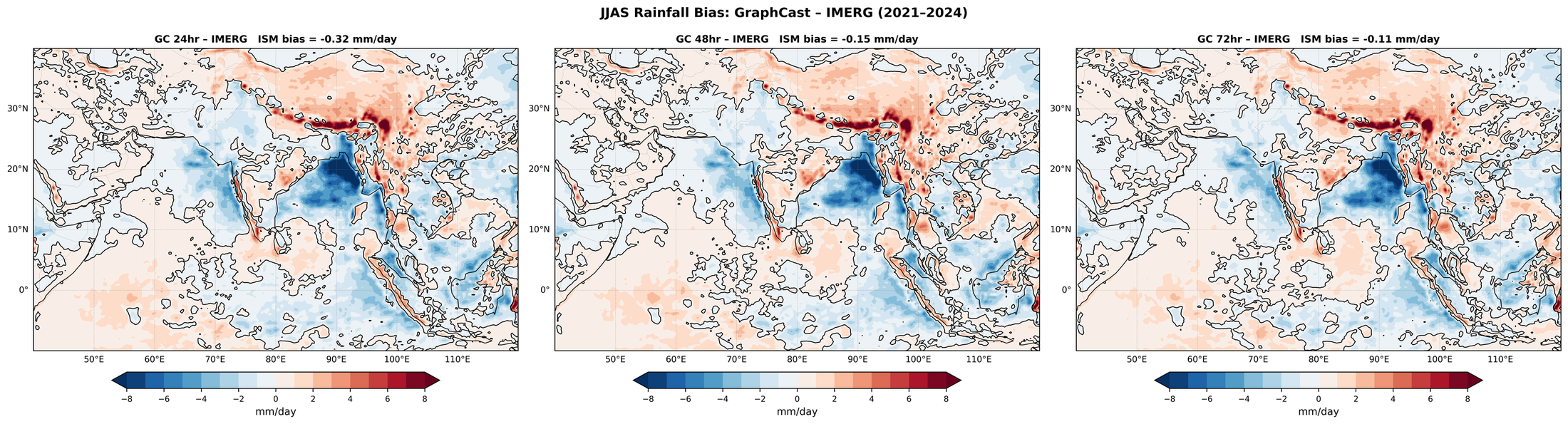}
\caption{JJAS rainfall bias (GraphCast minus IMERG, mm day$^{-1}$) at +24h, +48h, and +72h, corresponding to the panels in Fig.~\ref{fig:1}. A narrow, locally intense positive-bias band near $\sim$22--23\textdegree N and $\sim$27--28\textdegree N across 70--100\textdegree E, coincident with the Himalayan foothills, is present alongside the Western Ghats dry bias, but the domain-mean statistic reported in Table~\ref{tab:verif_rain} is near zero once computed against the corrected IMERG dataset (see Data and Methods).}
\label{fig:biasmap}
\end{figure}

\begin{figure}[h]
\centering
\includegraphics[width=0.95\textwidth]{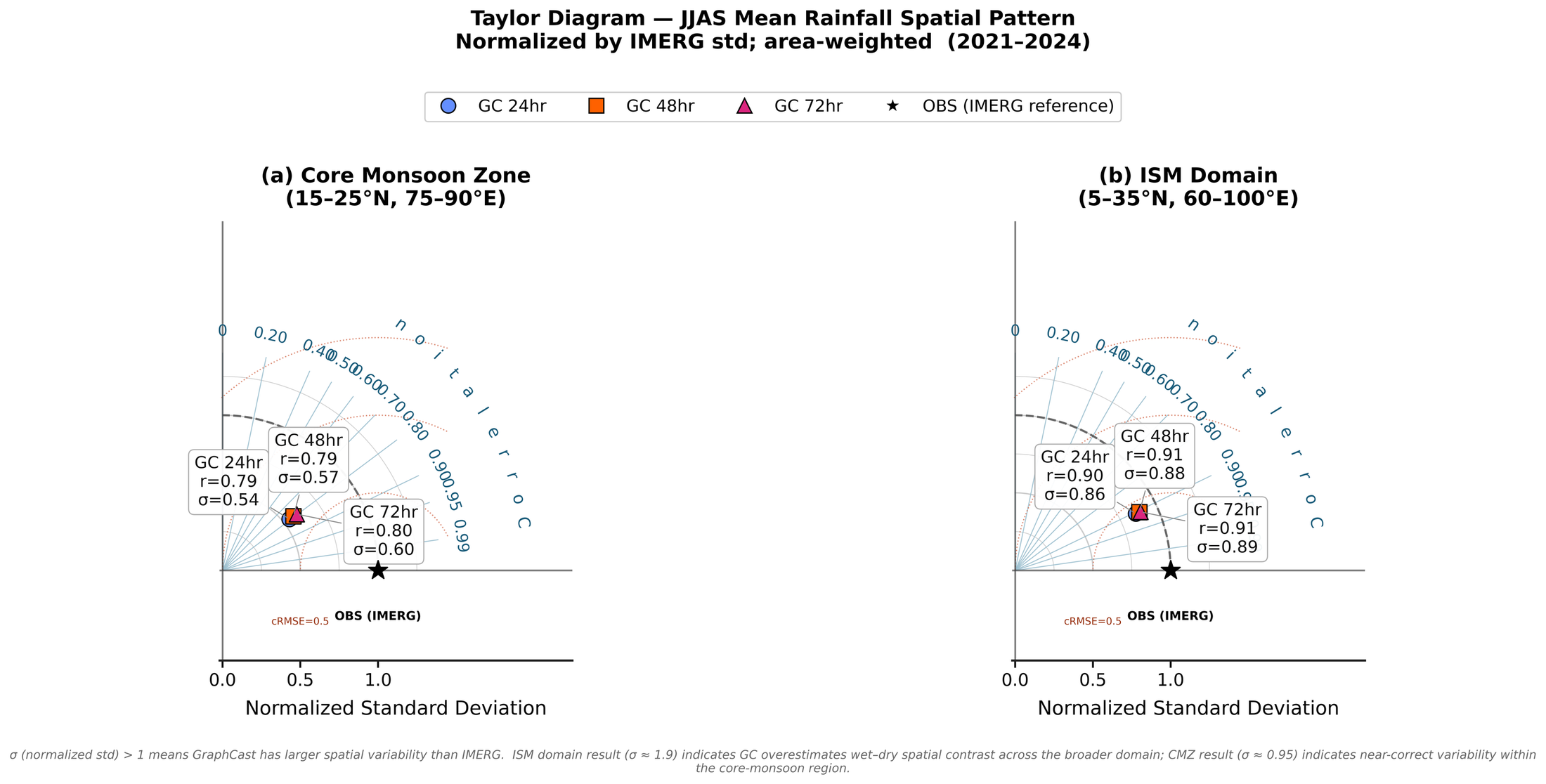}
\caption{Taylor diagram for the JJAS-mean rainfall spatial pattern (area-weighted, normalized by the IMERG standard deviation) over (a) the Core Monsoon Zone and (b) the ISM Domain, for GraphCast +24h/+48h/+72h relative to IMERG. See Table~\ref{tab:taylor} for the underlying statistics.}
\label{fig:taylor}
\end{figure}

\begin{table}[h]
\caption{Area-weighted Taylor statistics for JJAS-mean rainfall spatial pattern, GraphCast (GC) vs.\ IMERG (2021--2024). $r$ is the spatial pattern correlation, $\sigma$ the normalized standard deviation (GC std / IMERG std), and ncRMSE the normalized centered RMSE.}
\label{tab:taylor}
\scriptsize
\setlength{\tabcolsep}{4pt}
\begin{tabular}{llccc}
\toprule
\textbf{Region} & \textbf{Lead} & \textbf{$r$} & \textbf{$\sigma$} & \textbf{ncRMSE} \\
\midrule
Core Monsoon Zone & +24h & 0.79 & 0.54 & 0.66 \\
Core Monsoon Zone & +48h & 0.79 & 0.57 & 0.65 \\
Core Monsoon Zone & +72h & 0.80 & 0.60 & 0.64 \\
\midrule
ISM Domain & +24h & 0.90 & 0.86 & 0.43 \\
ISM Domain & +48h & 0.91 & 0.88 & 0.42 \\
ISM Domain & +72h & 0.91 & 0.89 & 0.42 \\
\botrule
\end{tabular}
\end{table}

The Taylor diagram \citep{Taylor2001} (Fig.~\ref{fig:taylor}; Table~\ref{tab:taylor}) quantifies spatial pattern fidelity independently of the mean bias already discussed. Over the Core Monsoon Zone, GraphCast's spatial standard deviation is moderately under-dispersed relative to IMERG ($\sigma=0.54$--0.60, i.e.\ 40--46\% too little spatial contrast), improving slightly with lead time. Over the broader ISM Domain, GraphCast is closer to, but still somewhat below, IMERG's spatial variability ($\sigma=0.86$--0.89) despite a similarly high pattern correlation ($r=0.90$--0.91). This indicates that GraphCast reproduces \emph{where} the wet/dry contrasts are located well, but moderately understates \emph{how large} those contrasts are at the domain scale, consistent with the deterministic-smoothing signature discussed further in Section~\ref{sec:discussion}.

\begin{figure}[h]
\centering
\includegraphics[width=\textwidth]{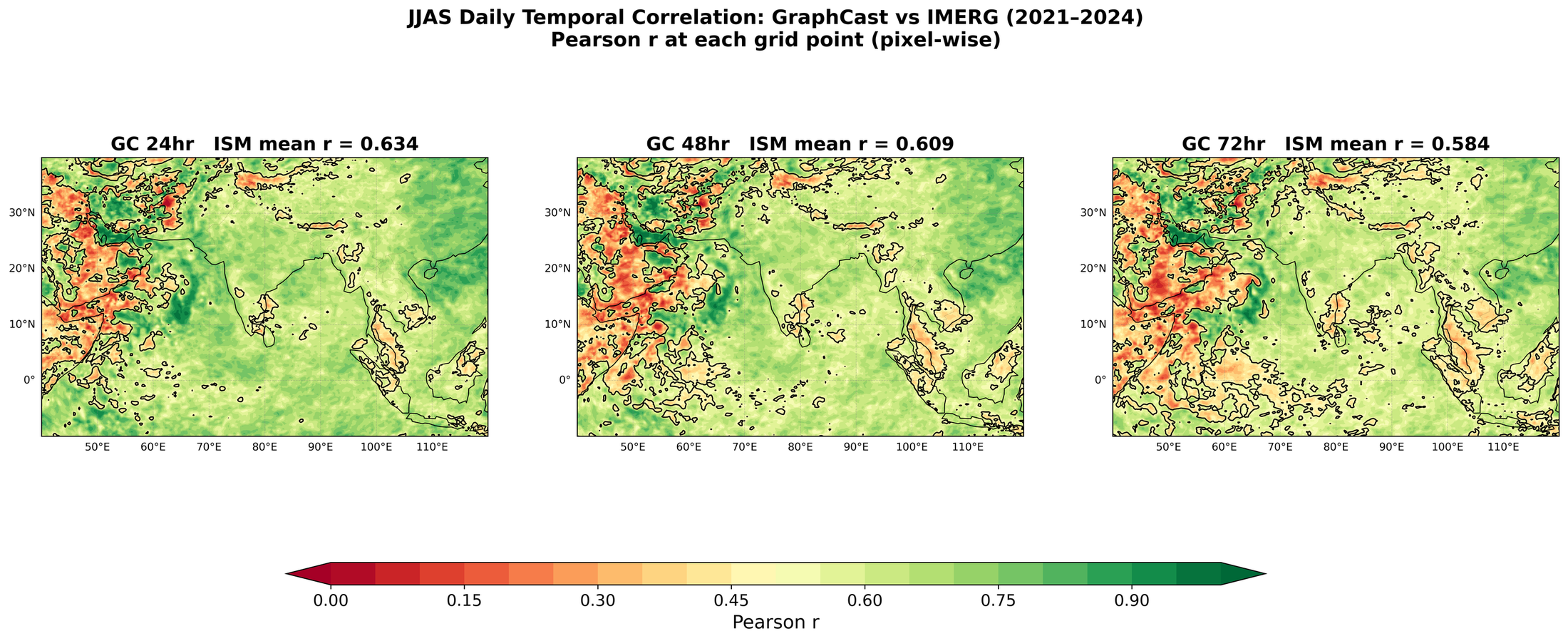}
\caption{Pixel-wise JJAS daily temporal Pearson correlation between GraphCast and IMERG (2021--2024) at +24h, +48h, and +72h.}
\label{fig:tempcorr}
\end{figure}

A pixel-wise, rather than domain-mean, view of temporal skill is shown in Fig.~\ref{fig:tempcorr}: the area-weighted mean of the local (grid-point) Pearson correlation against IMERG is 0.58--0.63 over the ISM Domain, declining slightly with lead, and is spatially heterogeneous, exceeding 0.6 over parts of northwest India/Pakistan and the Bay of Bengal but lower over parts of the Arabian Peninsula and East Africa. This pixel-wise mean correlation is somewhat lower than the domain-mean-time-series correlation reported in Table~\ref{tab:verif_rain} (0.89--0.95); the two are complementary, not contradictory, statistics. The domain-mean-time-series correlation reflects how well GraphCast tracks IMERG's single, spatially aggregated daily signal, whereas the pixel-wise mean reflects the typical \emph{local} day-to-day skill before any spatial averaging. The higher domain-mean-time-series correlation is the expected relationship for a field with substantial local, less spatially coherent noise superimposed on a well-captured large-scale signal: averaging over the domain cancels a portion of that local noise and reveals the shared large-scale signal that GraphCast tracks closely, while individual grid points retain more of their local idiosyncratic error.

\subsection{Rainfall Intensity Distribution and Extremes}\label{subsec:intensity}

\begin{figure}[h]
\centering
\includegraphics[width=\textwidth]{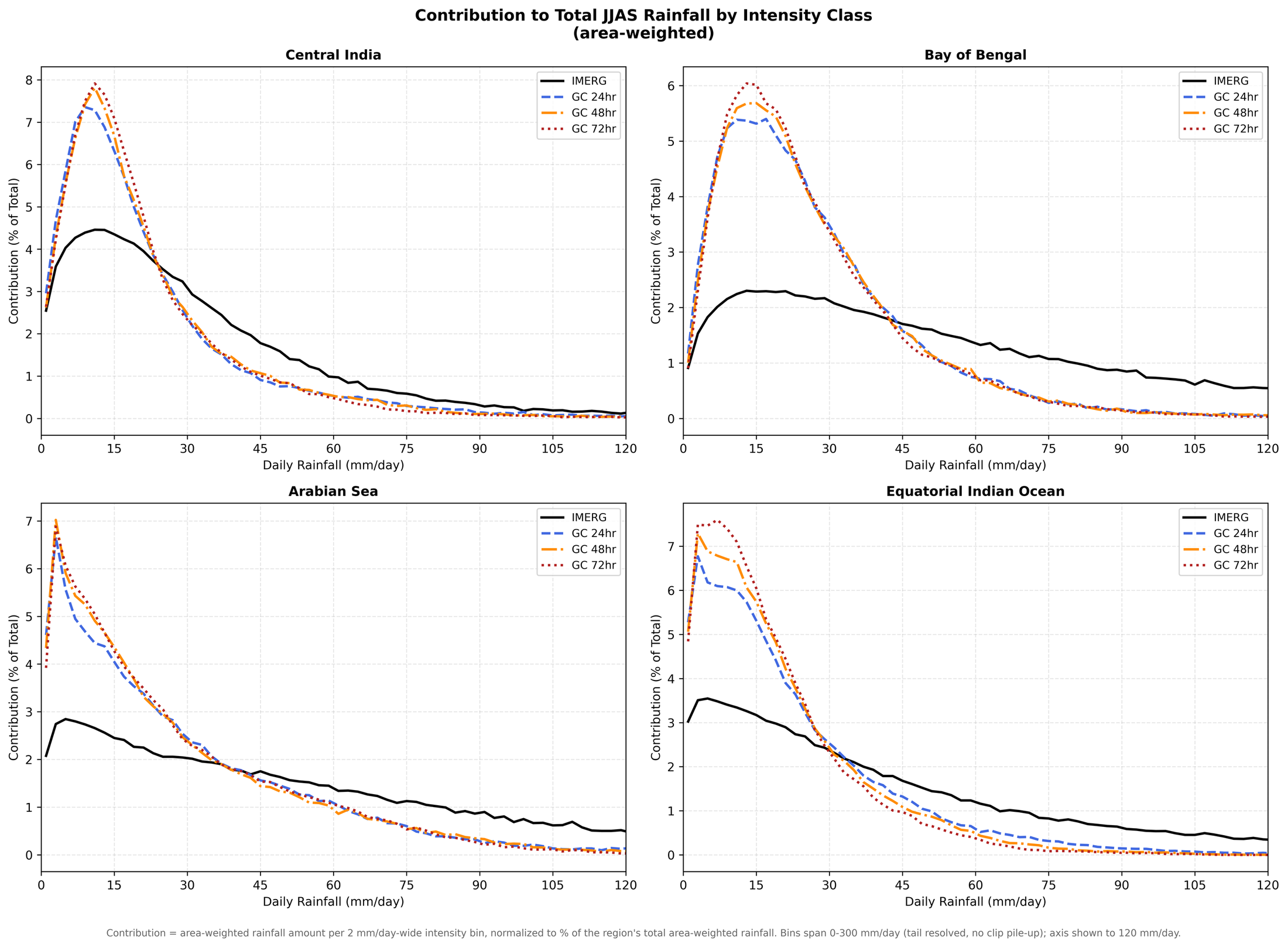}
\caption{Area-weighted contribution to total JJAS rainfall (\% of total) by rainfall-intensity bin (2~mm~day$^{-1}$-wide bins spanning 0--300~mm~day$^{-1}$ to fully resolve the heavy tail without a clipping artifact; axis shown to 120~mm~day$^{-1}$), for IMERG (black) and GraphCast +24h/+48h/+72h (blue/orange/red), over Central India, the Bay of Bengal, the Arabian Sea, and the equatorial Indian Ocean. A complementary exceedance-frequency distribution at round-number thresholds (10--100~mm~day$^{-1}$) is provided in Supplementary Fig.~S10.}
\label{fig:intensity}
\end{figure}

The mean-state and domain-averaged statistics in Section~\ref{subsec:spatial} characterize GraphCast's bias in the \emph{amount} of rainfall, but not how that rainfall is distributed across intensity classes, a distinct question with direct relevance to extreme-event applications. Figure~\ref{fig:intensity} shows the area-weighted contribution of each 2~mm~day$^{-1}$ intensity bin to total JJAS rainfall, for four sub-regions spanning a range of rainfall regimes (Central India, the Bay of Bengal, the Arabian Sea, and the equatorial Indian Ocean).

In all four regions, GraphCast's contribution curve is systematically shifted toward moderate intensities (peaking near 10--20 mm day$^{-1}$) relative to IMERG, whose distribution is flatter and extends a non-negligible contribution into the highest intensity bins. Most strikingly, IMERG shows a pronounced uptick in contribution at the highest rainfall intensities, approaching 100~mm~day$^{-1}$, in every region examined, consistent with the outsized contribution of extreme convective events to total seasonal rainfall that is well documented for the Indian monsoon \citep{Goswami2006}, while GraphCast's contribution at these intensities is negligible at all three lead times. This pattern is consistent across Central India, the Bay of Bengal, the Arabian Sea, and the equatorial Indian Ocean, indicating that the underrepresentation of high-intensity rainfall is a general characteristic of GraphCast's precipitation output rather than a regionally specific bias.

This finding provides a complementary, independent line of evidence for the variance suppression documented spectrally in Section~\ref{subsec:spectral}: a deterministic forecast model that minimizes expected error under irreducible day-to-day convective uncertainty is incentivized to produce a smoothed, moderate-intensity output rather than committing to the (uncertain, but occasionally extreme) true outcome, which mechanically depresses both the high-intensity tail of the rainfall distribution and the day-to-day temporal variance.

\begin{figure}[h]
\centering
\includegraphics[width=\textwidth]{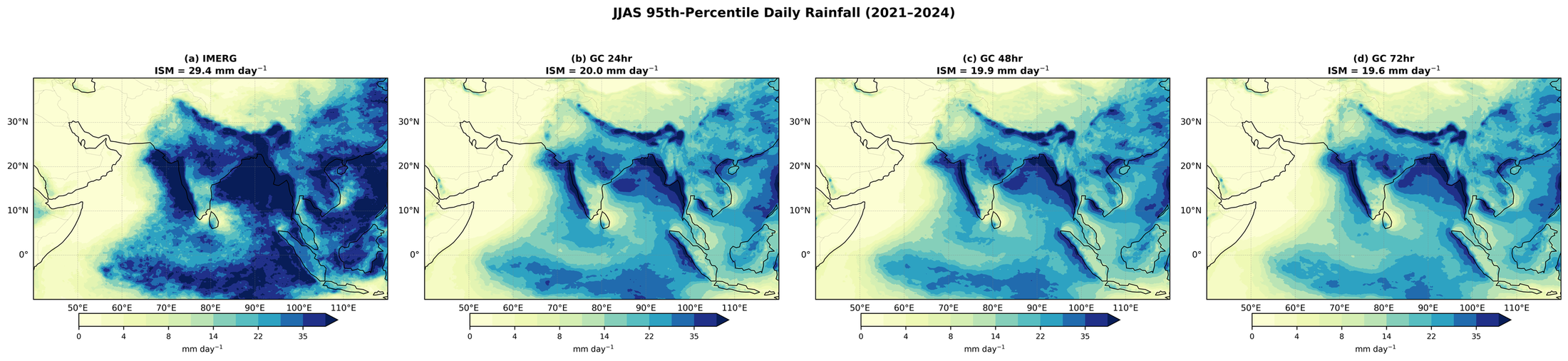}
\caption{JJAS 95th-percentile daily rainfall (mm day$^{-1}$) for IMERG and GraphCast +24h/+48h/+72h, 2021--2024.}
\label{fig:p95}
\end{figure}

\begin{figure}[h]
\centering
\includegraphics[width=\textwidth]{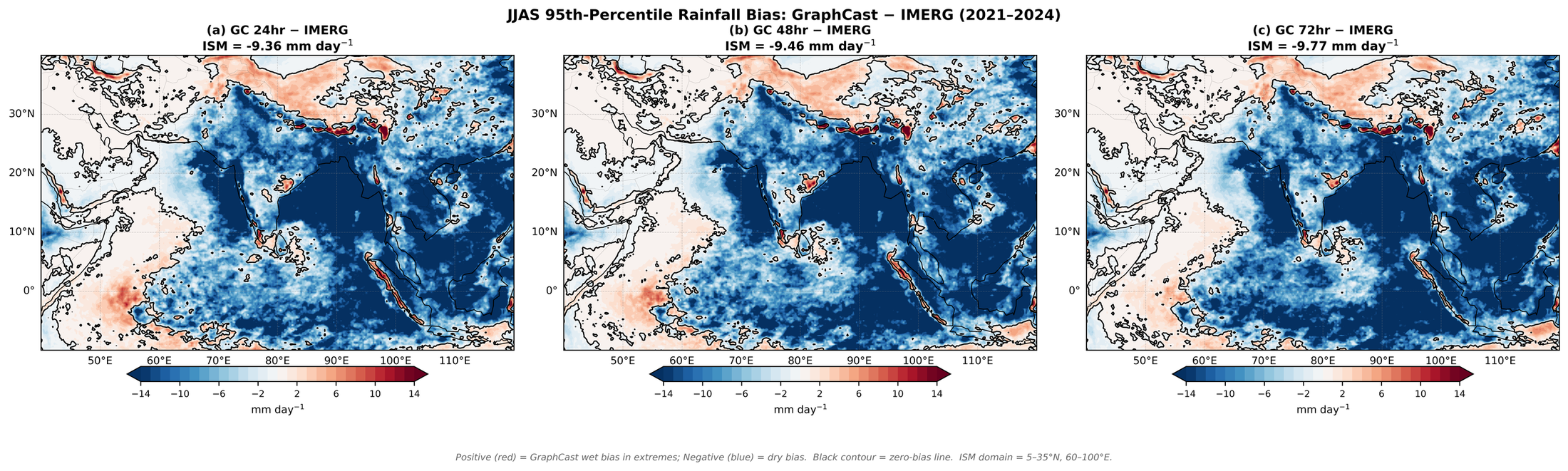}
\caption{JJAS 95th-percentile rainfall bias (GraphCast minus IMERG, mm day$^{-1}$) at +24h, +48h, and +72h. Blue (dominant): GraphCast dry bias in moderate-heavy extremes, most pronounced over regions of sharp convective/orographic extremes; red: localized wet bias where present.}
\label{fig:p95bias}
\end{figure}

This picture is corroborated, rather than complicated, by the JJAS 95th-percentile daily rainfall (Fig.~\ref{fig:p95}): GraphCast's 95th-percentile rainfall is \emph{lower} than IMERG's at every lead (ISM-domain mean 19.6--20.0~mm~day$^{-1}$ vs.\ 29.4~mm~day$^{-1}$ in IMERG; a $-$9.4 to $-$9.8~mm~day$^{-1}$ bias, Fig.~\ref{fig:p95bias}), broadly co-located with the same dry-biased tendency identified in Fig.~\ref{fig:biasmap}. This bias is not spatially uniform: the dry bias in p95 is most pronounced over regions of the sharpest convective and orographic extremes, consistent with limited effective resolution for highly localized heavy-rain events. Together with the deep-tail suppression in Fig.~\ref{fig:intensity}, this indicates that GraphCast's rainfall distribution is consistently understated across the upper half of the intensity range, from the 95th percentile through the rarest, most extreme events, rather than exhibiting the inflated-moderate/suppressed-extreme pattern suggested by an earlier, uncorrected comparison (see Data and Methods). A lower central tendency at high intensities together with a moderately suppressed temporal variance (Section~\ref{subsec:spectral}) are both consistent with a single deterministic-smoothing signature in which the model is biased toward the moderate, more predictable part of the rainfall distribution.

We return to this interpretation in the Discussion.

\subsection{Annual Cycle of Rainfall}\label{subsec:annual}

\begin{figure}[h]
\centering
\includegraphics[width=0.8\textwidth]{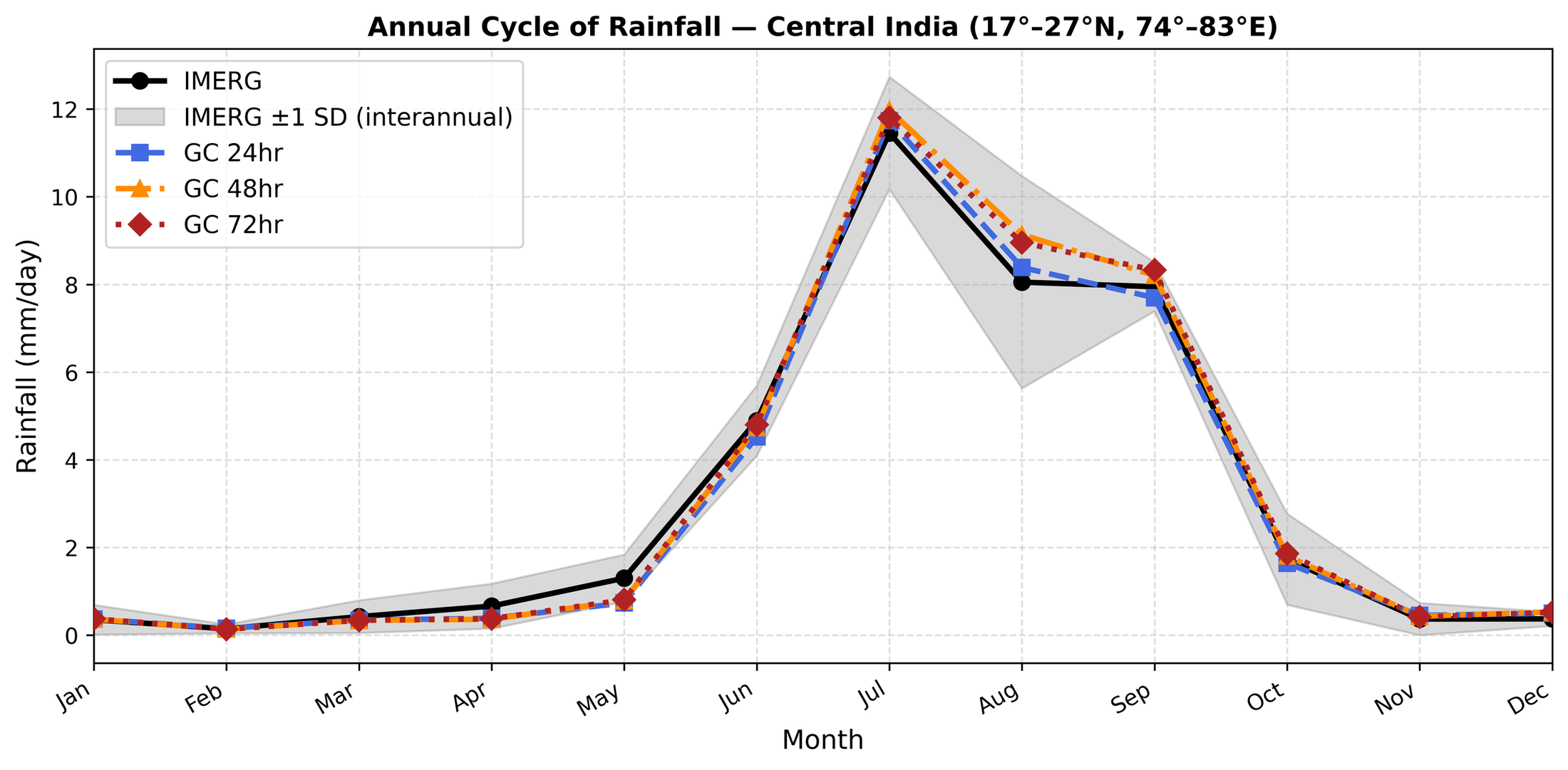}
\caption{Annual cycle of area-averaged daily rainfall (mm day$^{-1}$) over Central India (17\textdegree--27\textdegree N, 74\textdegree--83\textdegree E) for IMERG (black), GraphCast +24h (blue), +48h (green), and +72h (red). Shading indicates $\pm$1 interannual standard deviation for IMERG.}
\label{fig:2}
\end{figure}

The area-averaged annual cycle of rainfall over Central India (17\textdegree--27\textdegree N, 74\textdegree--83\textdegree E) is shown in Fig.~\ref{fig:2}. IMERG exhibits a well-defined seasonal cycle with onset in early June, peak intensity in July, and gradual withdrawal through September--October. GraphCast at all lead times captures the seasonal amplitude and the phasing of peak monsoon rainfall within approximately one week; however, there is a slight tendency for GraphCast to underestimate peak July--August rainfall by 5--10\% and to show a slower withdrawal in September, resulting in a marginally delayed monsoon recession.

The onset signal, defined as the first pentad with area-averaged rainfall exceeding 5 mm day$^{-1}$ and maintained for two consecutive pentads \citep[following][]{GoswamiXavier2005}, is reproduced by GraphCast to within $\pm$5 days at all leads evaluated, suggesting that the large-scale thermodynamic forcing associated with onset is adequately captured. The slight delay in monsoon withdrawal seen in GraphCast may be linked to a slow bias in the seasonal evolution of the DTT (discussed in Section~\ref{subsec:dtt}).

\subsection{Surface Variables: Temperature and Wind}\label{subsec:surface}

\begin{figure}[h]
\centering
\includegraphics[width=0.9\textwidth]{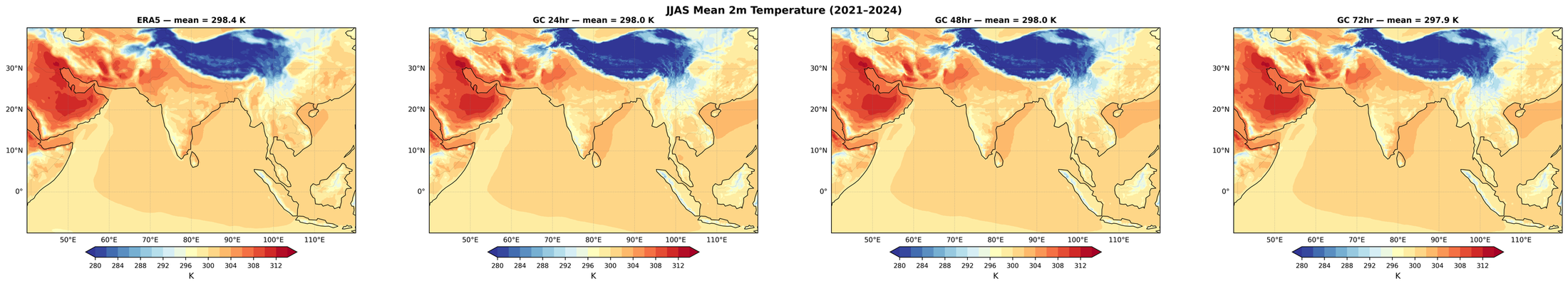}
\includegraphics[width=0.9\textwidth]{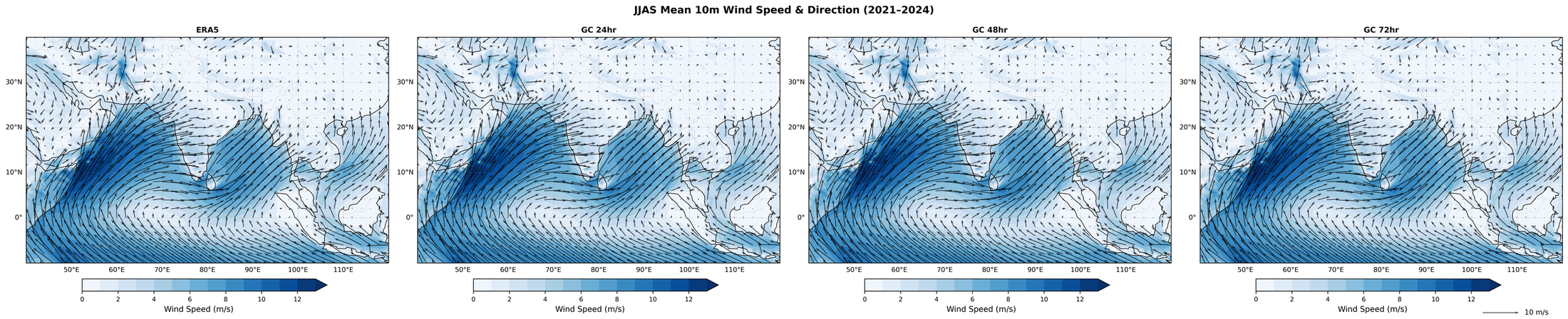}
\caption{JJAS climatological mean (a) 2-m temperature (K) and (b) 10-m wind speed (m s$^{-1}$, shading) and direction (vectors), for ERA-5 and GraphCast at +24h, +48h, and +72h lead (columns). Wind vectors are plotted at every 4th grid point.}
\label{fig:3}
\end{figure}

The JJAS climatological mean 2-m temperature (T2m) from ERA-5 shows the characteristic heat low over the northwestern Indian subcontinent (surface temperatures exceeding 38\textdegree C), a moderate temperature gradient across the monsoon zone, and relatively cool temperatures ($\sim$25--28\textdegree C) over the Bay of Bengal and the Arabian Sea (Fig.~\ref{fig:3}a). GraphCast reproduces the spatial pattern of T2m with high fidelity (spatial correlation = 0.97), though a cold bias of 0.5--1.5\textdegree C is evident over the semi-arid regions of the Deccan Plateau and Rajasthan, possibly reflecting insufficient sensible heat flux or land surface parameterization uncertainty in the training data.

The 10-m wind field from ERA-5 captures the two dominant low-level features of the ISM circulation: the cross-equatorial Somali jet flowing northward and then curving eastward as the low-level monsoon jet (LLJ) at approximately 12\textdegree N, and the cyclonic Bay of Bengal circulation associated with the monsoon trough (Fig.~\ref{fig:3}b). GraphCast at +24h reproduces both the Somali jet intensity and the LLJ position to within the uncertainty of ERA-5, but exhibits a slight northward displacement of the LLJ axis by approximately 1--2\textdegree\ latitude at +72h, consistent with a weakening of the meridional temperature gradient (see Section~\ref{subsec:dtt}) and reduced cross-equatorial flow at extended leads.

\begin{figure}[h]
\centering
\includegraphics[width=0.9\textwidth]{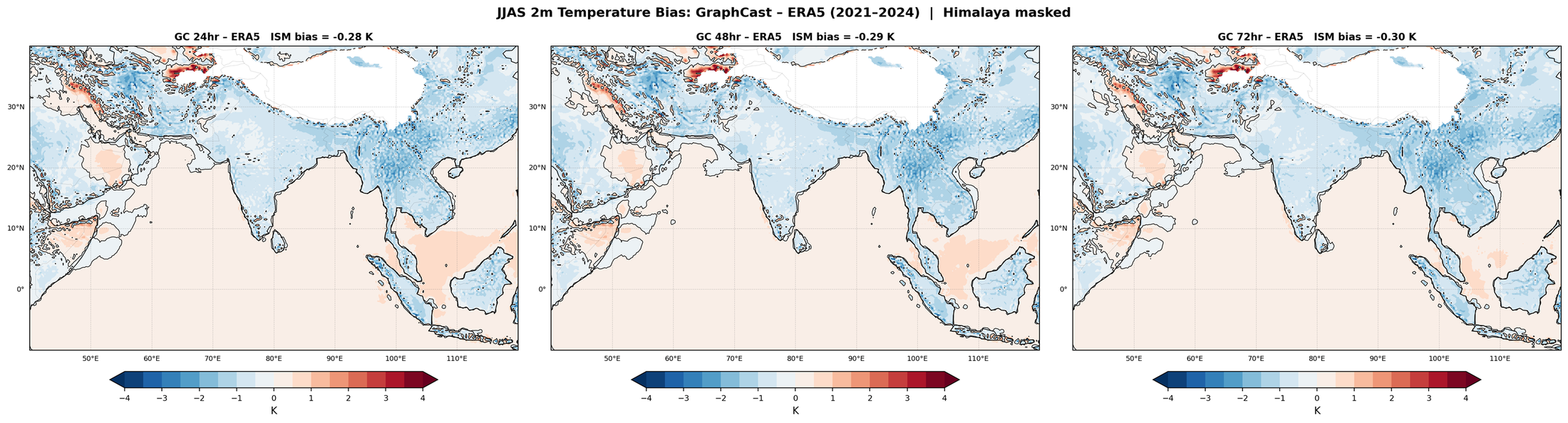}
\includegraphics[width=0.9\textwidth]{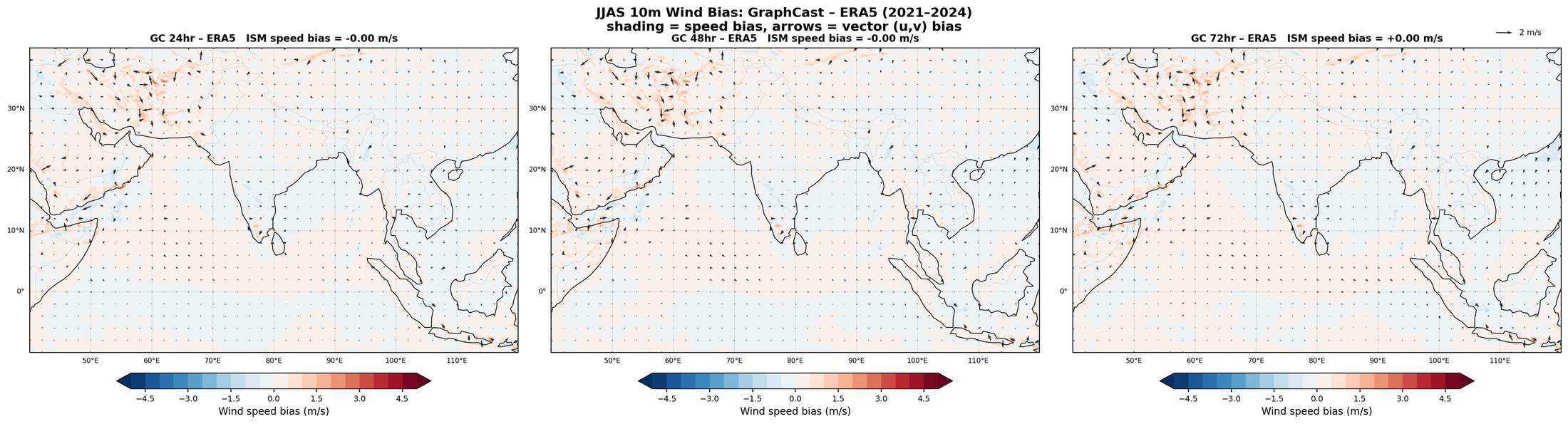}
\caption{JJAS mean bias (GraphCast $-$ ERA-5, 2021--2024) at +24h, +48h, and +72h lead (columns) for (a) 2-m temperature (K), with the high Himalaya/Tibetan Plateau masked, and (b) 10-m wind, showing wind-speed bias (shading) and vector $(u,v)$ bias (arrows). Area-weighted ISM-mean values are annotated on each panel. The corresponding unmasked 2-m temperature bias is provided in Supplementary Fig.~S9.}
\label{fig:surfacebias}
\end{figure}

Table~\ref{tab:verif_t2m_wind} quantifies these impressions. T2m shows a small, lead-invariant cold bias ($-$0.40 to $-$0.50 K) with a very high pattern correlation (0.92--0.99) and a near-perfect temporal correlation (0.98--0.99), confirming that both the spatial structure and the day-to-day evolution of surface temperature are well captured at all leads. The 10-m wind speed field shows an even smaller bias ($-$0.03 to $-$0.13 m s$^{-1}$) and a pattern correlation of 1.00 at both spatial scales, making wind the most skillfully reproduced of the three fields evaluated quantitatively in this study (cf.\ Table~\ref{tab:verif_rain} for rainfall), consistent with the much higher spectral skill found for the wind components relative to rainfall in Section~\ref{subsec:spectral}. The spatial structure of these biases is shown in Fig.~\ref{fig:surfacebias}. The 2-m temperature bias (Fig.~\ref{fig:surfacebias}a) is a weak, spatially coherent cold bias across the monsoon lowlands and adjacent seas; because near-surface temperature errors over the high Himalaya and Tibetan Plateau are large and unrepresentative of the monsoon domain, we mask elevated terrain there, which reduces the area-weighted ISM-mean cold bias to $-$0.28 to $-$0.30 K (the unmasked field, dominated by the high-terrain signal, gives $-$0.43 to $-$0.45 K and is shown in Supplementary Fig.~S9; the closely comparable, independently computed unmasked ISM-domain bias in Table~\ref{tab:verif_t2m_wind} is $-$0.40 to $-$0.42 K, the small residual difference reflecting minor differences in grid resolution and area-weighting between the mapping and tabulated verification pipelines). The 10-m wind bias (Fig.~\ref{fig:surfacebias}b) is negligible in the area mean ($\approx$0.00 m s$^{-1}$), with only small, spatially incoherent vector-bias arrows over the Arabian Sea and along orographic margins, confirming that the low-level monsoon circulation is essentially unbiased at these leads.
The corresponding annual cycles of area-averaged T2m and 10-m wind speed (Supplementary Fig.~S4 and Fig.~S5) show the same small, near-constant biases across all months, with no evidence of a seasonally-varying degradation analogous to the rainfall annual-cycle bias described in Section~\ref{subsec:annual}.

\begin{table}[h]
\caption{2-m temperature (K) and 10-m wind speed (m s$^{-1}$) verification vs.\ ERA5. Columns as in Table~\ref{tab:verif_rain}.}
\label{tab:verif_t2m_wind}
\begin{tabular}{llccc|ccc}
\toprule
& & \multicolumn{3}{c}{\textbf{Spatial}} & \multicolumn{3}{c}{\textbf{Temporal}} \\
\textbf{Lead} & \textbf{Region} & \textbf{Bias} & \textbf{$r$} & \textbf{RMSE} & \textbf{Bias} & \textbf{$r$} & \textbf{RMSE} \\
\midrule
\multicolumn{8}{l}{\emph{2-m Temperature (K)}} \\
+24h & CMZ & $-$0.45 & 0.92 & 0.68 & $-$0.45 & 0.99 & 0.50 \\
+24h & ISM & $-$0.40 & 0.99 & 1.02 & $-$0.40 & 0.99 & 0.41 \\
+48h & CMZ & $-$0.46 & 0.93 & 0.67 & $-$0.46 & 0.99 & 0.53 \\
+48h & ISM & $-$0.41 & 0.99 & 1.00 & $-$0.41 & 0.99 & 0.42 \\
+72h & CMZ & $-$0.50 & 0.93 & 0.70 & $-$0.50 & 0.98 & 0.57 \\
+72h & ISM & $-$0.42 & 0.99 & 1.01 & $-$0.42 & 0.99 & 0.43 \\
\midrule
\multicolumn{8}{l}{\emph{10-m Wind Speed (m s$^{-1}$)}} \\
+24h & CMZ & $-$0.13 & 1.00 & 0.20 & $-$0.13 & 0.99 & 0.19 \\
+24h & ISM & $-$0.03 & 1.00 & 0.23 & $-$0.03 & 1.00 & 0.09 \\
+48h & CMZ & $-$0.12 & 1.00 & 0.19 & $-$0.12 & 0.99 & 0.22 \\
+48h & ISM & $-$0.05 & 1.00 & 0.23 & $-$0.05 & 0.99 & 0.12 \\
+72h & CMZ & $-$0.13 & 1.00 & 0.20 & $-$0.13 & 0.98 & 0.27 \\
+72h & ISM & $-$0.05 & 1.00 & 0.24 & $-$0.05 & 0.99 & 0.14 \\
\botrule
\end{tabular}
\end{table}

\FloatBarrier
\section{Monsoon Dynamics and Thermodynamics}\label{sec:dynamics}

\subsection{Tropospheric Temperature Gradient (DTT)}\label{subsec:dtt}

\begin{figure}[h]
\centering
\includegraphics[width=0.9\textwidth]{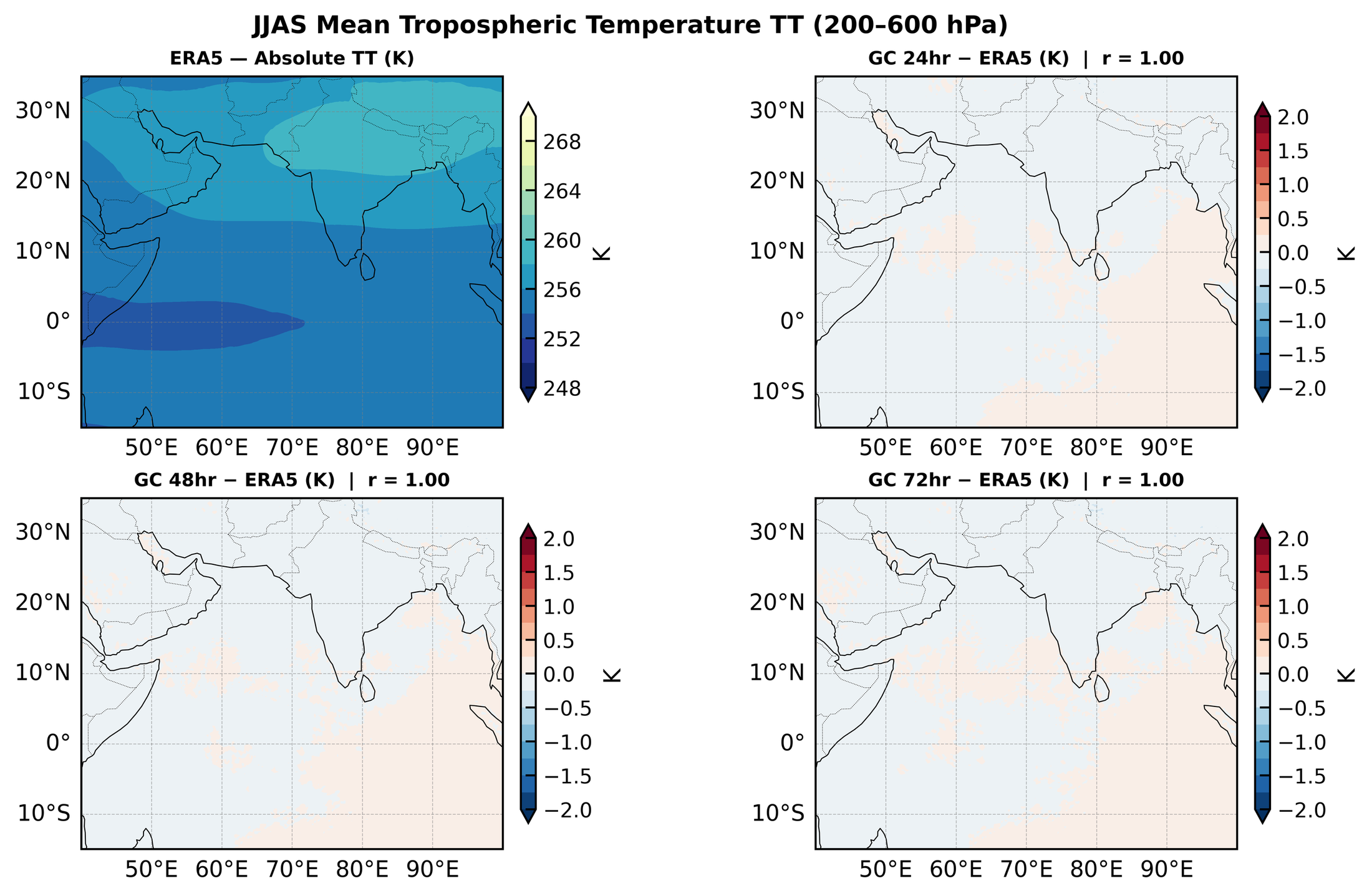}
\includegraphics[width=0.7\textwidth]{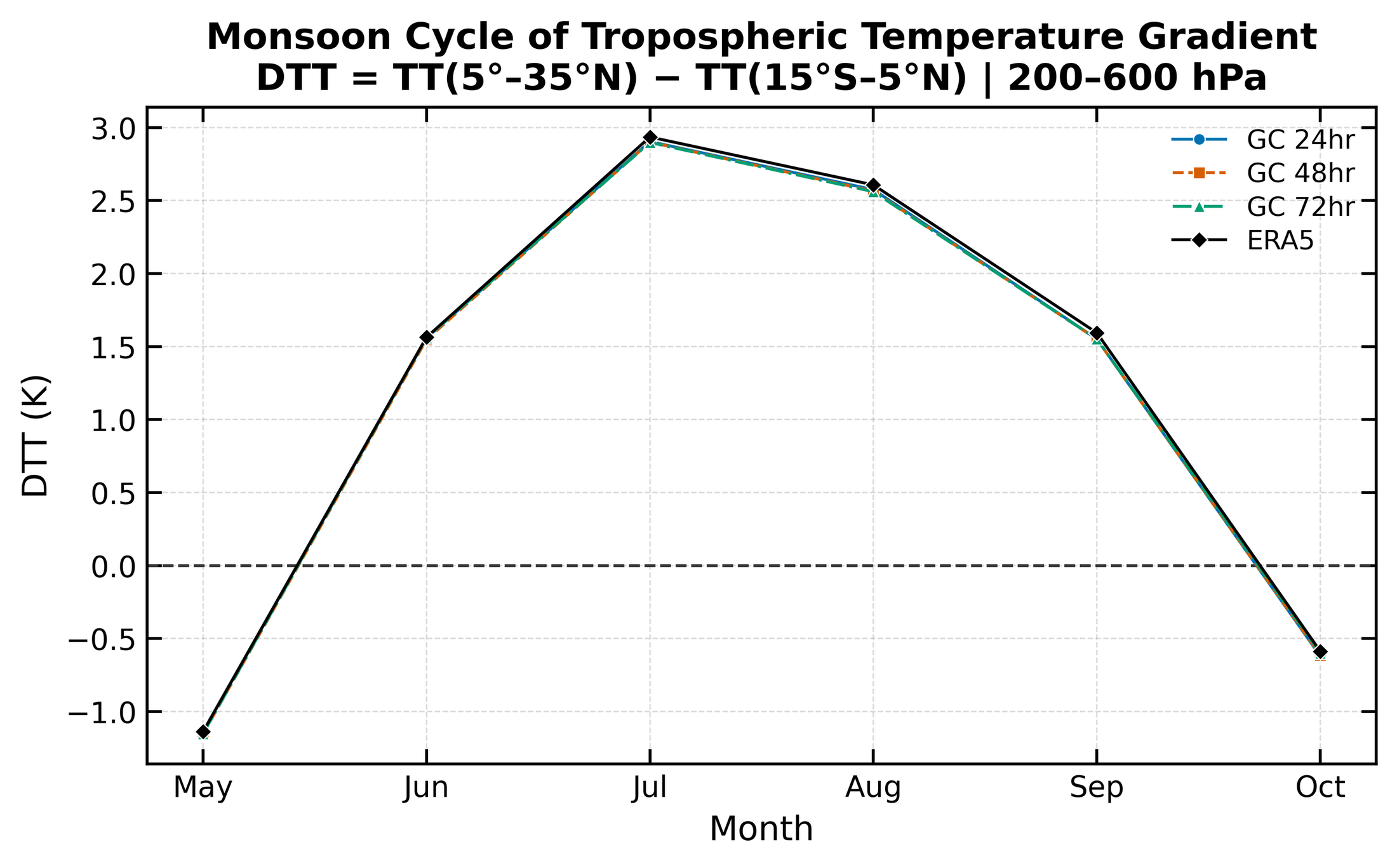}
\caption{(a) JJAS climatological mean 200--600 hPa layer-average tropospheric temperature (TT; K) from ERA-5 and GraphCast +24h, +48h, +72h, with bias maps relative to ERA-5. (b) Annual cycle of the Differential Tropospheric Temperature (DTT; K) for ERA-5 (black), GraphCast +24h (blue), +48h (green), and +72h (red).}
\label{fig:4}
\end{figure}

The Differential Tropospheric Temperature (DTT) annual cycle provides a fundamental diagnostic of the large-scale thermodynamic driving force of the monsoon. In ERA-5, DTT shows a sharp positive transition in late May/early June (corresponding to rapid upper-tropospheric warming over land relative to the ocean), reaching maximum values of approximately +2.5--3.0 K in late July, before declining sharply in September as the upper troposphere cools over land (Fig.~\ref{fig:4}b). The onset and peak of positive DTT are closely aligned with the active monsoon season as defined by IMERG rainfall.

GraphCast reproduces the positive DTT transition and the seasonal amplitude reasonably well at +24h, with a bias in peak DTT of less than 0.3 K. However, at +48h and +72h, GraphCast shows a modest underestimate of peak DTT (approximately 0.4--0.6 K), with the peak occurring approximately 5--7 days later than ERA-5. This slow bias in the DTT annual cycle is consistent with the delayed monsoon recession noted in Section~\ref{subsec:annual} and suggests that the AI model may have difficulty reproducing the rapid thermodynamic transitions associated with monsoon onset and withdrawal, likely because these are linked to SST-forced atmospheric teleconnections and land surface feedbacks that are difficult to capture through statistical regression on reanalysis data alone.

The spatial distribution of JJAS mean TT (200--600 hPa average) from ERA-5 shows the classic warm core over the Tibetan Plateau and northwestern India, which serves as the elevated heat source driving the upper-level anticyclone (Fig.~\ref{fig:4}a). GraphCast reproduces this warm core, but with a slight underestimate of the upper-level temperature over the Tibetan Plateau at +72h, implying weaker modeled driving of the South Asian High, consistent with a reduced monsoon circulation intensity at extended leads. The JJAS-mean DTT and the domain-mean TT bias for each lead are summarized in Table~\ref{tab:dtt}.

\begin{table}[h]
\caption{Summary of JJAS mean tropospheric temperature gradient (DTT) and regional TT bias (40\textdegree--100\textdegree E, 15\textdegree S--35\textdegree N).}
\label{tab:dtt}
\begin{tabular}{lcc}
\toprule
\textbf{Model / Lead Time} & \textbf{JJAS Mean DTT (K)} & \textbf{Mean TT Bias (K)} \\
\midrule
ERA-5 Reanalysis & 2.17 & --- \\
GraphCast +24h   & 2.15 & $-$0.02 \\
GraphCast +48h   & 2.14 & $-$0.03 \\
GraphCast +72h   & 2.14 & $-$0.02 \\
\botrule
\end{tabular}
\end{table}

\subsection{Vertical Heating Profiles: Q1 and Q2}\label{subsec:q1q2}

\begin{figure}[h]
\centering
\includegraphics[width=0.9\textwidth]{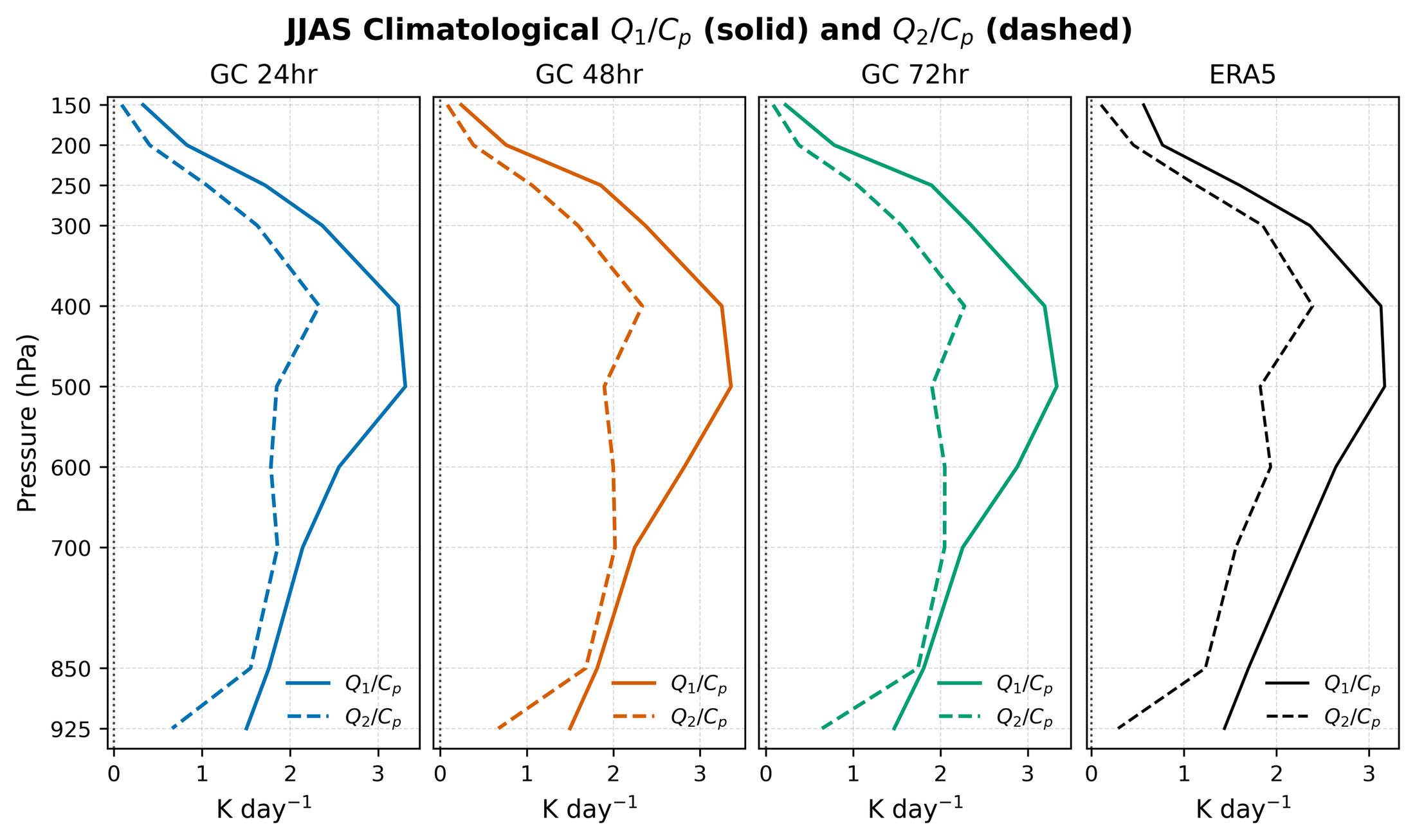}
\includegraphics[width=0.9\textwidth]{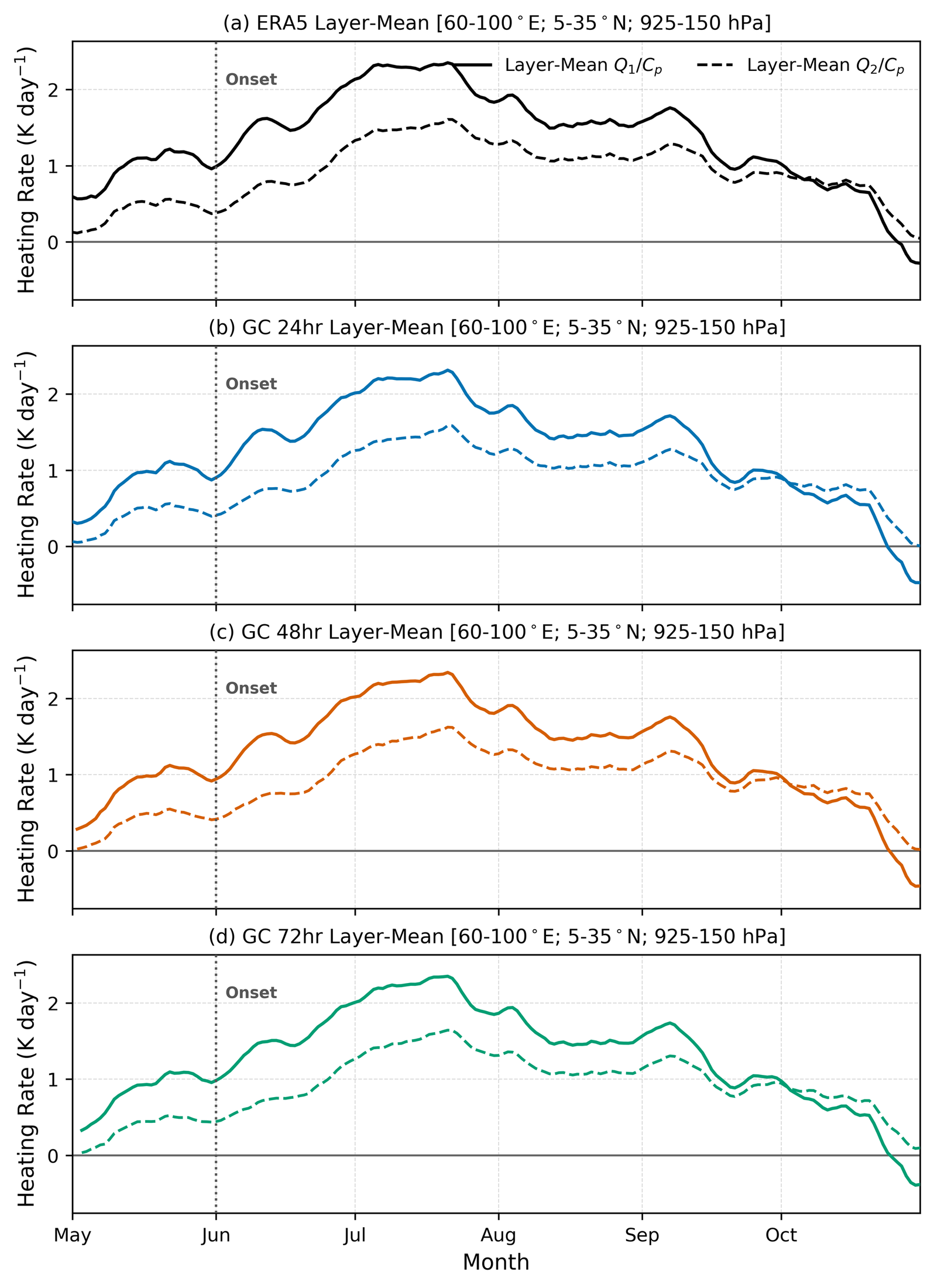}
\caption{Apparent heat source Q1 (solid) and apparent moisture sink Q2 (dashed), both in K day$^{-1}$. (a--b) JJAS-mean vertical profiles, area-averaged over the monsoon core region (10\textdegree--25\textdegree N, 70\textdegree--90\textdegree E). (c--d) Seasonal evolution of the 925--150 hPa layer-mean, area-averaged over the broader Indian summer monsoon domain (5\textdegree--35\textdegree N, 60\textdegree--100\textdegree E). ERA-5 is shown in black; GraphCast +24h, +48h, +72h in blue, orange, and green respectively.}
\label{fig:5}
\end{figure}

The apparent heat source Q1 and apparent moisture sink Q2 vertical profiles provide the thermodynamic ``fingerprint'' of convection and are a stringent test of any model's ability to represent deep tropical convection \citep{Yanai1973}. Q1 and Q2 are computed from the full Yanai budget (local tendency, horizontal advection, and vertical advection terms) evaluated on native 6-hourly fields, which preserves the full amplitude of the local tendency term relative to a daily-mean formulation. Area-averaged over the monsoon core region (10\textdegree--25\textdegree N, 70\textdegree--90\textdegree E) for JJAS, ERA-5 Q1 shows a top-heavy profile with maximum heating near 400--500 hPa ($\sim$3.1--3.2 K day$^{-1}$), reflecting the dominance of deep stratiform precipitation during the active monsoon (Fig.~\ref{fig:5}a). The Q2 profile exhibits a moisture-sink maximum near 400 hPa ($\sim$2.4 K day$^{-1}$), with Q1 and Q2 of similar magnitude and vertical structure throughout the column (Fig.~\ref{fig:5}b), the classic signature of deep cumulus convection first identified by \citet{Yanai1973} and confirmed specifically for the Bay of Bengal sector of the Asian summer monsoon by \citet{YanaiTomita1998}, who likewise report Q1 and Q2 ``similar to each other'' and both strongly positive during the active monsoon season.

GraphCast at +24h, +48h, and +72h reproduces the shape and magnitude of both profiles closely, including the levels of peak heating (500 hPa) and peak moisture-sink strength (400 hPa). Q1 in the 850--925 hPa layer (1.76--1.81 K day$^{-1}$ across the three leads) matches or slightly exceeds ERA-5 (1.44--1.70 K day$^{-1}$) at the same levels, indicating that GraphCast's boundary-layer heating is well represented. A modest difference appears in Q2 near the surface, where GraphCast's moisture sink at 925 hPa (0.64--0.67 K day$^{-1}$) runs somewhat higher than ERA-5's (0.29 K day$^{-1}$), suggesting a slight overestimate of boundary-layer moisture consumption in GraphCast relative to ERA-5 at this single level; this is a second-order feature against the otherwise close agreement across the rest of the column.

The seasonal evolution of layer-mean (925--150 hPa) Q1 and Q2 (Fig.~\ref{fig:5}c,d, averaged over 5\textdegree--35\textdegree N, 60\textdegree--100\textdegree E) shows both quantities rising from near-zero in early May, strengthening through onset (early June) to a broad maximum in July--August, and declining through the withdrawal phase (September--October). GraphCast reproduces this seasonal cycle closely at all three leads, with Q1 and Q2 remaining positive and tracking each other throughout the season in both ERA-5 and GraphCast. This co-varying, same-sign relationship is the expected signature of deep cumulus convection \citep{Yanai1973} and is independently confirmed for this same region by the NCEP--NCAR-reanalysis-based climatology of \citet{YanaiTomita1998}, the AIRS/MERRA observational analysis of \citet{Wong2011}, and a recent ERA-5-based composite analysis of Indian monsoon depressions \citep{Goswami2026}, all of which report Q1 and Q2 strengthening together rather than in opposition. We discuss this finding further, including its relation to \citet{Mukhopadhyay2010}, in Section~\ref{sec:discussion}.

\subsection{Meridional Structure and Wind Shear}\label{subsec:shear}

The meridional cross-section of JJAS mean zonal wind from ERA-5 reveals the characteristic bi-directional shear of the ISM: low-level westerlies (the LLJ peaking near 850 hPa) and upper-level easterlies (the Tropical Easterly Jet, TEJ, peaking near 150--200 hPa). The easterly wind shear (U200~--~U850) provides a bulk measure of this vertical structure and is closely correlated with monsoon intensity on interannual timescales \citep{WebsterYang1992}.

\begin{figure}[h]
\centering
\includegraphics[width=0.9\textwidth]{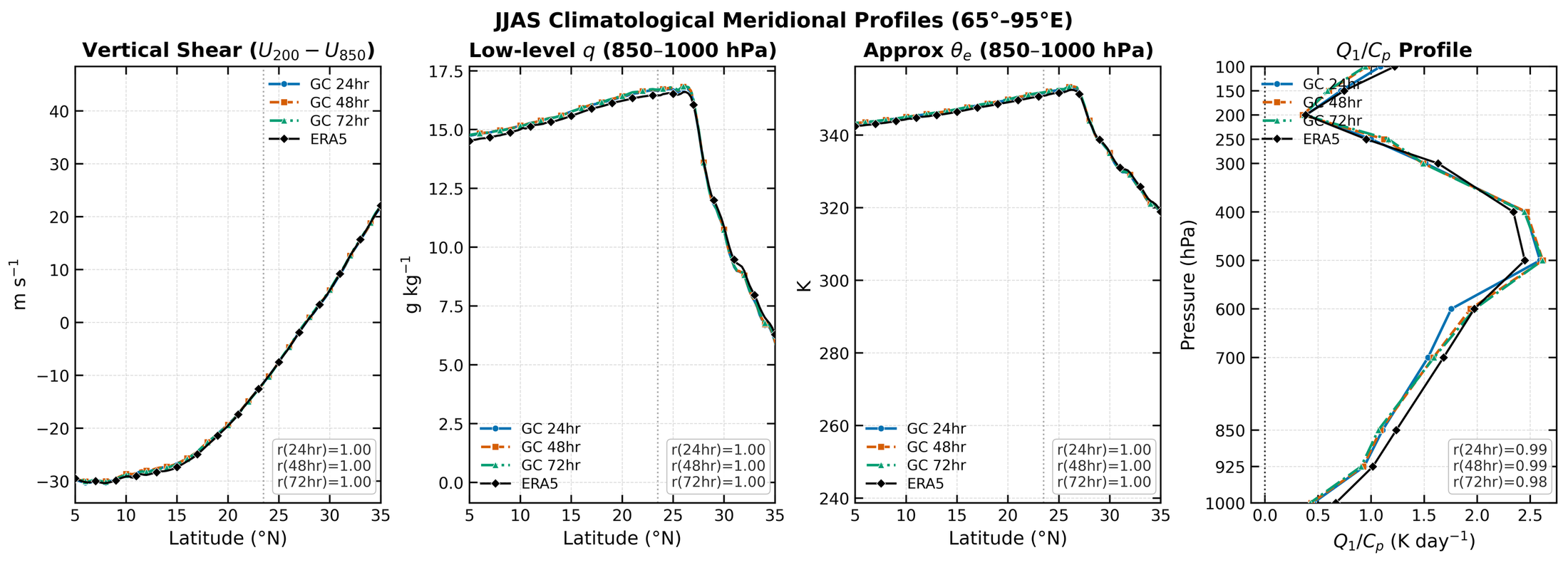}
\includegraphics[width=0.7\textwidth]{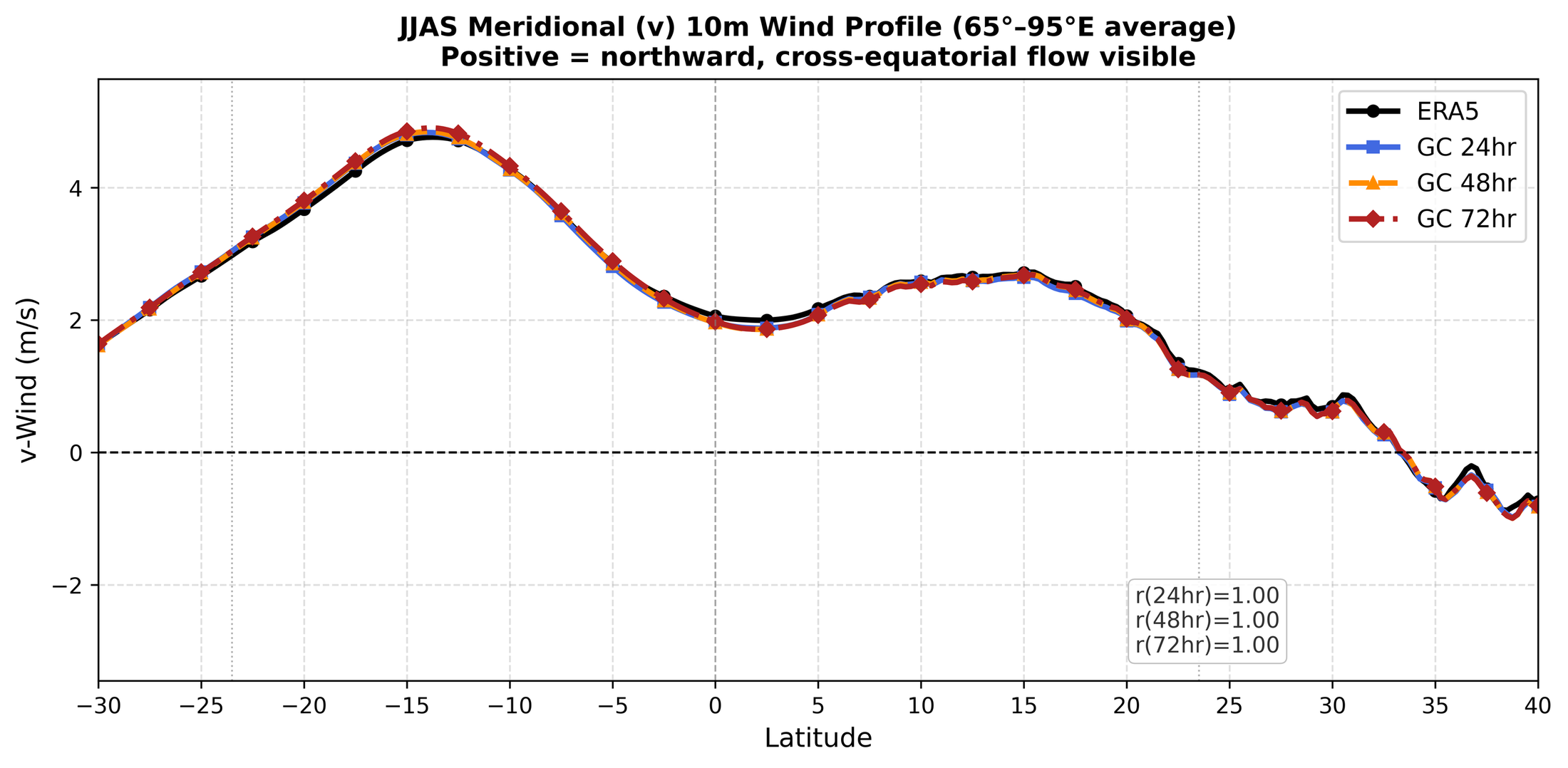}
\caption{JJAS-mean meridional structure of the monsoon circulation and thermodynamics, averaged over 65\textdegree--95\textdegree E, for ERA-5 and GraphCast +24h/+48h/+72h as a function of latitude: (a) vertical wind shear ($U_{200}-U_{850}$), (b) low-level specific humidity (850--1000 hPa), (c) low-level equivalent potential temperature $\theta_e$ (850--1000 hPa), and (d) the JJAS-mean $Q_1/C_p$ vertical profile, shown for reference alongside the latitude-dependent diagnostics in (a--c). (e) Meridional (v) component of the 10-m wind as a function of latitude over the same longitude band, showing the cross-equatorial flow into the monsoon trough. Pearson correlation coefficients ($r$) between each GraphCast lead and ERA-5 are annotated in each panel.}
\label{fig:6}
\end{figure}

GraphCast reproduces the meridional structure of the cross-equatorial flow with good fidelity at +24h (Fig.~\ref{fig:6}). The LLJ core position and intensity are captured within 10--15\% of ERA-5, though the TEJ is slightly weaker at +48h and +72h leads (approximately 5--10\% reduction in peak easterly wind), consistent with the weakened upper-level thermodynamic driving noted in the TT analysis. The cross-equatorial Somali jet, which feeds moisture into the Arabian Sea branch of the LLJ, shows no systematic bias at +24h but exhibits a modest weakening at +72h, contributing to reduced moisture flux convergence over the monsoon core region and the associated dry bias noted in Section~\ref{subsec:spatial}. Domain-averaged JJAS thermodynamic and dynamic statistics for ERA-5 and each GraphCast lead are summarized in Table~\ref{tab:stats}, and the corresponding forecast skill (pattern correlation, RMSE, and bias) of the $Q_1/C_p$ profile and the $U_{200}-U_{850}$ wind shear relative to ERA-5 is reported in Table~\ref{tab:skill}.

\begin{table}[h]
\caption{Domain-averaged JJAS thermodynamic and dynamic statistics (5\textdegree--35\textdegree N, 60\textdegree--100\textdegree E).}
\label{tab:stats}
\begin{tabular}{lcccc}
\toprule
\textbf{Statistic} & \textbf{ERA-5} & \textbf{GC +24h} & \textbf{GC +48h} & \textbf{GC +72h} \\
\midrule
$Q_1/C_p$ at 400--500 hPa (K day$^{-1}$) & 2.47 & 2.55 & 2.56 & 2.55 \\
$Q_2/C_p$ at 400--500 hPa (K day$^{-1}$) & 1.61 & 1.59 & 1.60 & 1.60 \\
Level of peak $Q_1$ (hPa) & 500 & 500 & 500 & 500 \\
Level of peak $Q_2$ (hPa) & 400 & 400 & 400 & 400 \\
\midrule
Peak easterly wind shear (m s$^{-1}$) & $-$30.39 & $-$30.34 & $-$30.12 & $-$30.11 \\
Peak low-level $q$ (g kg$^{-1}$) & 16.60 & 16.75 & 16.82 & 16.84 \\
Peak low-level $\theta_e$ (K) & 352.43 & 353.05 & 353.34 & 353.41 \\
\botrule
\end{tabular}
\end{table}

\begin{table}[h]
\caption{Statistical skill of GraphCast forecasts relative to ERA-5 reanalysis.}
\label{tab:skill}
\begin{tabular}{lcccccc}
\toprule
& \multicolumn{3}{c}{\textbf{$Q_1/C_p$ profile}} & \multicolumn{3}{c}{\textbf{Wind shear ($U_{200}-U_{850}$)}} \\
\cmidrule(lr){2-4} \cmidrule(lr){5-7}
\textbf{Lead} & \textbf{PCC} & \textbf{RMSE} & \textbf{Bias} & \textbf{PCC} & \textbf{RMSE} & \textbf{Bias} \\
\midrule
GC +24h & 0.993 & 0.11 & $+$0.01 & 1.000 & 0.34 & $+$0.18 \\
GC +48h & 0.992 & 0.16 & $+$0.06 & 1.000 & 0.38 & $+$0.26 \\
GC +72h & 0.989 & 0.17 & $+$0.06 & 1.000 & 0.31 & $+$0.23 \\
\botrule
\end{tabular}
\end{table}

\FloatBarrier
\section{Rainfall Variability and Propagation}\label{sec:variability}

\subsection{Intraseasonal Variability (ISV)}\label{subsec:isv}

\begin{figure}[h]
\centering
\includegraphics[width=0.9\textwidth]{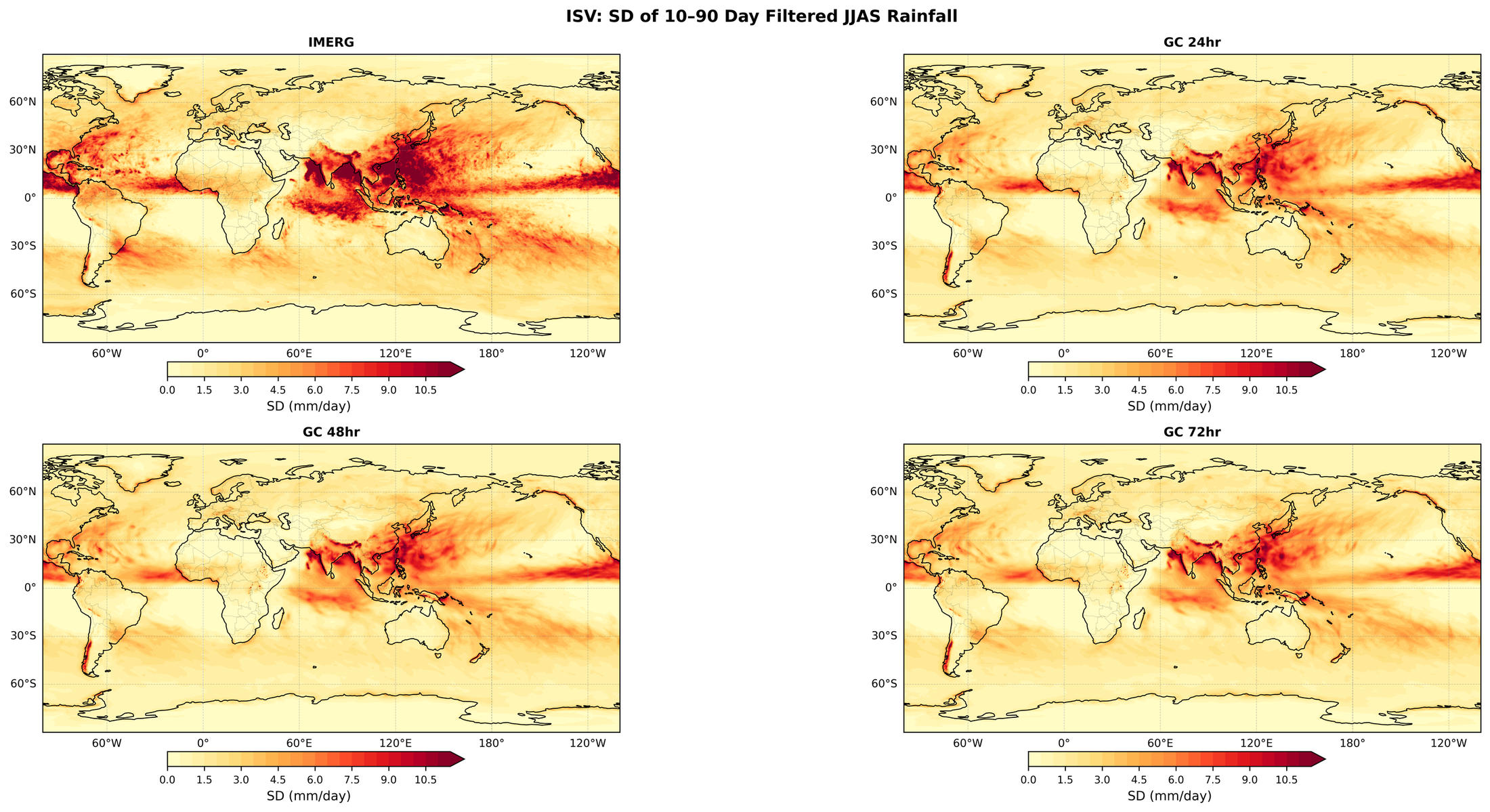}
\includegraphics[width=0.9\textwidth]{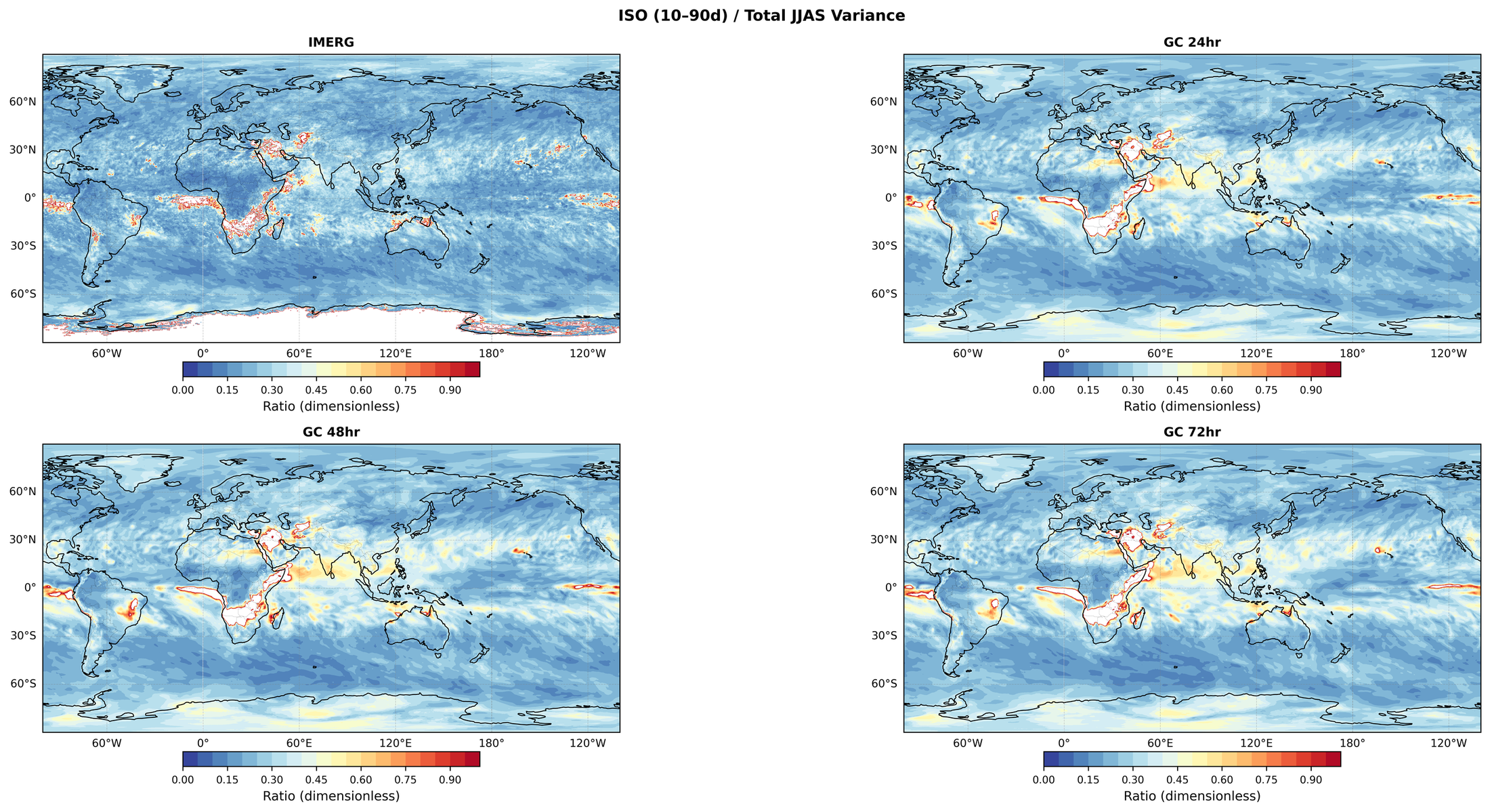}
\caption{(a) Standard deviation of 10--90 day bandpass-filtered daily rainfall (mm day$^{-1}$) for IMERG (top) and GraphCast +24h/+48h/+72h. (b) Ratio of 10--90 day intraseasonal rainfall variance to total JJAS rainfall variance for IMERG (top) and GraphCast +24h/+48h/+72h. See text for why elevated ratios over marginal-rainfall regions in GraphCast likely reflect disproportionate seasonal-band variance loss rather than ISO amplification.}
\label{fig:7}
\end{figure}

The standard deviation of 10--90 day bandpass-filtered daily rainfall provides a measure of the amplitude of intraseasonal variability (ISV) associated with the BSISO, MJO, and quasi-biweekly modes. In IMERG, the ISV amplitude is highest over the Bay of Bengal and over the monsoon core region, reflecting the active-break cycle of the ISM (Fig.~\ref{fig:7}a). GraphCast at +24h shows a broadly similar spatial pattern of ISV amplitude but with a visibly reduced magnitude almost everywhere, consistent with the spectral variance suppression quantified in Section~\ref{subsec:spectral}.

The ratio of 10--90 day variance to total JJAS rainfall variance (Fig.~\ref{fig:7}b) requires careful interpretation: GraphCast shows locally elevated ratios relative to IMERG over several marginal-rainfall, low-total-variance regions (e.g.\ the Horn of Africa, the Arabian Peninsula margins, and parts of the Sahel), even though the PSD-based band-integrated power ratios (Section~\ref{subsec:spectral}) show a fairly uniform suppression across timescales (0.53--0.66 from the synoptic to the annual band) rather than a markedly uneven one. A locally elevated ratio here does not, by itself, indicate that GraphCast amplifies intraseasonal variability in an absolute sense: because the ratio is a quotient of intraseasonal-band variance to total variance, a region where GraphCast's total (denominator) variance is sharply reduced relative to IMERG's, as is generically the case over these low-rainfall, high climatological-noise regions, can show an elevated ratio even when its intraseasonal-band variance is itself suppressed in an absolute sense, provided it is suppressed somewhat less than the total. We therefore interpret the locally elevated ratios over marginal-rainfall regions as reflecting this denominator effect rather than a genuine, region-specific amplification of intraseasonal rainfall variability, consistent with the broadly suppressed, not amplified, ISV amplitude documented directly in Fig.~\ref{fig:7}a.

This ISV amplitude suppression likely reflects a combination of factors: the deterministic nature of GraphCast precluding the representation of stochastic convective triggering, the model's difficulty in sustaining coherent, slowly evolving moisture anomalies on intraseasonal timescales, and the absence of air--sea coupling, which is known to amplify BSISO propagation through SST--convection feedbacks \citep{Fu2003}.

\begin{figure}[h]
\centering
\includegraphics[width=0.9\textwidth]{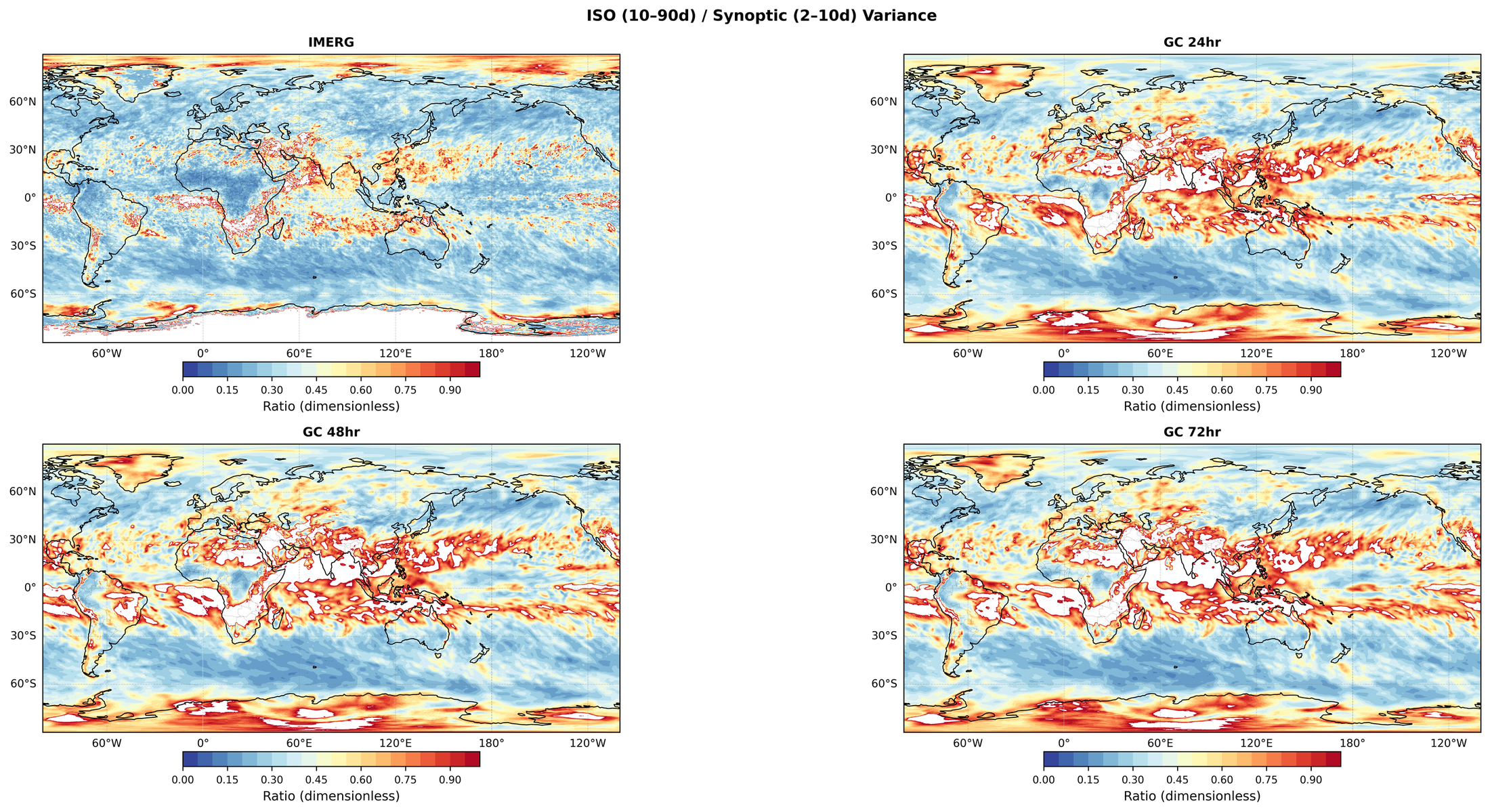}
\caption{Ratio of 10--90 day intraseasonal rainfall variance to synoptic-band (2--10 day) rainfall variance, for IMERG and GraphCast +24h/+48h/+72h. Provided alongside Fig.~\ref{fig:7}b as a corroborating decomposition; see text for why elevated ratios over marginal-rainfall regions reflect a collapsing denominator (total/seasonal-band variance) rather than genuine intraseasonal amplification.}
\label{fig:ratio_iso_syn}
\end{figure}

\begin{figure}[h]
\centering
\includegraphics[width=0.9\textwidth]{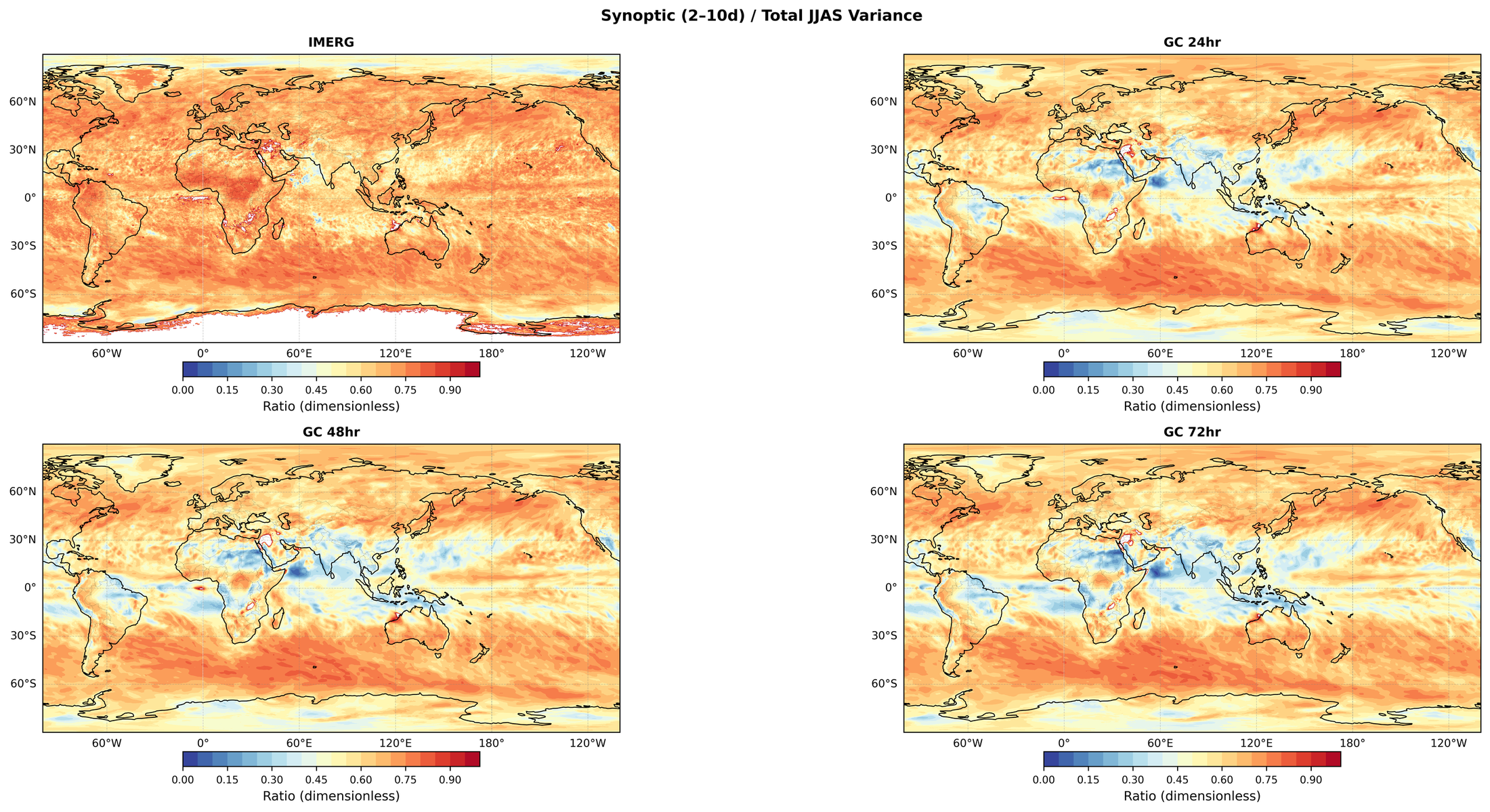}
\caption{Ratio of synoptic-band (2--10 day) rainfall variance to total JJAS rainfall variance, for IMERG and GraphCast +24h/+48h/+72h. Provided alongside Fig.~\ref{fig:7}b and Fig.~\ref{fig:ratio_iso_syn} as a corroborating decomposition.}
\label{fig:ratio_syn_total}
\end{figure}

\subsection{Northward and Eastward Propagation}\label{subsec:propagation}

\begin{figure}[h]
\centering
\includegraphics[width=0.9\textwidth]{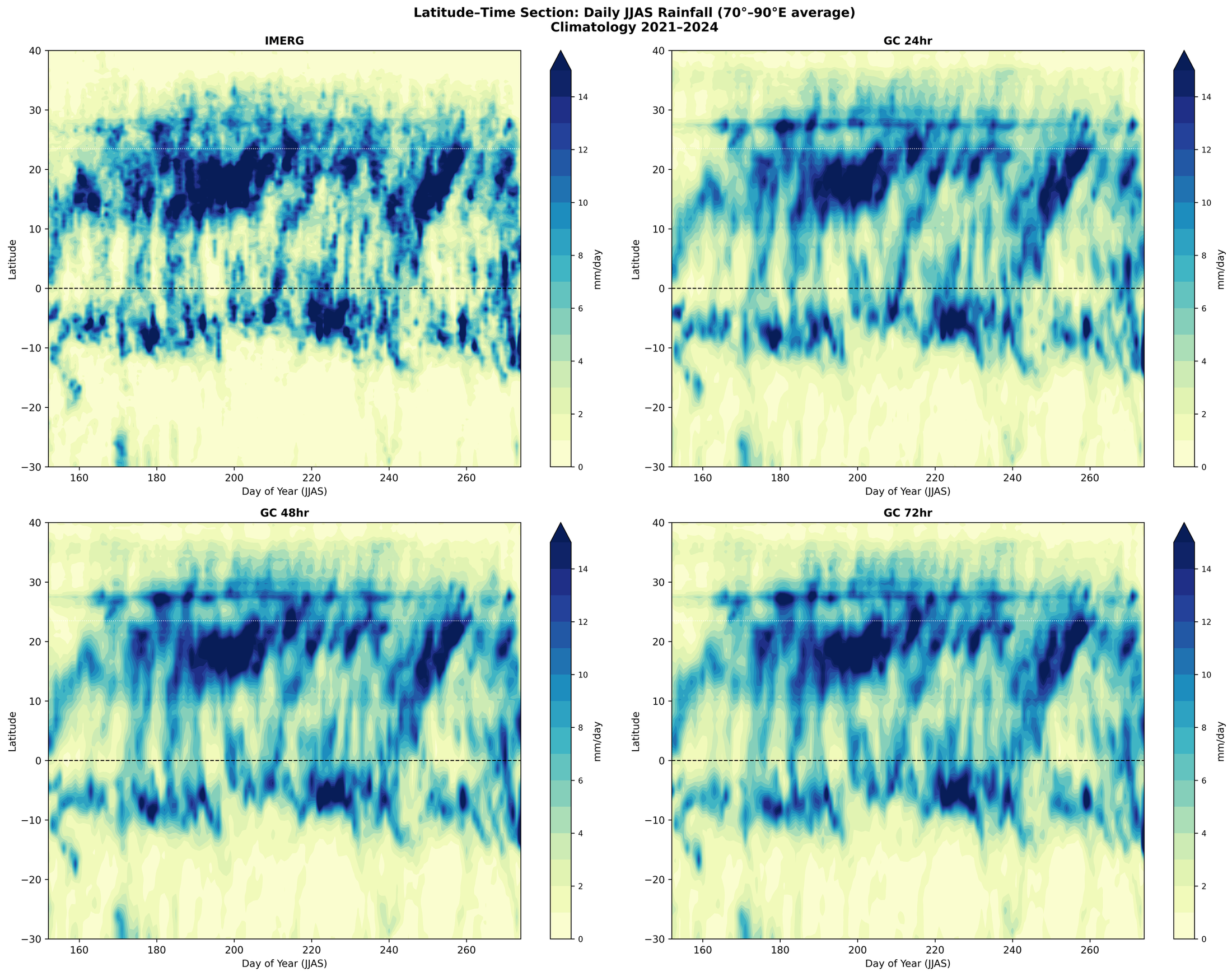}
\includegraphics[width=0.9\textwidth]{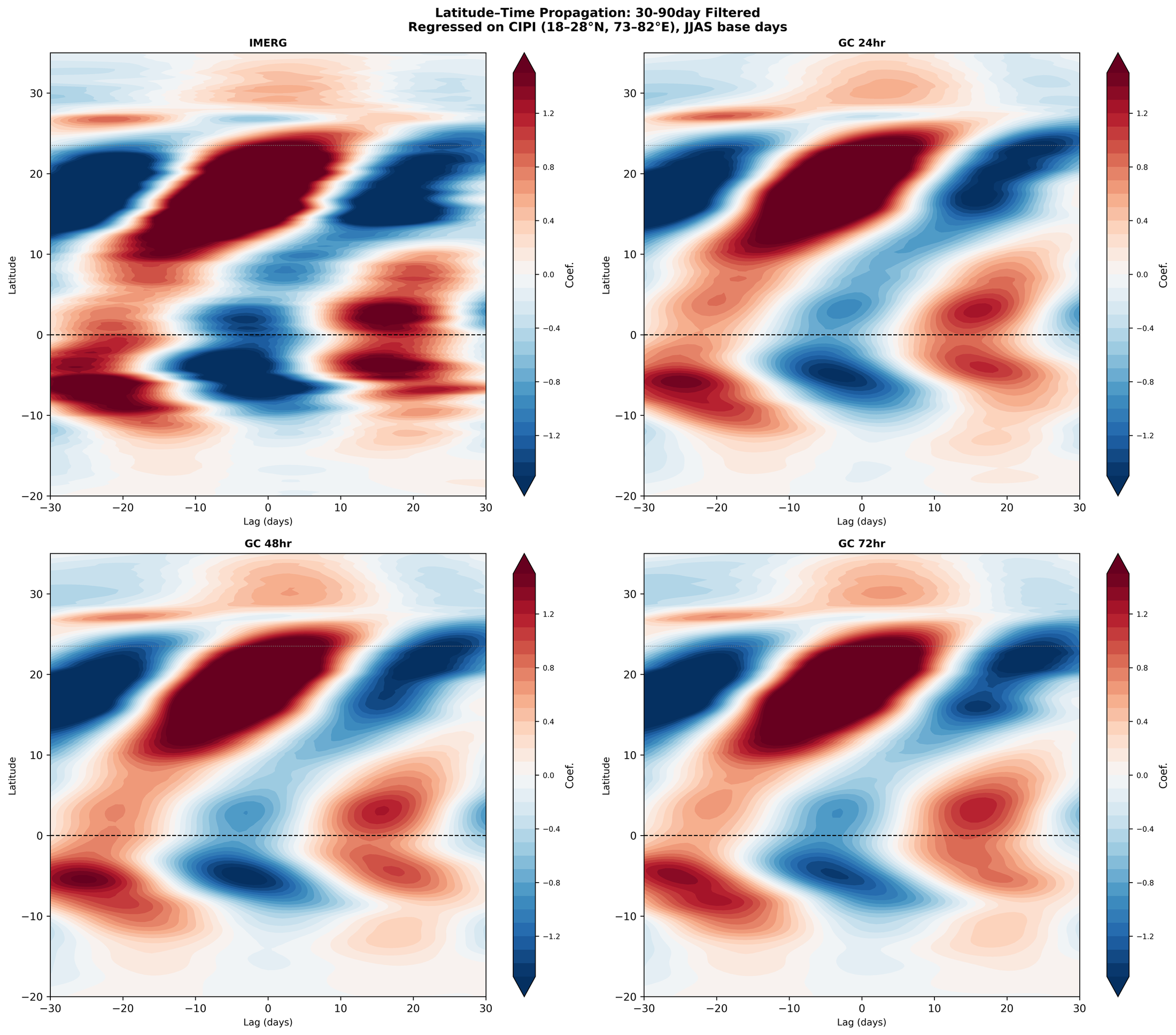}
\caption{Hovm\"oller diagrams (latitude--time) of 30--90 day bandpass-filtered daily rainfall anomalies (mm day$^{-1}$) averaged over 70\textdegree--90\textdegree E for (a) IMERG and (b--d) GraphCast +24h/+48h/+72h, composited across 2021--2024. Positive (negative) anomalies are shown in red (blue). Note the thin, nearly time-invariant features at $\sim$22--23\textdegree N and $\sim$27--28\textdegree N in the GraphCast panels (see text).}
\label{fig:8}
\end{figure}

The Hovm\"oller diagram (latitude--time) of 30--90 day filtered rainfall averaged over 70\textdegree--90\textdegree E is widely regarded as the definitive benchmark for evaluating monsoon ISV in models \citep{Lee2013,Waliser2009}. In IMERG, coherent northward-propagating rainbands are visible from the equatorial Indian Ocean (approximately 5\textdegree S) to 25\textdegree N over approximately 30--40 days, representing the canonical BSISO signal (Fig.~\ref{fig:8}a). The phase speed of northward propagation is approximately 0.8--1.0\textdegree\ day$^{-1}$.

GraphCast at +24h reproduces the general character of northward propagation, with discernible rainband features propagating from the equatorial region toward 15--20\textdegree N. However, the propagation is less coherent than observed: the phase speed is slightly underestimated ($\sim$0.6--0.7\textdegree\ day$^{-1}$), and the rainbands exhibit weaker amplitude and less organized spatial structure. At +72h, the northward propagation signal is further degraded, with intraseasonal energy appearing more stationary in places. We additionally note the presence of thin, nearly time-invariant horizontal features at approximately 22--23\textdegree N and 27--28\textdegree N in the GraphCast panels of Fig.~\ref{fig:8}, absent from IMERG; these coincide with regions of strong climatological orographic precipitation (the Himalayan foothills and adjoining terrain) and warrant further investigation as a potential terrain-locked smoothing signature independent of the domain-mean statistics discussed in Section~\ref{sec:discussion}.

\begin{figure}[h]
\centering
\includegraphics[width=0.9\textwidth]{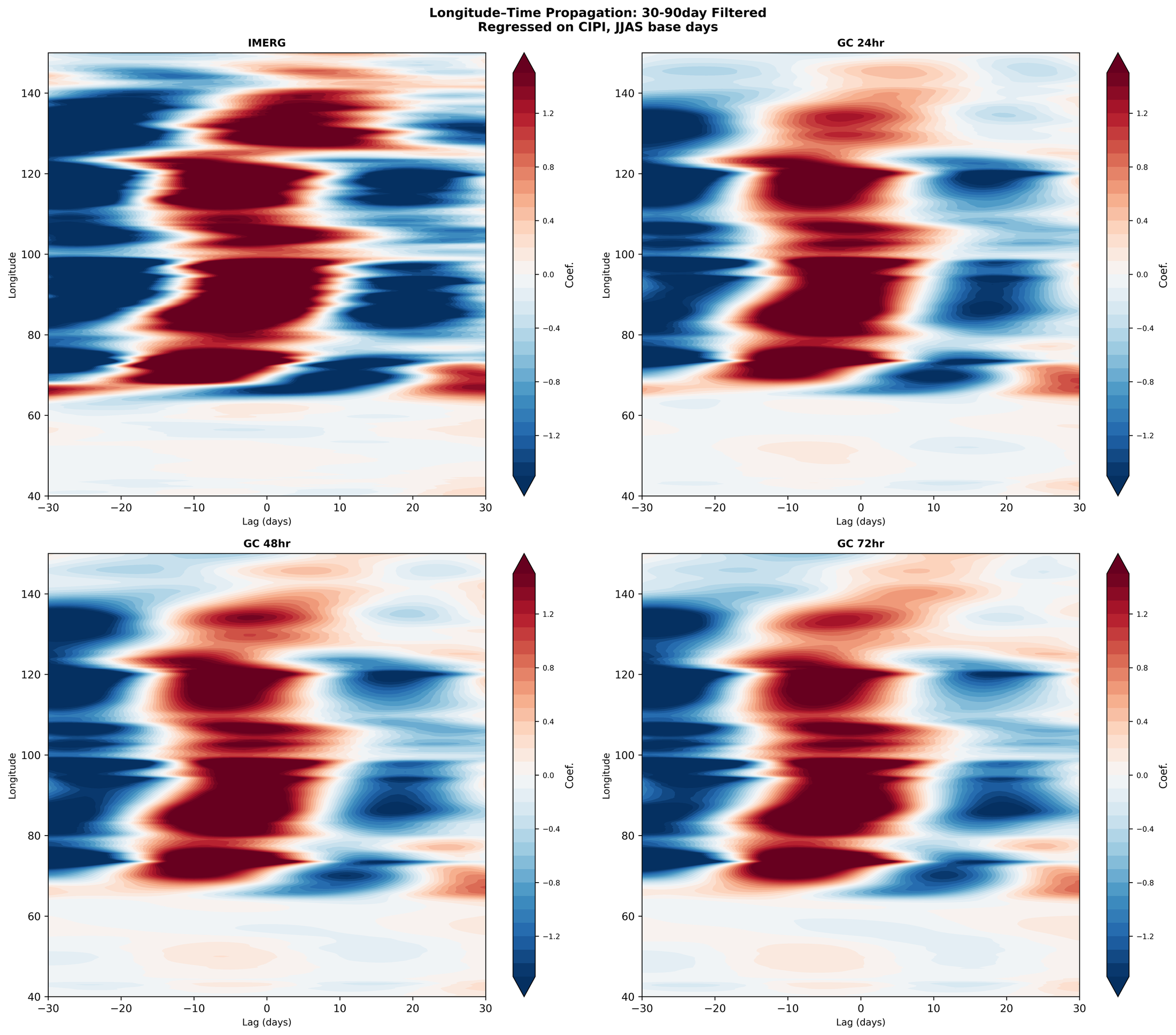}
\caption{Longitude--time regression of 30--90 day bandpass-filtered rainfall anomalies onto a Central Indian Precipitation Index (CIPI) base-day series, for IMERG and GraphCast +24h/+48h/+72h.}
\label{fig:9}
\end{figure}

The eastward-propagating envelope is characterized independently using a longitude--time regression of 30--90 day filtered rainfall onto a Central Indian Precipitation Index (CIPI) base-day series (Fig.~\ref{fig:9}), which isolates the canonical eastward-propagating signal from the noisier longitude--time Hovm\"oller of any single realization. IMERG shows a coherent eastward-propagating envelope crossing the Maritime Continent (60--150\textdegree E) over roughly 20--30 lagged days, consistent with the MJO/BSISO eastward group velocity. GraphCast reproduces this envelope's structure and approximate phase speed closely at all three lead times, including the leading/trailing out-of-phase anomalies at $\pm$20 days, in clear contrast to the northward-propagation degradation described above. This asymmetry, with eastward propagation preserved better than northward propagation, is discussed further in Section~\ref{sec:discussion}. The quasi-biweekly (10--20 day) band shows a qualitatively similar propagation structure in both latitude--time and longitude--time sections (Supplementary Fig.~S2--S3).

\subsection{Spectral Analysis}\label{subsec:spectral}

\begin{figure}[h]
\centering
\includegraphics[width=\textwidth]{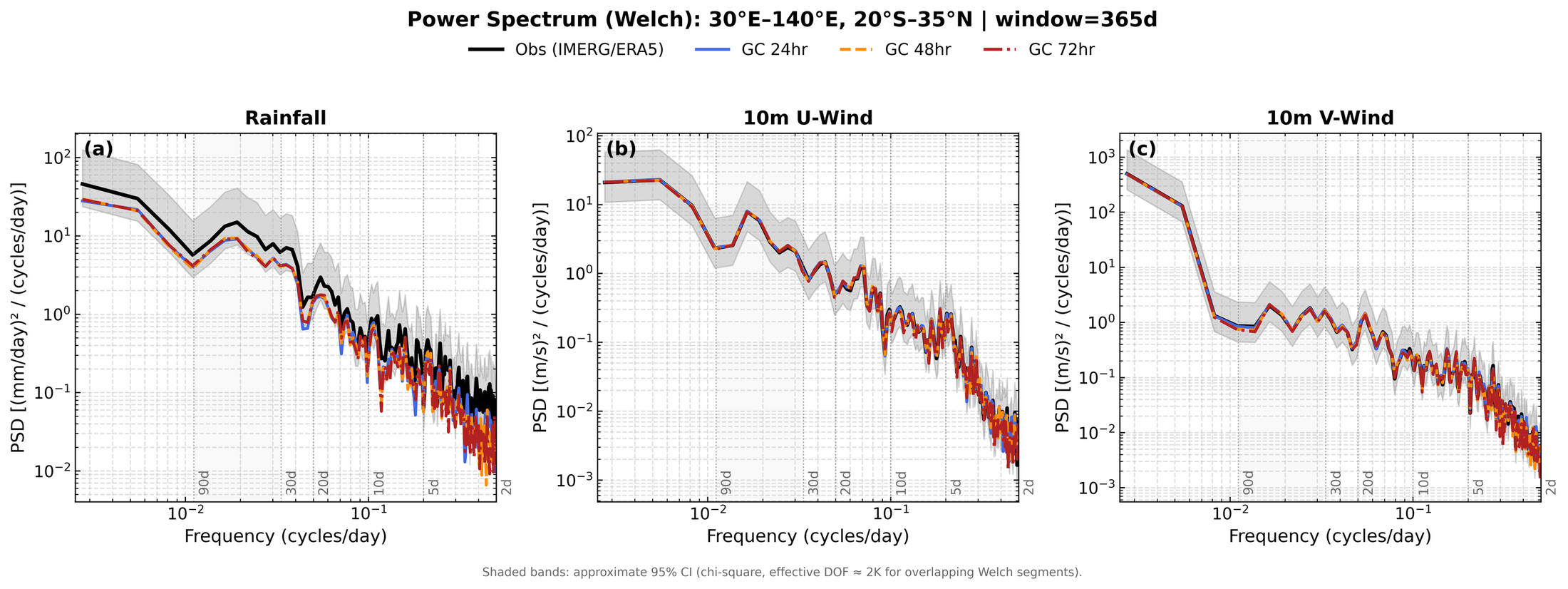}
\includegraphics[width=\textwidth]{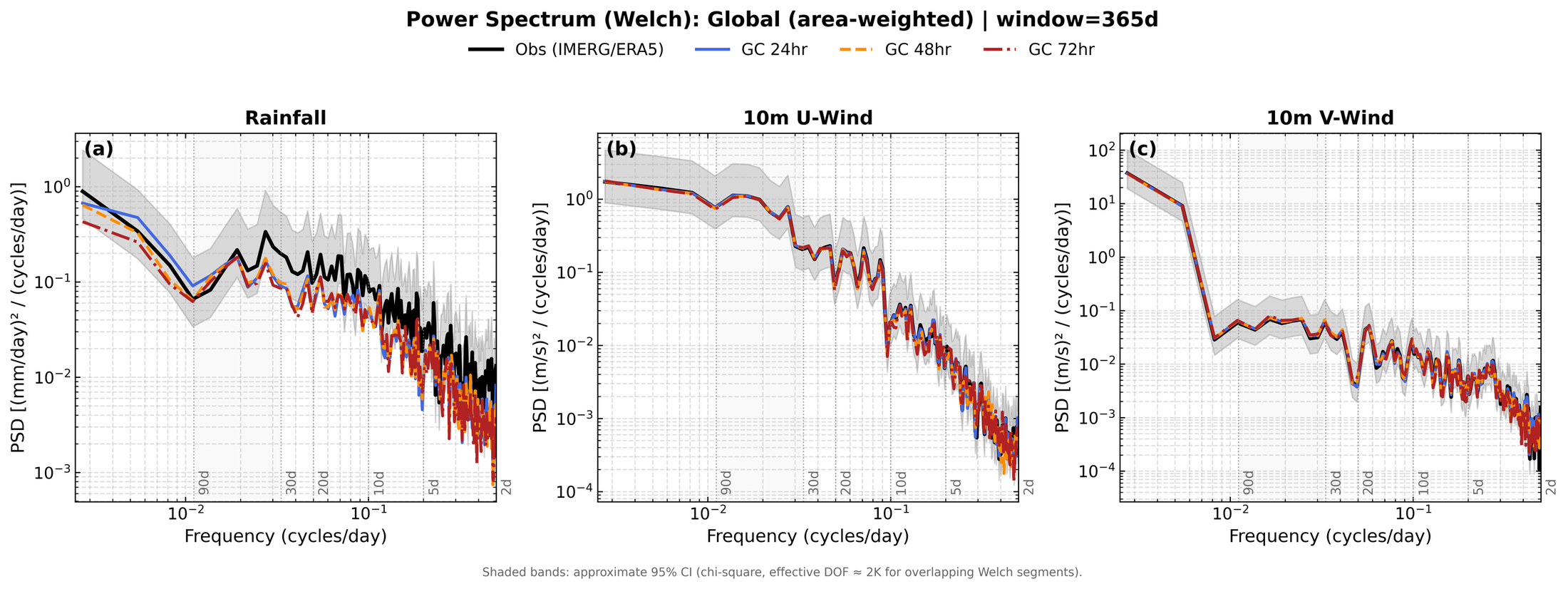}
\caption{Power spectral density of the area-weighted domain-mean daily time series of rainfall (a), 10-m U-wind (b), and 10-m V-wind (c), for the regional ISM domain (top; 30\textdegree E--140\textdegree E, 20\textdegree S--35\textdegree N) and globally (bottom), for IMERG/ERA5 (black) and GraphCast +24h/+48h/+72h (blue/orange/red). Shaded bands: approximate 95\% CI (chi-square, effective DOF $\approx 2K$ for overlapping Welch segments; see Section~\ref{subsec:metrics}).}
\label{fig:psd}
\end{figure}

The power spectral density (PSD) of the area-weighted domain-mean daily rainfall and 10-m wind time series over the regional ISM domain (30\textdegree E--140\textdegree E, 20\textdegree S--35\textdegree N) provides a quantitative characterization of the distribution of variance across timescales (Fig.~\ref{fig:psd}; Tables~\ref{tab:psd_rain}--\ref{tab:psd_u10}). IMERG rainfall power decreases gradually from interannual/seasonal frequencies down to a broad shoulder near the synoptic band, consistent with the known structure of ISM rainfall variability \citep{Krishnamurti1980}. GraphCast's rainfall spectrum sits moderately below IMERG's across most of the resolved spectrum: the grid-mapped Total Var Ratio is 0.63 at all three lead times (Table~\ref{tab:psd_rain}), meaning GraphCast's rainfall time series retains roughly 63\% of IMERG's total temporal variance regardless of lead. This suppression is fairly uniform across bands, ranging from 0.53--0.58 (Synoptic) to 0.65--0.66 (Annual), with the Seasonal, MJO, Subseasonal, and QBW bands all falling within approximately 0.58--0.66; unlike a single dominant band, the moderate variance deficit is spread broadly and evenly across timescales. The log-spectral shape correlation (0.96--0.97) is high, meaning GraphCast captures the overall \emph{shape} of the spectral decay well, while the Spectral NSE score declines from 0.73 at +24h to 0.63 at +48h and 0.59 at +72h, quantifying a modest, monotonic degradation in spectral fidelity with lead time superimposed on the moderate, lead-invariant amplitude suppression. The spectral slope over the 2--30 day band steepens from $-$1.75 (+24h) to $-$1.93 (+72h), versus $-$1.56 in IMERG (Table~\ref{tab:psd_slopes}), indicating that GraphCast loses relatively more power at the higher-frequency end of this band as lead time increases.

The 10-m wind components show a markedly different picture: both U10 and V10 achieve a Total Var Ratio of 0.98--1.01 and a Spectral NSE of 0.97--0.99 at all leads (Table~\ref{tab:psd_rain}), confirming that GraphCast reproduces both the amplitude and the detailed spectral shape of the wind field with high fidelity across the entire resolved frequency range, in sharp contrast to rainfall. This dichotomy, near-perfect spectral fidelity for the dynamical wind field alongside severe, broadband suppression for rainfall, is a central, recurring pattern in this study (cf.\ Sections~\ref{subsec:intensity} and \ref{subsec:spatial}) and is discussed mechanistically in Section~\ref{sec:discussion}. The supplementary wind-speed PSD (Supplementary Fig.~S1), computed from the area-averaged wind-speed \emph{magnitude} time series rather than from a combination of the U/V component spectra (see Section~\ref{subsec:metrics}), shows the same high fidelity as the individual components.

\begin{table}[h]
\caption{PSD spectral skill metrics (regional domain). Shape Corr, Log RMSE, Log MAE, and Spectral NSE (bounded above by 1; not bounded below) are computed on log10 PSD over resolved frequencies ($f\geq1/180$ day$^{-1}$). Total Var Ratio is a grid-mapped power ratio (model PSD log-log interpolated onto the observational frequency grid before both are integrated on that common grid; see Section~\ref{subsec:metrics}), not a strict Parseval-consistent variance decomposition.}
\label{tab:psd_rain}
\begin{tabular}{llccccc}
\toprule
\textbf{Var.} & \textbf{Lead} & \textbf{Shape Corr} & \textbf{Log RMSE} & \textbf{Log MAE} & \textbf{Spectral NSE} & \textbf{Total Var} \\
\midrule
Rain & +24h & 0.97 & 0.31 & 0.27 & 0.73 & 0.63 \\
Rain & +48h & 0.96 & 0.37 & 0.32 & 0.63 & 0.63 \\
Rain & +72h & 0.96 & 0.39 & 0.32 & 0.59 & 0.63 \\
\midrule
U10  & +24h & 1.00 & 0.06 & 0.05 & 0.99 & 1.01 \\
U10  & +48h & 0.99 & 0.09 & 0.07 & 0.97 & 0.99 \\
U10  & +72h & 0.99 & 0.10 & 0.08 & 0.97 & 0.98 \\
\midrule
V10  & +24h & 1.00 & 0.05 & 0.04 & 0.99 & 1.00 \\
V10  & +48h & 0.99 & 0.08 & 0.06 & 0.98 & 0.99 \\
V10  & +72h & 0.99 & 0.09 & 0.07 & 0.97 & 0.98 \\
\botrule
\end{tabular}
\end{table}

\begin{table}[h]
\caption{PSD band-integrated power ratios (regional domain; grid-mapped, see Table~\ref{tab:psd_rain} caption and Section~\ref{subsec:metrics}). Bands: Annual (180--365d), Seasonal (90--180d), MJO (30--90d), Subseasonal (20--30d), QBW (10--20d), Synoptic (2--10d).}
\label{tab:psd_u10}
\begin{tabular}{llcccccc}
\toprule
\textbf{Var.} & \textbf{Lead} & \textbf{Annual} & \textbf{Seasonal} & \textbf{MJO} & \textbf{Subseas.} & \textbf{QBW} & \textbf{Synoptic} \\
\midrule
Rain & +24h & 0.65 & 0.65 & 0.62 & 0.58 & 0.64 & 0.58 \\
Rain & +48h & 0.66 & 0.62 & 0.64 & 0.61 & 0.63 & 0.54 \\
Rain & +72h & 0.66 & 0.66 & 0.62 & 0.60 & 0.66 & 0.53 \\
\midrule
U10  & +24h & 1.01 & 1.00 & 0.99 & 0.99 & 0.98 & 0.99 \\
U10  & +48h & 1.00 & 0.99 & 0.97 & 0.97 & 0.96 & 0.97 \\
U10  & +72h & 0.99 & 0.98 & 0.96 & 0.96 & 0.95 & 0.96 \\
\midrule
V10  & +24h & 1.00 & 1.00 & 0.99 & 0.99 & 0.99 & 0.99 \\
V10  & +48h & 1.00 & 0.99 & 0.98 & 0.97 & 0.97 & 0.97 \\
V10  & +72h & 0.99 & 0.98 & 0.97 & 0.96 & 0.96 & 0.96 \\
\botrule
\end{tabular}
\end{table}

\begin{table}[h]
\caption{Spectral slope (2--30 day band, log-log linear fit) comparison.}
\label{tab:psd_slopes}
\begin{tabular}{lccc}
\toprule
\textbf{Product} & \textbf{Rainfall} & \textbf{U10} & \textbf{V10} \\
\midrule
Obs   & $-$1.54 & $-$2.05 & $-$2.07 \\
GC +24h & $-$1.75 & $-$2.06 & $-$2.08 \\
GC +48h & $-$1.89 & $-$2.10 & $-$2.12 \\
GC +72h & $-$1.93 & $-$2.12 & $-$2.14 \\
\botrule
\end{tabular}
\end{table}

These spectral results are consistent with the ISV analysis in Section~\ref{subsec:isv} and the intensity-distribution results in Section~\ref{subsec:intensity}, and collectively indicate that GraphCast's rainfall variance is suppressed broadly across timescales rather than in any single band, while the dynamical wind field is reproduced with near-observational spectral fidelity. The systematic suppression of rainfall spectral power at subseasonal-to-seasonal frequencies is a significant limitation for extended-range prediction applications, where the BSISO provides the primary source of subseasonal predictability.

\FloatBarrier
\section{Discussion}\label{sec:discussion}

The results of this study reveal a coherent pattern for the Indian Summer Monsoon that can be organized around three themes: (i) a moderate, fairly uniform suppression of rainfall variance and high-intensity rainfall that is consistent across independent diagnostics (spectral, intensity-distribution, and spatial-variability), including a dry bias at both the orographic (Western Ghats) and domain-mean scales; (ii) a closely reproduced large-scale thermodynamic structure, with the Q1/Q2 vertical profiles, seasonal evolution, and DTT annual cycle all tracking ERA-5 well at every lead; and (iii) an asymmetric degradation of subseasonal propagation that is more severe in the northward (BSISO) than the eastward (MJO/Kelvin-wave-like) component.

\subsection*{A deterministic-smoothing signature in GraphCast precipitation}

Three independent diagnostics converge on the same underlying picture for rainfall. The intensity-distribution analysis (Section~\ref{subsec:intensity}) shows GraphCast's distribution right-shifted relative to IMERG: the moderate-heavy 95th-percentile threshold is biased dry ($-$9.4 to $-$9.8~mm~day$^{-1}$ domain-mean) and the rarest, most extreme events (the highest 1--2\% of the intensity range) are even more systematically underrepresented across every sub-region examined. The spectral analysis (Section~\ref{subsec:spectral}) shows GraphCast retaining roughly 63\% of IMERG's total rainfall temporal variance, moderately and fairly evenly suppressed across every timescale from synoptic to seasonal. The ISV amplitude maps (Section~\ref{subsec:isv}) show the same suppression spatially. We interpret this convergent pattern as a deterministic-model smoothing (regression-to-the-mean) signature, consistent with the reduced small-scale spectral energy and loss of physical fidelity documented for other current ML weather-prediction models \citep{Bonavita2024}. A model trained to minimize expected error under genuine day-to-day forecast uncertainty in tropical convection is incentivized toward a smoothed, moderate-amplitude output rather than committing to the uncertain, occasionally extreme, true realization, which simultaneously depresses high-intensity and extreme rainfall and lowers the temporal variance at every timescale where that uncertainty is high. This mechanism is well established for deterministic, MSE-trained data-driven forecast models more broadly. \citet{Subich2025} show directly for GraphCast that standard mean-squared-error training blurs fine-scale structure, reducing its effective resolution from $\sim$1250~km to $\sim$160~km, and that a modified loss function restores sharpness and improves surface-wind and tropical-cyclone extremes, while \citet{Gupta2025} report that current global AI weather models, evaluated against observations during the South Asian monsoon specifically, systematically underpredict extreme rainfall despite strong skill on large-scale variables, the same qualitative pattern, in the same region, found here. This interpretation is also consistent with the markedly higher spectral and intensity fidelity of the 10-m wind field (Section~\ref{subsec:spectral}), which is governed by larger-scale, more deterministically predictable dynamics than convective rainfall.

\subsection*{A note on an IMERG data-processing correction}

An earlier version of this analysis used an IMERG regridding step that was subsequently found to systematically suppress observed rainfall magnitude (by roughly half, more severely in high-rainfall regions than in the global mean) and to be missing substantial fractions of 2022 and 2024 due to incomplete data retrieval, both verified and corrected against the native-resolution IMERG V07 archive before this manuscript was finalized (see Data and Methods). That earlier, uncorrected comparison had suggested a large domain-averaged wet bias in GraphCast's rainfall that appeared, at face value, to sit awkwardly alongside the orographic-scale dry bias over the Western Ghats discussed above. Against the corrected IMERG dataset, this apparent conflict does not arise: GraphCast's domain-mean rainfall bias is small and of the same (dry) sign as the orographic-scale bias at both the Core Monsoon Zone and ISM Domain scales (Section~\ref{subsec:spatial}), consistent with a single, moderate underrepresentation of rainfall intensity rather than two biases of opposite sign requiring separate explanations. We report this correction transparently because it materially changed several of this study's quantitative conclusions, and because the broader lesson, that an observational baseline can itself be the source of an apparent model bias, is relevant to other AI-weather-model evaluation studies that rely on a single regridded satellite product without independent cross-validation against the native-resolution source.

\subsection*{Thermodynamic structure and the Western Ghats dry bias}

The large-scale thermodynamic structure of the monsoon is well reproduced: the Q1/Q2 vertical profiles and seasonal evolution (Section~\ref{subsec:q1q2}) and the DTT annual cycle (Section~\ref{subsec:dtt}) both closely track ERA-5 at all three leads. The orographic-scale dry bias over the Western Ghats (Section~\ref{subsec:spatial}) is therefore better understood not as a distinct thermodynamic deficiency but as a regional expression of the same deterministic-smoothing mechanism identified above: orographically forced precipitation over the Western Ghats is spatially concentrated and high-intensity by nature, and is consequently among the features most susceptible to the regression-to-the-mean suppression that depresses high-intensity rainfall more broadly across the domain.

\subsection*{Asymmetric subseasonal propagation: northward vs.\ eastward}

The propagation results in Section~\ref{subsec:propagation} reveal an asymmetry not previously highlighted in the AI weather model literature for this domain: GraphCast's eastward-propagating (MJO/Kelvin-wave-like) signal, evaluated via the CIPI-regressed longitude--time envelope, is preserved with substantially higher fidelity than its northward-propagating (BSISO) counterpart, evaluated via the latitude--time Hovm\"oller. This is likely attributable to the absence of dynamic air--sea coupling in GraphCast, which prevents the positive feedback between warm SST anomalies and enhanced surface evaporation north of active convective cells, a documented amplifying mechanism for BSISO propagation specifically \citep{Fu2003,Klingaman2011}. The moisture-convergence feedback that drives northward propagation \citep{Jiang2004} is consequently weaker in GraphCast's simulated BSISO envelope than it would be with explicit ocean coupling. Eastward MJO propagation, by contrast, is governed more by large-scale circumnavigating circulation anomalies that may be better constrained by the large-scale dynamical fields GraphCast reproduces with high fidelity (Section~\ref{subsec:spectral}). These factors collectively suggest that improving AI monsoon prediction on intraseasonal timescales will require either explicit coupling with an ocean model, stochastic perturbation schemes, or new training approaches that weight the representation of slowly evolving tropical processes, and that northward-propagating BSISO dynamics specifically, rather than subseasonal variability in general, should be the priority target for such improvements.

\subsection*{Limitations}

It is important to contextualize these findings within the limitations of the study. The evaluation period (2021--2024) encompasses only four monsoon seasons, which may not fully capture the interannual variability of the ISM. Additionally, the comparison of GraphCast against ERA-5 atmospheric variables is not entirely independent, since GraphCast is trained on ERA-5; biases in upper-level wind or temperature fields relative to ERA-5 may reflect genuine model error or, alternatively, the model's divergence from the reanalysis under real initial conditions derived from an independent analysis. The use of IMERG as an independent precipitation benchmark partially mitigates this concern for the rainfall evaluation. The PSD-based spectral metrics in Section~\ref{subsec:spectral} are computed from the area-weighted domain-mean time series (Section~\ref{subsec:metrics}) and therefore characterize coherent, domain-scale variability rather than a spatial average of point-wise spectral behavior; a grid-point-wise spectral analysis is a natural extension for future work. The approximate 95\% confidence intervals on the Welch PSD estimates (Section~\ref{subsec:metrics}) should be interpreted as indicative shading rather than an exact, calibrated inferential interval, given the non-independence of overlapping spectral segments.

A further point of comparison concerns the sign relationship between Q1 and Q2 during active monsoon. \citet{Mukhopadhyay2010} report Q1 and Q2 of opposite sign in their ECMWF (ERA-Interim) reanalysis over the same domain (5\textdegree--35\textdegree N, 60\textdegree--100\textdegree E; their Fig.~14a), whereas our ERA-5 and GraphCast results show Q1 and Q2 of the same sign throughout the annual cycle, strengthening together from onset through the monsoon peak (Fig.~\ref{fig:5}c,d). This co-varying relationship is the standard signature of deep cumulus convection established by \citet{Yanai1973} and independently confirmed for this same monsoon system by \citet{YanaiTomita1998} using NCEP--NCAR reanalysis, by \citet{Wong2011} using AIRS/MERRA observations, and by \citet{Goswami2026} using ERA-5 composites of Indian monsoon depressions, all of whom report Q1 and Q2 strengthening together rather than in opposition. Notably, \citet{Xavier2007}, whom \citet{Mukhopadhyay2010} cite in support of an opposite-sign Q1--Q2 relationship, themselves report the same co-varying behavior in their full-column budget (``large positive values of Q1 are accompanied by large values of Q2''; their Fig.~9); the opposite-sign result they also report is obtained from a separate diagnostic integrated only over 600--200~hPa, constructed for an unrelated purpose (defining monsoon onset from the tropospheric temperature gradient) rather than as a full-column moisture budget. Given this body of independent, monsoon-specific evidence, we interpret the co-varying Q1--Q2 relationship in our ERA-5 and GraphCast results as the physically expected one, and the discrepancy with \citet{Mukhopadhyay2010}'s Fig.~14a as most likely reflecting differences specific to that figure rather than a general feature of the reanalysis-based Q1/Q2 climatology.

Looking ahead, the availability of GraphCast probabilistic ensemble products (64 members) from 2025 onward, alongside diffusion-based probabilistic ML systems such as GenCast \citep{Price2025}, offers a significant opportunity to examine whether ensemble spread can compensate for the variance suppression identified here, and whether probabilistic skill scores (CRPS, reliability diagrams) reveal additional sources of forecast uncertainty over the monsoon domain. Such analyses are deferred to future work.

\FloatBarrier
\section{Conclusions}\label{sec:conclusions}

This study presents the first systematic climatological evaluation of Google's GraphCast AI weather prediction system for the Indian Summer Monsoon, using deterministic hindcasts initialized at 00 UTC for 2021--2024 against ERA-5 reanalysis and IMERG observational precipitation. The principal findings are as follows:

\begin{enumerate}
\item GraphCast reproduces the broad spatial organization of ISM mean-state rainfall, with an area-weighted spatial pattern correlation against IMERG of 0.79--0.80 over the Core Monsoon Zone and 0.90--0.91 over the full ISM Domain (Table~\ref{tab:verif_rain}). This pattern fidelity coexists with a small domain-averaged dry bias ($-$0.7 to $-$1.2~mm~day$^{-1}$ Core Monsoon Zone; $-$0.15 to $-$0.36~mm~day$^{-1}$ ISM Domain) and a localized dry bias at the Western Ghats orographic maximum specifically (approximately 15--20\% at +24h, amplifying to 25--30\% at +72h): two biases of the same sign operating at different spatial scales, discussed further in Section~\ref{sec:discussion}.

\item GraphCast's rainfall temporal variance is moderately, and fairly uniformly, suppressed across timescales (grid-mapped Total Var Ratio $\approx$0.63 relative to IMERG at all leads; Table~\ref{tab:psd_rain}), with band ratios ranging narrowly from 0.53--0.58 (Synoptic) to 0.65--0.66 (Annual). This is corroborated independently by reduced ISV amplitude in the spatial standard-deviation maps (Section~\ref{subsec:isv}) and by a rainfall intensity distribution that is right-shifted relative to IMERG (Section~\ref{subsec:intensity}): the moderate-heavy 95th-percentile threshold is biased dry ($-$9.4 to $-$9.8~mm~day$^{-1}$ domain-mean), while the rarest, most extreme events are even more systematically underrepresented across all four sub-regions examined. By contrast, the 10-m wind field is reproduced with near-observational spectral fidelity (Total Var Ratio 0.98--1.01; Spectral NSE 0.97--0.99) and a near-zero bias (Table~\ref{tab:verif_t2m_wind}), establishing a clear dichotomy between GraphCast's strong dynamical and moderately weaker precipitation-variance skill.

\item The Taylor diagram analysis (Fig.~\ref{fig:taylor}, Table~\ref{tab:taylor}) shows that GraphCast moderately under-disperses the spatial pattern of JJAS-mean rainfall at the Core-Monsoon-Zone scale ($\sigma=0.54$--0.60, i.e.\ 40--46\% too little spatial contrast) and is closer to, though still somewhat below, IMERG's variability at the ISM-Domain scale ($\sigma=0.86$--0.89) despite a high pattern correlation ($r=0.90$--0.91). Pixel-wise daily temporal correlation (Fig.~\ref{fig:tempcorr}; area-weighted mean 0.58--0.63) is somewhat lower than the domain-mean-time-series correlation (Table~\ref{tab:verif_rain}; 0.89--0.95), the expected relationship when substantial local, less spatially-coherent noise is superimposed on a well-captured large-scale signal that domain averaging reveals more cleanly.

\item The thermodynamic structure of the monsoon, as characterized by Q1/Q2 vertical profiles and seasonal evolution, is closely reproduced by GraphCast at all three leads, including the levels of peak heating and moisture-sink strength and the co-varying, same-sign relationship between Q1 and Q2 expected for deep cumulus convection \citep{Yanai1973,YanaiTomita1998}; this is accompanied by a modest underestimate of the DTT annual cycle amplitude at extended leads.

\item Subseasonal propagation is degraded asymmetrically: the northward-propagating BSISO signal (latitude--time Hovm\"oller, Section~\ref{subsec:propagation}) is substantially weaker and less coherent in GraphCast than in IMERG, particularly at +48h/+72h, whereas the eastward-propagating MJO/Kelvin-wave-like envelope (CIPI-regressed longitude--time structure, Section~\ref{subsec:propagation}) is preserved with markedly higher fidelity at all three lead times. This asymmetry, identified here for the first time in an AI weather model evaluation of the ISM, suggests that northward BSISO dynamics specifically, rather than subseasonal predictability in general, are the priority target for model improvement.

\item GraphCast demonstrates stronger skill at synoptic timescales (3--7 days) than at seasonal-to-intraseasonal timescales, adequately capturing the synoptic-band spectral peak and monsoon depressions/transient disturbances. The surface wind field (10~m) is well represented at all leads, with only minor degradation at +72h manifested as a slight northward displacement of the LLJ axis.
\end{enumerate}

These results provide benchmark diagnostics for the operational application of GraphCast over South Asia and highlight specific physical processes and scales, namely shallow convection, orographic-scale vs.\ domain-scale precipitation bias, rainfall intensity extremes, and northward- vs.\ eastward-propagating subseasonal modes, where targeted improvements are most needed. The convergence of three independent diagnostics (intensity distribution, spectral variance, and spatial ISV amplitude) on a single deterministic-smoothing interpretation for the rainfall biases, contrasted with the high fidelity of the dynamical wind field, is in our view the most actionable finding of this study for the AI weather model development community. They also motivate systematic investigation of the upcoming probabilistic GraphCast ensemble products to assess whether ensemble-based prediction can recover the variance and propagation skill found lacking in the deterministic output evaluated here.

\backmatter

\bmhead{Acknowledgments}
GraphCast forecast data were obtained from Google's GraphCast. ERA-5 data were obtained from the Copernicus Climate Change Service. IMERG Final Run precipitation data were provided by NASA.

\section*{Declarations}

\begin{itemize}
\item \textbf{Funding:} Not applicable.
\item \textbf{Conflict of interest/Competing interests:} The authors declare no conflicts of interest.
\item \textbf{Ethics approval:} Not applicable.
\item \textbf{Consent to participate:} Not applicable.
\item \textbf{Consent for publication:} Not applicable.
\item \textbf{Availability of data and materials:} ERA-5 reanalysis data are freely available from the Copernicus Climate Data Store (\url{https://cds.climate.copernicus.eu/}). IMERG Final Run Version 07 precipitation data are available from the NASA Goddard Earth Sciences Data and Information Services Center (GES DISC; \url{https://disc.gsfc.nasa.gov/}). GraphCast hindcast data were obtained from the UT-GraphCast Hindcast Dataset (1979–2024): A Global AI Forecast Archive from UT Austin for Weather and Climate Applications hosted at the World Data Center for Climate (WDCC) (\url{https://www.wdc-climate.de/ui/entry?acronym=graphcast}) \citep{Sudharsan2024}.
\item \textbf{Code availability:} Analysis code is available upon reasonable request to the corresponding author.
\item \textbf{Authors' contributions:} Somnath Luitel: Formal analysis; Methodology, Visualization; Writing -- original draft; Writing -- review \& editing; Data curation. Parthasarathi Mukhopadhyay: Conceptualization; Methodology, Writing -- review \& editing. Manmeet Singh: Formal analysis; Visualization; Writing -- review \& editing. Sandeep Juneja: Writing -- review \& editing. Saptarishi Dhanuka: Writing -- review \& editing.
\end{itemize}

\bibliography{references}

\end{document}


\title{Supplementary Material: Evaluating the Fidelity of GraphCast AI Forecasts for the Indian Summer Monsoon: A Climatological Assessment Against ERA-5 Reanalysis and IMERG Observations}

\author[1]{\fnm{Somnath} \sur{Luitel}}
\author[1]{\fnm{Manmeet} \sur{Singh}}
\author*[2]{\fnm{Parthasarathi} \sur{Mukhopadhyay}}\email{parthasarathi64@gmail.com}
\author[2]{\fnm{Sandeep} \sur{Juneja}}
\author[2]{\fnm{Saptarishi} \sur{Dhanuka}}

\affil[1]{\orgdiv{Department of Earth, Environmental, and Atmospheric Sciences}, \orgname{Western Kentucky University}, \orgaddress{\city{Bowling Green}, \state{KY}, \country{USA}}}
\affil[2]{\orgname{Ashoka University}, \orgaddress{\city{Sonipat}, \state{Haryana}, \country{India}}}

\maketitle

\noindent This Supplementary Material accompanies the main text and provides corroborating figures referenced therein. Figure and table numbers follow the S$n$ convention (Fig.~S1, Table~S1, etc.) and are distinct from the main-text numbering.

\section*{Supplementary Figures}

\begin{figure}[h]
\centering
\includegraphics[width=0.95\textwidth]{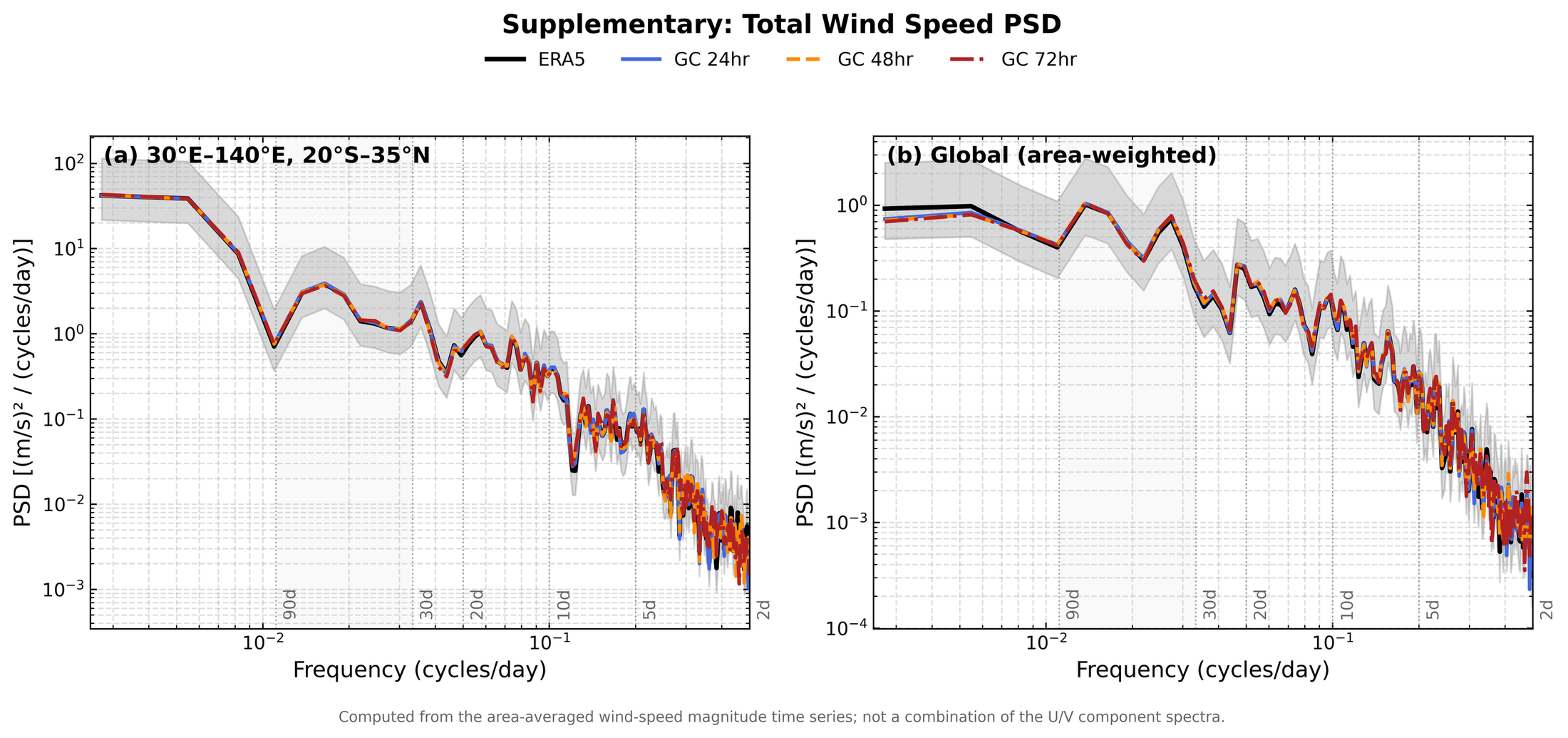}
\caption{Power spectral density of the area-weighted domain-mean 10-m wind-speed \emph{magnitude} ($\sqrt{u^2+v^2}$) time series, for the regional ISM domain (a) and globally (b), for ERA5 (black) and GraphCast +24h/+48h/+72h. Computed directly from the wind-speed-magnitude time series rather than by combining the U/V component spectra (see main-text Section~2.3); referenced in main-text Section~5.3.}
\label{sfig:windspeed_psd}
\end{figure}

\begin{figure}[h]
\centering
\includegraphics[width=0.9\textwidth]{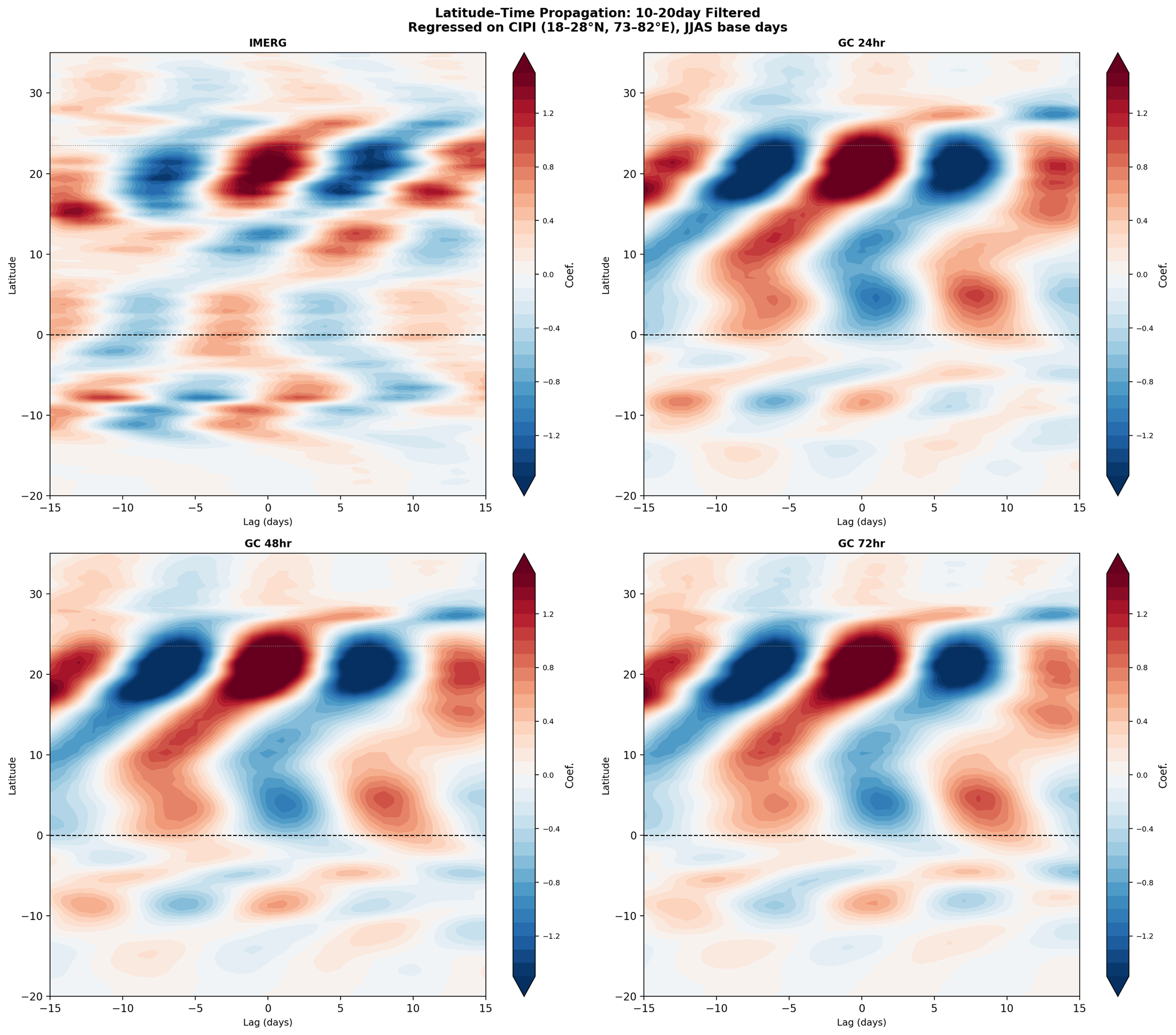}
\caption{Latitude--time Hovm\"oller diagram of 10--20 day (quasi-biweekly) bandpass-filtered daily rainfall anomalies (mm day$^{-1}$) averaged over 70\textdegree--90\textdegree E, for IMERG and GraphCast +24h/+48h/+72h. Referenced in main-text Section~5.2.}
\label{sfig:qbw_lattime}
\end{figure}

\begin{figure}[h]
\centering
\includegraphics[width=0.9\textwidth]{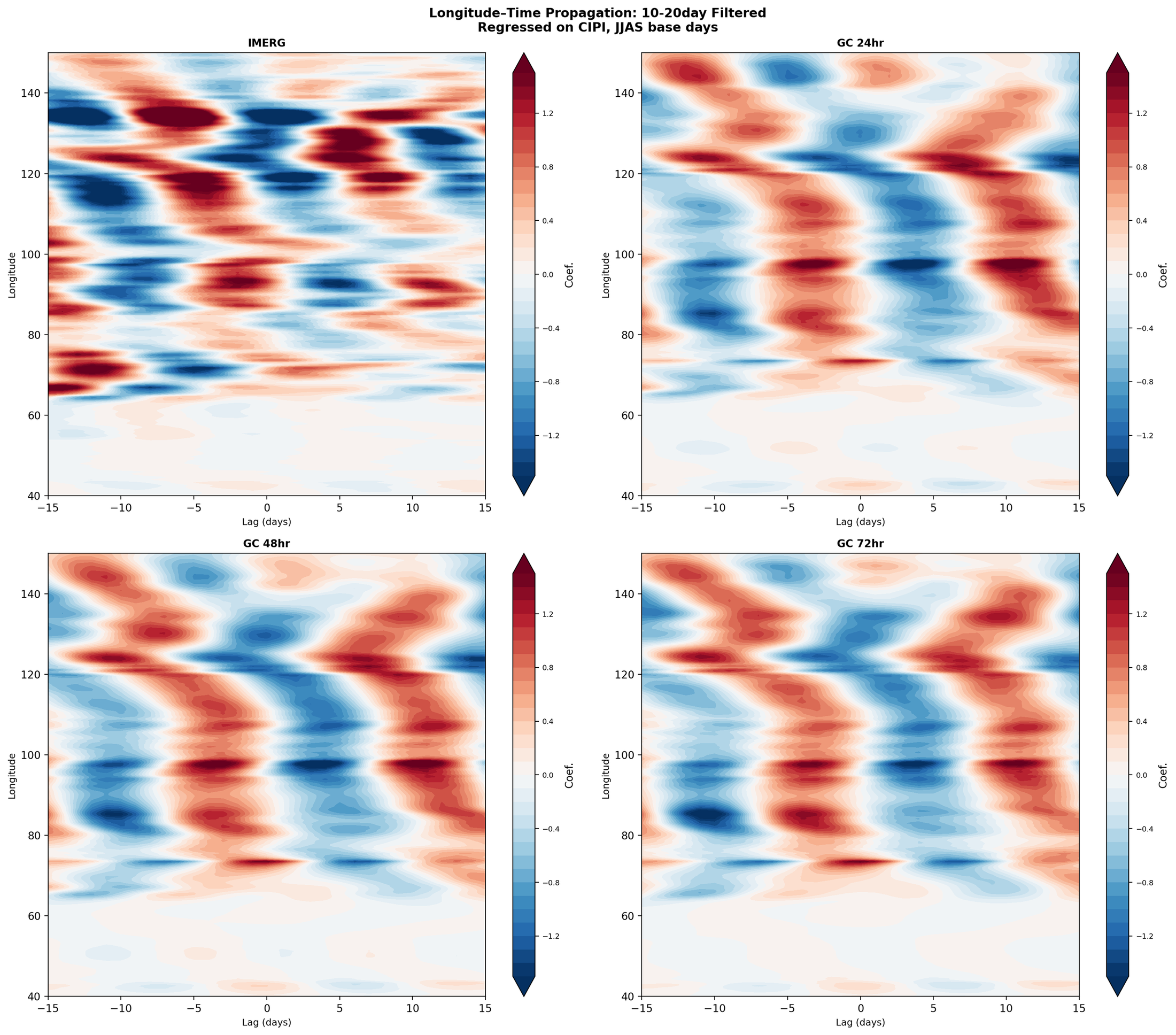}
\caption{Longitude--time regression of 10--20 day (quasi-biweekly) bandpass-filtered rainfall anomalies onto a Central Indian Precipitation Index (CIPI) base-day series, for IMERG and GraphCast +24h/+48h/+72h. Referenced in main-text Section~5.2.}
\label{sfig:qbw_lontime}
\end{figure}

\begin{figure}[h]
\centering
\includegraphics[width=0.85\textwidth]{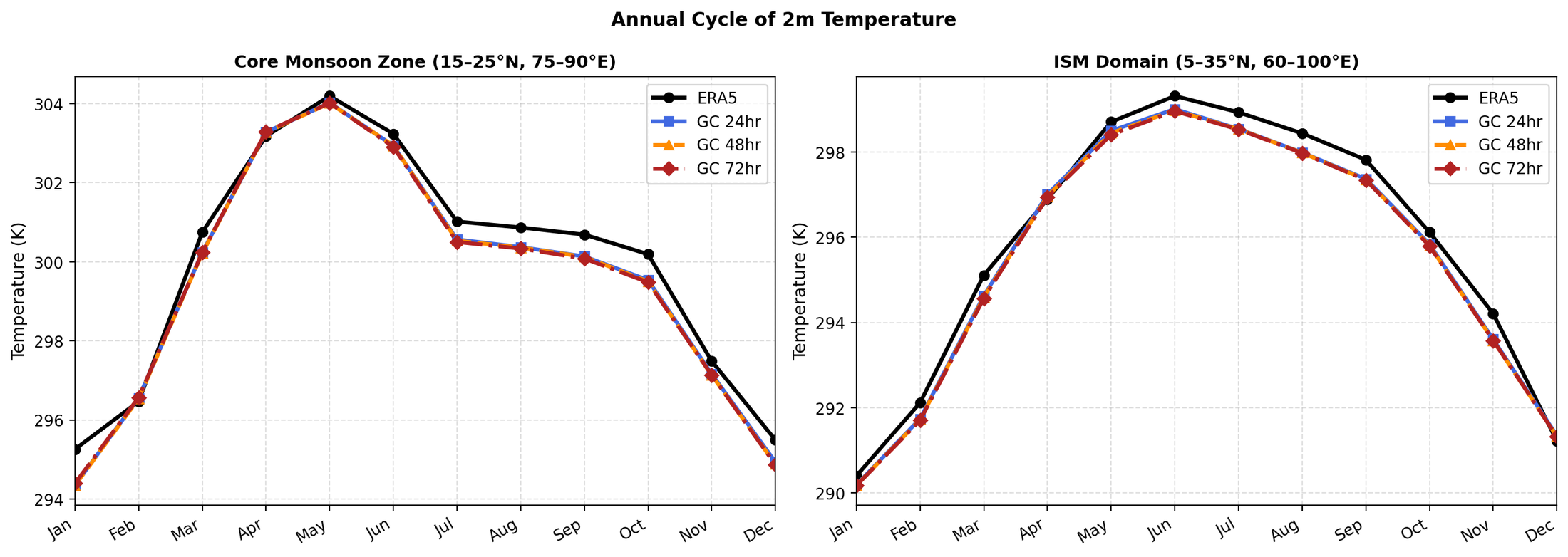}
\caption{Annual cycle of area-averaged 2-m temperature (K) over the Core Monsoon Zone (15--25\textdegree N, 75--90\textdegree E) and the ISM Domain (5--35\textdegree N, 60--100\textdegree E), for ERA5 and GraphCast +24h/+48h/+72h. Referenced in main-text Section~3.4.}
\label{sfig:t2m_annual}
\end{figure}

\begin{figure}[h]
\centering
\includegraphics[width=\textwidth]{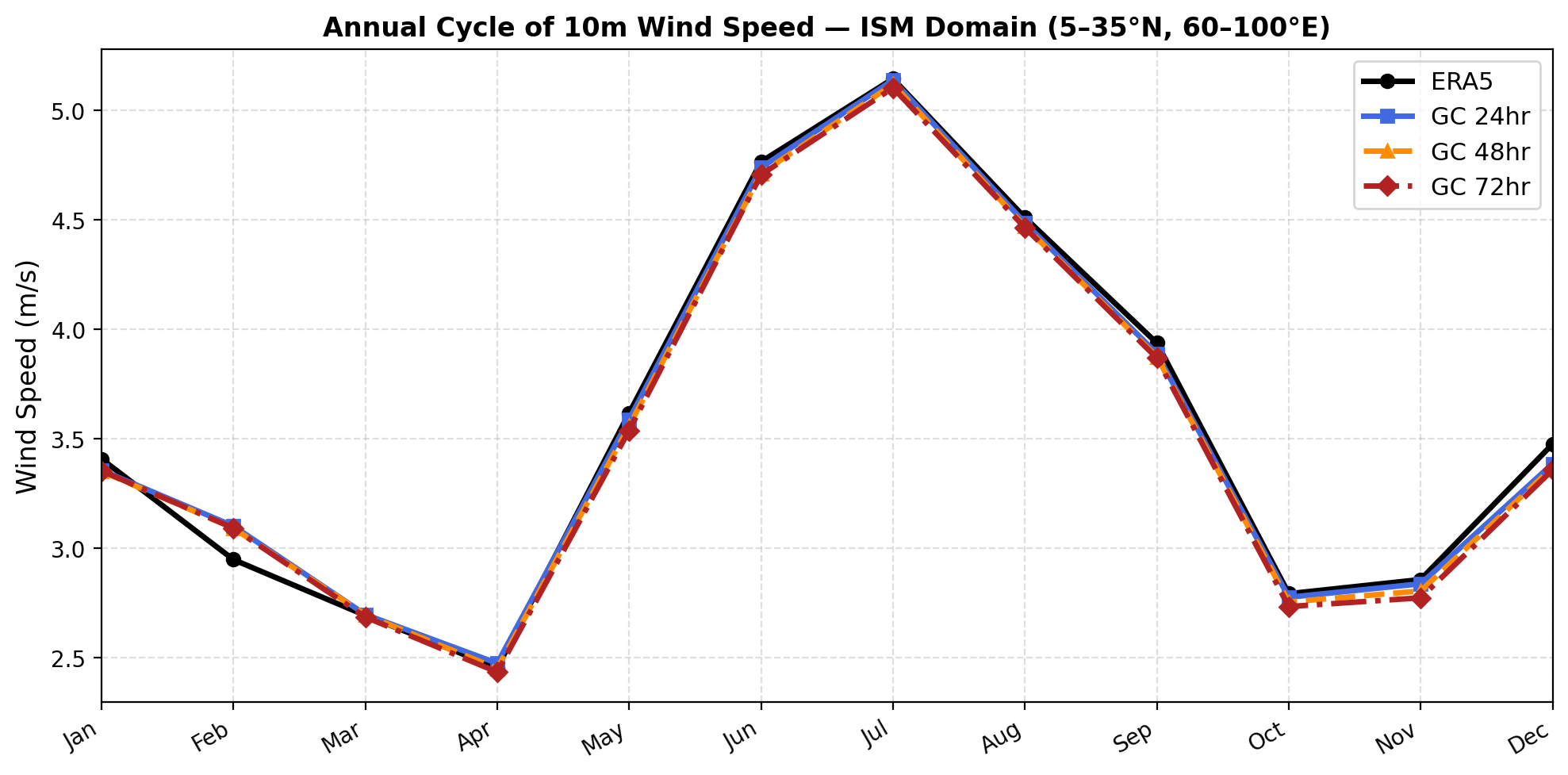}
\caption{Annual cycle of area-averaged 10-m wind speed (m s$^{-1}$) over the Core Monsoon Zone and the ISM Domain, for ERA5 and GraphCast +24h/+48h/+72h. Referenced in main-text Section~3.4.}
\label{sfig:wind_annual}
\end{figure}

\begin{figure}[h]
\centering
\includegraphics[width=0.95\textwidth]{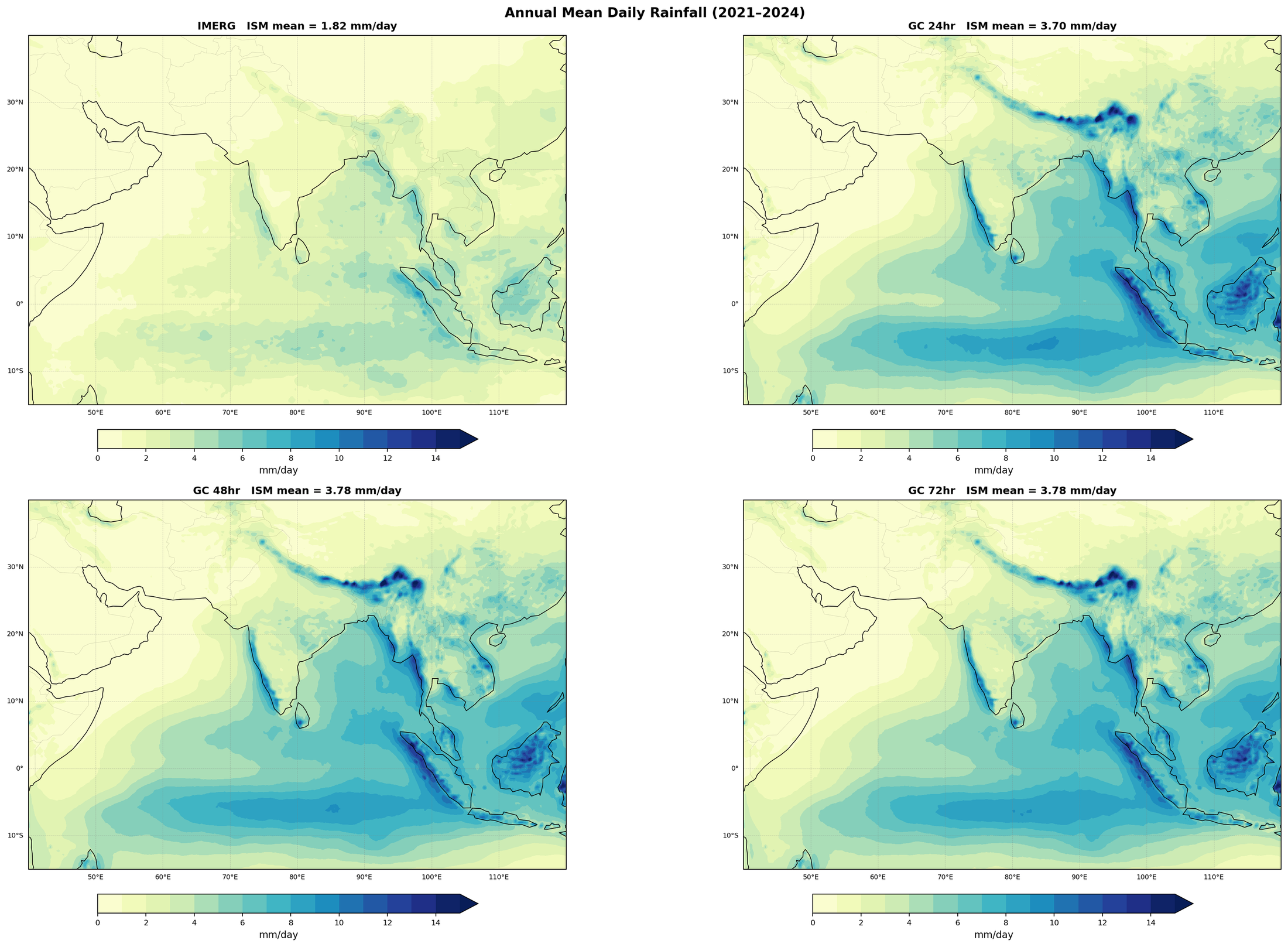}
\caption{Annual (all-season, not JJAS-only) mean daily rainfall (mm day$^{-1}$), 2021--2024, for IMERG and GraphCast +24h/+48h/+72h. Referenced in main-text Section~6 (Discussion): the narrow, intense band at $\sim$27--28\textdegree N is present in the annual mean, not only during JJAS.}
\label{sfig:annual_spatial}
\end{figure}

\begin{figure}[h]
\centering
\includegraphics[width=0.85\textwidth]{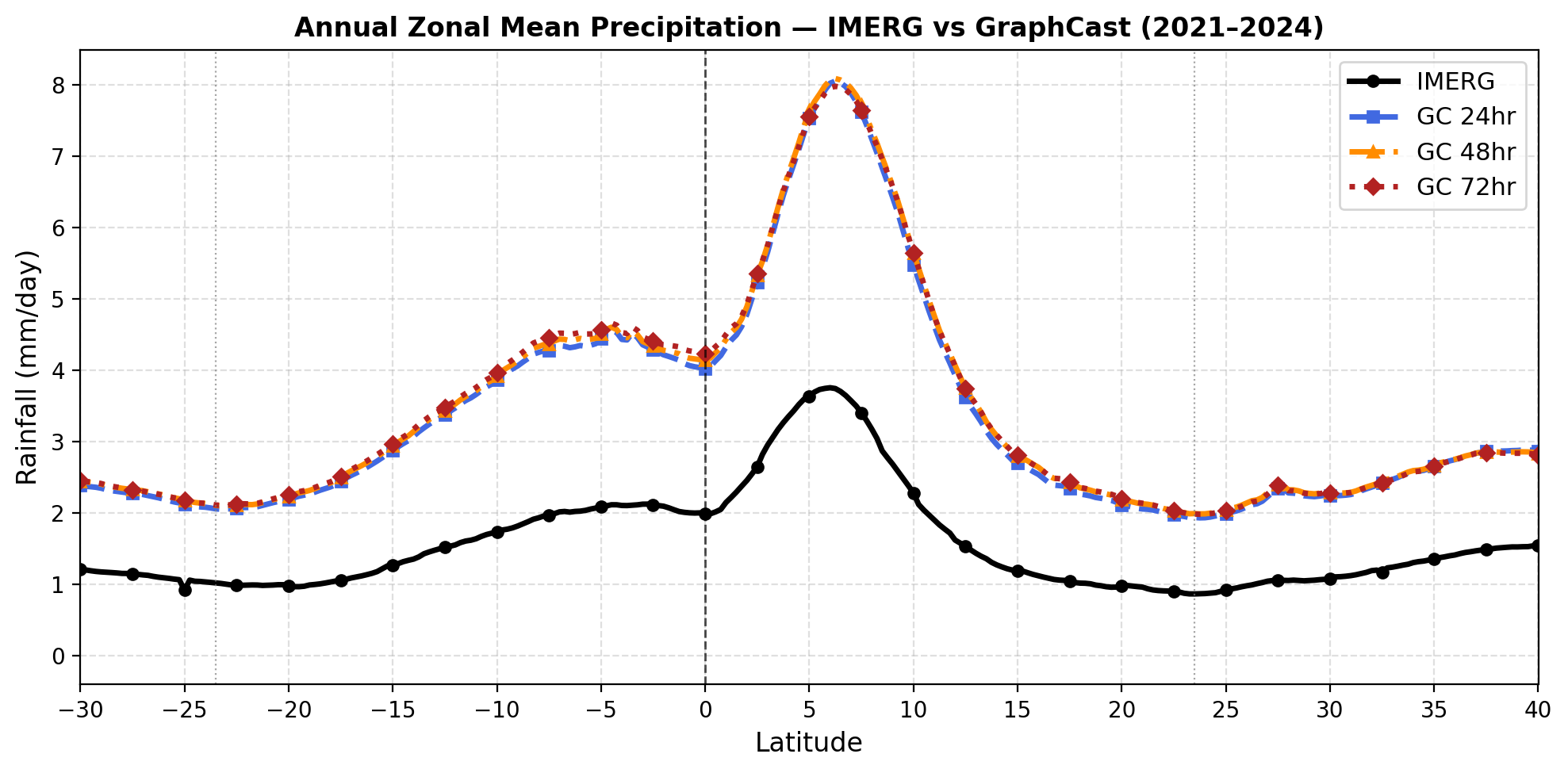}
\caption{Annual zonal-mean precipitation (mm day$^{-1}$) by latitude, IMERG vs.\ GraphCast +24h/+48h/+72h, 2021--2024. Referenced in main-text Section~6 (Discussion): note the roughly uniform additive offset between GraphCast and IMERG across nearly all latitudes, including subtropical/desert latitudes with little genuine convective uncertainty.}
\label{sfig:zonal_annual}
\end{figure}

\begin{figure}[h]
\centering
\includegraphics[width=\textwidth]{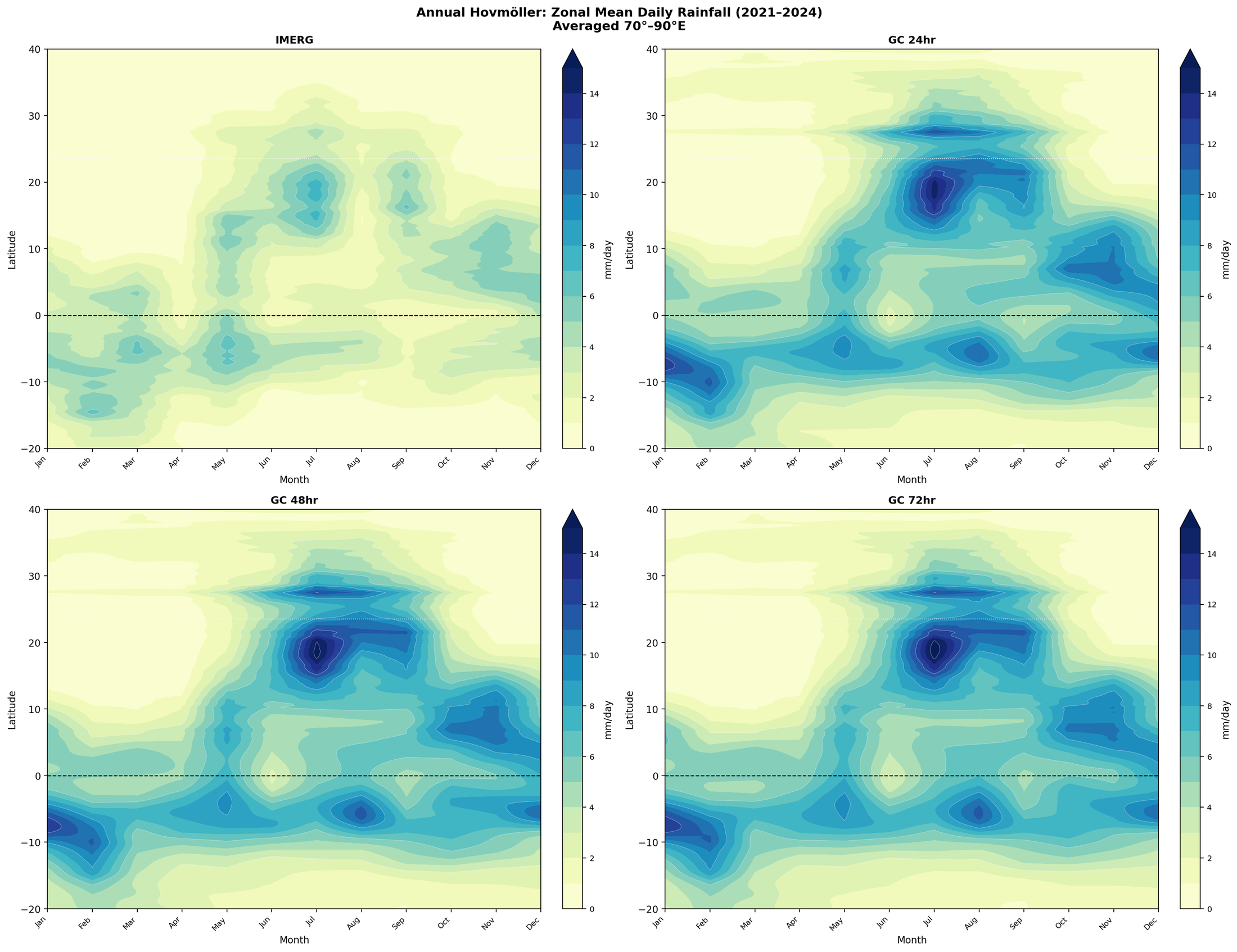}
\caption{Annual Hovm\"oller diagram of zonal-mean daily rainfall (mm day$^{-1}$, averaged 70--90\textdegree E), IMERG vs.\ GraphCast +24h/+48h/+72h, 2021--2024. Referenced in main-text Section~6 (Discussion): the $\sim$27--28\textdegree N feature is visible (faintly) outside JJAS and intensifies during the monsoon season rather than appearing only when monsoon convection is active.}
\label{sfig:annual_hovmoller}
\end{figure}

\begin{figure}[h]
\centering
\includegraphics[width=\textwidth]{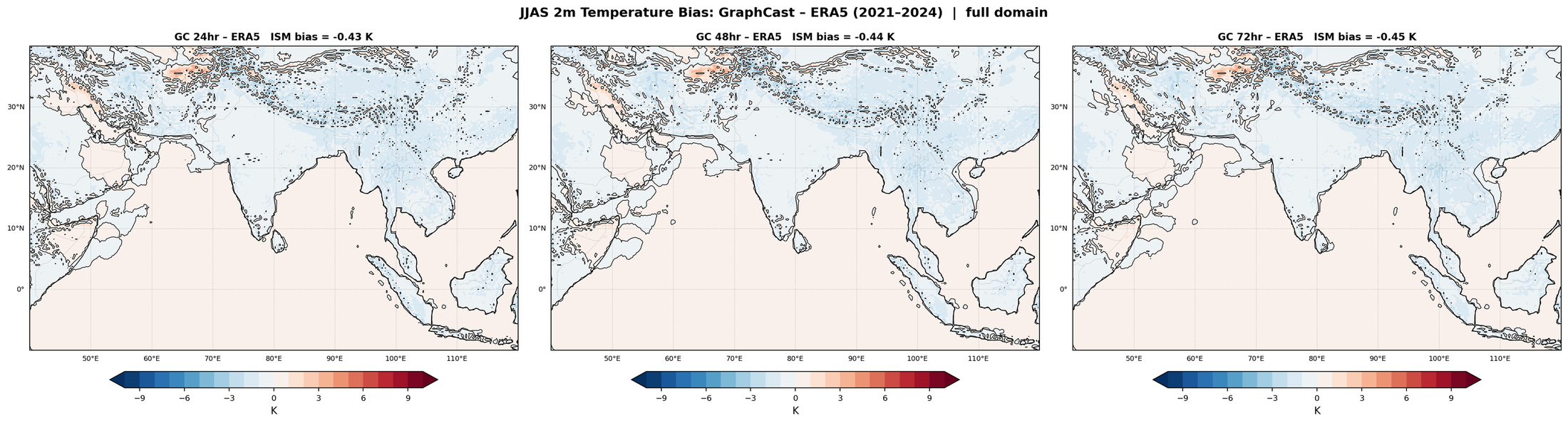}
\caption{JJAS mean 2-m temperature bias (GraphCast $-$ ERA-5, K, 2021--2024) at +24h/+48h/+72h lead over the full domain, i.e.\ without the high-terrain masking applied to the monsoon-domain 2-m temperature bias map in the main text (Section~3.4, Fig.~11a). The large near-surface cold bias over the Himalaya and Tibetan Plateau dominates the unmasked area mean ($-$0.43 to $-$0.45 K), which is why the elevated terrain is masked when reporting the monsoon-domain temperature bias in the main text.}
\label{sfig:t2m_bias_full}
\end{figure}

\begin{figure}[h]
\centering
\includegraphics[width=\textwidth]{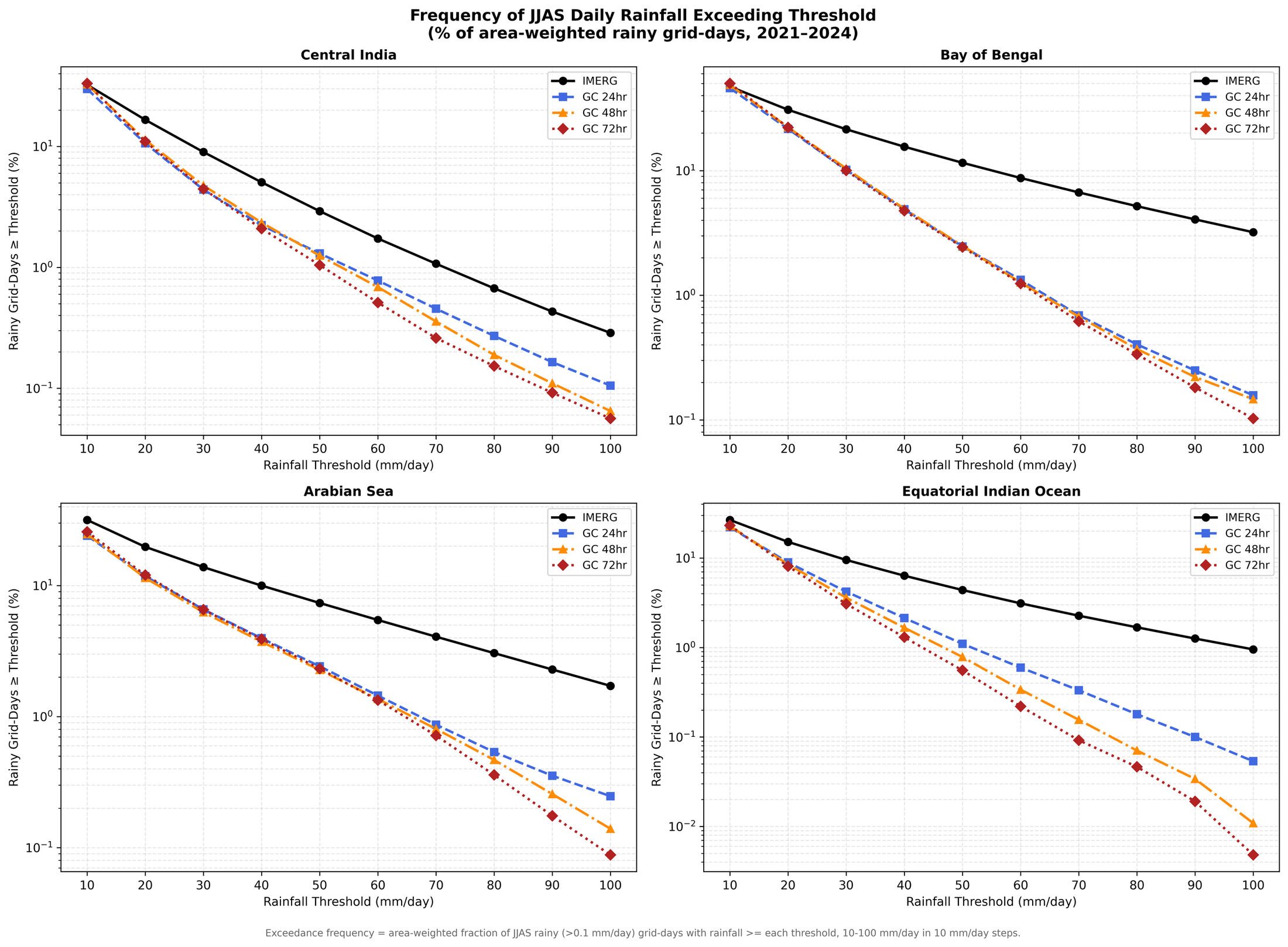}
\caption{Frequency of JJAS daily rainfall exceeding round-number thresholds (10--100~mm~day$^{-1}$ in 10~mm~day$^{-1}$ steps), expressed as the area-weighted percentage of rainy ($>$0.1~mm~day$^{-1}$) grid-days meeting or exceeding each threshold, for IMERG (black) and GraphCast +24h/+48h/+72h (blue/orange/red), over the same four regions as main-text Fig.~6 (Central India, Bay of Bengal, Arabian Sea, equatorial Indian Ocean). A complementary diagnostic to main-text Fig.~6: that figure shows which intensity bins contribute most to total rainfall \emph{volume}, while this figure shows how often rainfall of a given round-number magnitude actually occurs. Referenced in main-text Section~3.3.}
\label{sfig:exceedance_freq}
\end{figure}